# A Trustworthy Electronic Voting System for Australian Federal Elections

*Mark Eldridge*

# Contents

# 1 Introduction

Modern democratic governance is built upon a foundation of free and fair elections.

Trust in the election process requires that election systems are both fair and seen to be fair. As described by cryptographer and security expert Bruce Schneier: [1]

> "Elections serve two purposes. First, and most obvious, they are how we choose a winner. But second, and equally important, they convince the loser–and all the supporters–that he or she lost. To achieve the first purpose, the voting system must be fair and accurate. To achieve the second one, it must be *shown* to be fair and accurate."

In order to meet these objectives, a well designed election system requires both *accuracy* in its count, and *accessibility* to the average voter. The system must make it possible to prove that the result of an election is correct, and this proof must be accessible and understandable by the losing candidates and the voters who supported them.

Australia is a representative democracy with elections at a federal and state level occurring approximately every three to four years. Voting in these elections is compulsory, and candidates are selected using preferential voting systems.

Australia was the first country to introduce the secret ballot, which provides privacy to voters at polling places and preventing them from being coerced into voting for certain candidates. The voting system used for federal elections predominantly uses paper ballots and hand-counting, performed in the presence of candidate-appointed observers.

The existing system for determining election results in Australia is, for the most part, secure, accurate and understandable by the average voter. This thesis explores the design of electronic voting systems designed to achieve these same goals, while also improving on the existing system in the areas of counting speed, accuracy, tamper resistance and accessibility.

An electronic ballot can provide improved accessibility for certain voter demographics, such as the blind or vision-impaired. When using the existing paper-based voting system, these voters are required to rely on assistance to cast a vote, and are therefore not afforded a secret ballot. Electronic voting can provide these voters with the ability to vote without assistance.

An electronic voting system can also provide faster election results. An initial count can be performed in the seconds following the closure of polls, and the speed of final results is significantly improved even with the addition of paper re-counts or risk-limiting audits.

Electronic voting systems have seen limited use within Australian elections, most prominently for State Elections in Victoria (2014), New South Wales (2015) and Western Australia (2017), along with trials of electronic voting systems in federal elections in 2007.

This thesis presents an analysis of the iVote electronic voting system used for the 2017 Western Australian State Election (iVote WA), outlining a number of security risks introduced by the use of cloud-based distributed denial of service mitigation.

In addition, this thesis presents the results of a cross-sectional survey of Australian voters regarding levels of trust for three voting systems: the existing paper-based system used for Australian federal elections, the iVote WA system, and the vVote system used for the 2014 Victorian State Election.

The analysis of iVote, combined with the survey results, are used to inform a recommendation for future research and public policy regarding the use of electronic voting systems in Australian federal elections.

## Structure

The remainder of this thesis is organised as follows: Section 2 provides background information and an overview of related work; Section 3 provides an analysis of the iVote online voting system used for the WA State Election; Section 4 introduces a research plan for a cross-sectional survey of Australian voters regarding their level of trust in three different election systems; and Section 6 concludes.

# 2 Background and Related Work

This section provides necessary background information and context regarding electronic voting systems. Section 2.1 describes the formal definition of trusted and trustworthy systems. Section 2.2 provides design requirements for an well-designed voting system, definitions for electronic voting systems which will be used throughout this thesis, and an explanation of the unique security problems faced by electronic voting systems. Section 2.3 provides background to the voting system used for Australian federal elections, case studies of previous elections, and implications for the design of electronic voting systems.

Section 2.4 contains the first review of related work, with an overview of a seminal paper by Ronald Rivest. Section 2.5 continues with a paper by Stark and Wagner on Evidence-Based Elections. Section 2.6 contains case studies and related work on poll-site electronic voting systems (as defined in Section 2.2), and Section 2.7 provides the same for remote online voting systems.

Following this section, the reader should have a firm understanding of the necessary background and context regarding electronic voting systems.

## 2.1 Trustworthy Systems

In computer security, the concept of *trusted systems* is used to describe systems or components which are relied upon to meet a given security goal or policy [2]. For example, a network device such as a Data Diode may provide one-way transfer of data between networks of differing security classifications [3, 4]. The device is therefore a trusted system: if it does not behave as intended, the security assumptions governing it and other connected systems will be broken.

A *trustworthy system* is one where there is some level assurance that the trust placed in the system is well-founded [5]. For example, an analysis using formal methods may provide assurance that the system will behave as expected, and is therefore worthy of being trusted to meet a given security or safety goal [6].

When it comes to an election system, the concept of *trust* can also be applied using its general English definition: the ability of voters to have faith in the design and operation of the system and the election results produced by it. In order for an election system to be worthwhile, voters must trust that the result provided by the system is an accurate representation of the will of the voters.

For the purposes of this thesis, this latter usage of the word *trust* can be assumed if provided without additional qualification. Where the formal security definition of *trusted* or *trustworthy* are used, these will be identified as such.

## 2.2   Design of Election Systems

### 2.2.1   Characteristics of a Well-Designed Voting System

A well-designed voting system should provide the following four features [7, 8]:

Accuracy. The result of an election should accurately reflect the intent of the participating voters, and should be robust against errors introduced during the vote counting process.

Tamper Resistance. A voting system should be secure against attempts to directly or indirectly manipulate the outcome by any individual or group, including voters, candidates, or election officials.

Privacy and Anonymity. A voting system should be designed in such a way that a vote cannot be connected with the identity of the voter. If it was possible to link the identity of a voter and a completed ballot, they would be vulnerable to coercion methods such as bribery (vote buying) or threats. Anonymity protects voters from being coerced into voting for a particular candidate or party.

Accessibility. The voting system should make it possible for all eligible voters to cast a ballot without incurring undue inconvenience or hardship, regardless of age, background, or physical ability. In addition, the design of the voting system itself should be transparent, accessible and understandable by voters, such that the average voter can accept the result of the election has been fairly determined.

It is possible to define additional characteristics which are desirable in a voting system, but these four requirements will be used as a basis for the systems discussed in this thesis.

### 2.2.2   Possible Design Compromises

The design of a voting system can often result in a trade-off between the above features. For example, a common remote voting method is the *postal vote*, where a voter places a completed ballot in an envelope and mails it to a central location to be counted. This is an important measure for providing voting accessibility to expatriates, travellers, and other voters who cannot attend a physical polling place on the day of the election.

As a trade-off for this accessibility, postal voting has an adverse affect on the requirement for voter privacy and tamper resistance. The use of physical polling locations under the control of government officials (polling places) is designed to protect voter privacy and prevent the manipulation of paper ballots after they have been cast [9, 10]. Postal voting does not provide this same protection: as the marking and casting of the ballot takes place at an uncontrolled location such as the voter's home, it is therefore impossible to guarantee voter privacy, or that the ballot was not tampered with after being marked by the voter.

This balance between competing priorities has parallels with other aspects of computer security, where it is common for the design of electronic systems to encounter unavoidable trade-offs between security and usability [11, 12].

A well-designed voting system must identify where design priorities may compete, and should be transparent regarding any trade-offs made in favour of one feature over another.

### 2.2.3   Electronic Voting Systems

Electronic voting systems can be broadly divided into two categories [13]:

**Poll-Site Electronic Voting.** Voters still physically attend a polling place, but ballots are cast and counted electronically.

**Remote Online Voting.** Votes are typically cast remotely using a personal device, using the internet as a transfer method.

Examples of both categories of electronic voting system can be found in jurisdictions around the world, and a review of existing work on these systems is provided in Sections 2.6 and 2.7.

### 2.2.4   Voting: A Unique Security Problem

The requirements of a democratic election present a unique security and privacy problem [14]. Voters must be identified before voting, in order to ensure that only eligible voters can cast a ballot, and can do so only once. Simultaneously, voters must remain anonymous both during and after the election—if it were possible to link the identity of a voter with a completed ballot, it would be possible to bribe or coerce voters into voting for particular candidates.

The design of voting systems has evolved over centuries to incorporate security features that are now often taken for granted in modern elections: government-controlled polling places to provide protection against voter coercion [7]; the *secret* *ballot* to protect voter privacy [9]; candidate-appointed scrutineers to oversee voting and vote counting [15]; and independent bodies responsible for drawing electoral boundaries and managing voter registration [16].

Electronic voting is frequently seen as a logical next step for the design and execution of democratic elections, and it is often taken as given that the use of electronic systems will increase the speed and accuracy of vote counting [17, 18]. However, reliance on electronic systems (and the software which governs their operation) introduces additional complexity and risk to voting systems [19, 20, 21]. This issue is discussed in greater detail in Section 2.4.

Of the two broad categories of electronic voting systems, remote online voting receives a significant amount of attention. The problem of remote online voting is often seen by observers as equivalent in complexity to e-commerce or online banking [17, 18]. However, the security problems faced by these systems are not comparable to those faced by a voting system, primarily due to the requirement that the identity of voters remain anonymous, even from the voting system itself [22, 23, 24, 14].

The following foreword by Josh Benaloh from Microsoft Research succinctly summarises the unique security problems faced by elections [7]:

> One of the most difficult aspects of elections is their extraordinary trust and privacy requirements. Most security systems protect insiders from outsiders. When a credit card is transmitted to make an online purchase, the principal threats are from third-parties who might wish to compromise the privacy or the integrity of the transaction. In contrast, every participant in an election system is potentially malicious [...]
>
> Election officials should not know how individuals voted, and voters should not be able to show others how they voted. Indeed, the exceptional privacy requirement that voters be unable to reveal their votes—even if they desire to do so—makes voting especially difficult.

During the 2013 Australian federal election, a loss of almost 1,400 Western Australian Senate ballots resulted in a Special Election being held several months after polling day. As a result of this, the Federal Joint Standing Committee on Electoral Matters produced a number of reports regarding the conduct of the election.

In a foreword for a report assessing electronic voting options for future elections, the chair of the Committee, MP Tony Smith, noted [25]:

> After hearing from a range of experts, and surveying the international electoral landscapes it is clear to me that Australia is not in a position to

> introduce any large-scale system of electronic voting in the near future without catastrophically compromising our electoral integrity. [...]
>
> Given we complete so many transactions online, I am often asked why voting should be any different. My answer to that is that voting once every three years to determine our democratic destiny is not an everyday transaction.
>
> Not only do we have the right to a ballot; we have rightly enshrined within our system the right to a secret vote. Voting at a booth in a polling place guarantees this; voting over the internet threatens this.

When discussing the security and privacy issues associated with electronic voting systems, it is important to distinguish between poll-site electronic voting systems and remote online voting systems. The former retains the benefits of controlled polling places for casting ballots, while the latter suffers from fundamental privacy issues which are common to all remote voting methods [13]. These issues are discussed in greater detail in Sections 2.6 and 2.7.

### 2.2.5 Voter Trust and and the Perceived Legitimacy of Elections

A fundamental requirement of any voting system is to produce the correct outcome, such that the result of the election reflects the will of the voters who participated. In addition, the voting system must sufficiently convince voters that the result of the election did indeed reflect their will [1, 14].

Producing the correct outcome, therefore, can be considered a *necessary but not sufficient* condition for a well designed voting system. In order to properly convince the electorate that the election was legitimate, the system must be designed in such a way that voters can be satisfied their votes were received and counted correctly.

This aspect of voting systems—the need for voters to trust that the outcome was fairly determined—has gained increased prominence in the wake of the 2016 Presidential Elections in the United States [26, 27, 28], and the 2017 General Election in Kenya [29, 30]. In France and the Netherlands, use of electronic voting systems has been cancelled due to a perceived risk of attack by outside parties [31, 32].

In the Australian context, a substantial minority of Australians express doubts about the integrity of the electoral process and outcome [33], an issue which may threaten the perceived legitimacy of future elections. This issue is explored in greater detail in Section 2.3.5.

**Voting Systems as Critical Infrastructure**

On 6 January 2017, the U.S. Department of Homeland Security announced that U.S. election infrastructure would be designated as critical infrastructure, placing election systems in the same category as the energy sector and nuclear facilities [34, 35]. Within the international community, it is widely accepted that no nation-state should attack another country's critical infrastructure during peacetime [36].

This announcement followed recommendations for a change in classification from a number of prominent security experts including Shackelford et al., who noted that attacks against election infrastructure and voting systems do not need to successfully manipulate election results in order to be effective [21]:

> [T]here are an array of attack vectors that impact election security. And regardless of the success of hackers making use of these vulnerabilities, depending on the motivation of those involved, trust in the results may be undermined. Simply put, the attacker might not care who wins; the losing side believing that the election was stolen from them may be equally, if not more, valuable.

In Australia, security researchers including Roland Wen, Vanessa Teague, and Richard Buckland have recommended that electronic voting systems be required to meet higher standards for security and robustness than are required of commercial information technology systems. In their submission to the Inquiry into the 2010 federal election, they make the following recommendation [37]:

> E-election systems must be developed using best practices for mission critical systems rather than standard practices for commercial IT systems.

Systems designated as mission-critical are subject to significantly higher standards of design, engineering, and testing than those designed for commercial use, where cost efficiency is often a primary driver, and security is one of many secondary considerations [5, 6].

In recent years, the prevalence of highly publicised computer security breaches has reinforced the difficulty of protecting publicly-accessible information systems from sophisticated attackers. In addition, the nature of democratic elections and their impact on global geopolitics means an increased likelihood that an attack, if one occurs, will be instigated by a nation state attacker.

Given these issues, the accuracy and security of voting systems and their perception by voters is a topic of increasing relevance.

## 2.3   The Australian Federal Election System

### 2.3.1   Voting Systems of the Parliament of Australia

Australia is a parliamentary democracy with a bicameral parliament, consisting of the House of Representatives (the lower house) and the Senate (the upper house). The House of Representatives consists of 150 members elected from single-member electorates, which are designed to be approximately equal in population. The Senate consists of 76 members: 12 from each of the States and two from each Territory.

Elections in Australia have changed significantly since the first parliamentary elections were held in 1843 for the New South Wales Legislative Council. Australia was the first country to introduce the secret ballot (then known as the *Australian Ballot* due to its novelty), providing privacy to voters at polling places and preventing coercion of voters [7]. The body responsible for conducting federal elections in Australia is the Australian Electoral Commission (AEC), an independent statutory authority [16].

Within Australian parliaments, two types of voting system are currently in use [38]:

Preferential Voting  This voting system is used to elect members to the House of Representatives. Also known as Alternative Vote or Instant Runoff Voting (IRV), preferential voting allows a voter to mark candidates in order of preference. Votes are initially allocated to the first preference candidate, and then progressively allocated to subsequent candidates as earlier preferences are eliminated from the count.

Single Transferable Vote (STV)  Used for Senate elections, STV is a form of Proportional Representation, and intends to ensure that each party win seats in the Senate roughly in proportion to the percentage of voters who supported them.

Voting in Australian federal elections is compulsory for all enrolled citizens over 18 years of age, one of only 10 countries in the world to enforce such a system [39, 40]. The use of compulsory voting, combined with preferential and STV voting systems, results in a unique set of requirements for any electronic voting system designed to replace the existing paper-based system in Australia.

In particular, the voting method used for Senate elections is notable for the large ballot papers involved, due to the significant number of candidates. During the 2013 federal election, Senate ballot papers had to be limited to 1.02 metres in width by the Australian Electoral Commission, in order to ensure that voters could fit them in the polling booth [41, 42].

**Preferential Voting and Signature Attacks**

Because Australia's voting system involves preferential voting with many candidates, it is susceptible to a particular method of voter coercion known as a signature attack or "Italian attack", which targets preferential candidate selections in order to reveal the identities of voters [43, 44, 45].

To perform a signature attack, the attacker instructs a voter to cast a ballot with a particular sequence of preferences: beginning with the attacker's choice of candidate, followed by a particular ordering of candidates who have very low support rates. This is likely to result in a near-unique combination of preferences, thereby identifying the voters with a high likelihood. If the attacker is subsequently able to view completed ballots (for example, as an election scrutineer or volunteer), the attacker can identify the unique signature and thereby determine whether or not the voter followed their instructions.

The use of preferential voting with large candidate pools makes Australian elections particularly susceptible to signature attacks, and therefore has implications for the design of electronic voting systems for use in Australian elections. Although some techniques have been developed to provide protection against signature attacks [44], many existing electronic voting systems remain susceptible [46, 47, 48, 49]. This issue will be discussed in greater detail in Section 2.6.

**Voter Accessibility and the Secret Ballot**

A significant drawback of the existing paper-based voting system used for Australian federal elections is the fact that blind or visually impaired people are not afforded a secret ballot: they are required to seek help at a polling place and ask for a poll worker to complete a ballot on their behalf, or to dial a phone number and tell an operator which candidate they wish to vote for [13]. This has resulted in a number of calls for electronic voting to be implemented in order to allow blind and visually impaired persons access to a secret ballot [13, 50].

**Increasing use of Early Voting**

Early voting is defined as votes cast prior to election day, and is restricted to voters who meet one of several criteria defined by the AEC, including: being interstate or overseas, travelling, unable to leave their workplace to vote on election day, in hospital, or possessing a reasonable fear for their safety [51]. Early votes can be cast in person at an early voting centre, through a postal vote, or through a telephone voting system designed for vision-impaired voters.

During the 2013 federal election, over 3.7 million voters cast their ballots using early voting, comprising 27% of all voters. This number has increased from 15% and 18% in the 2007 and 2010 federal elections, respectively [52].

Of the 3.7 million early voting ballots cast in the 2013 federal election, 1.3 were postal votes, and 2,832 votes were cast via the AEC's existing telephone voting solution for voters with vision impairments [52]..

### 2.3.2   The 2007 Federal Election: Electronic Voting Trials

The 2007 Australian Federal Election was the first of its kind to allow any sort of electronic voting [13, 53].

Two trials were conducted: one of a poll-site electronic voting system for blind and low vision electors, and one of a remote online voting system for Australian Defence Force (ADF) personnel based overseas. The latter was conducted using the Defence Restricted Network (DRN), and was therefore not accessible from the wider internet [13, 53].

According to the Federal Joint Standing Committee on Electoral Matters (JSCEM), the outcome of the trial resulted in an average cost per vote cast of $2,597 for electronically assisted voting for blind and low vision electors, and $1,159 for remote online voting for ADF personnel via the DRN. This compared to an average cost per vote of $8.36 for ordinary paper ballots.

Based on the outcomes of these trials, the JSCEM made a recommendation that the systems not be continued at future federal elections [13, 53].

### 2.3.3   The 2013 Federal Election: Lost Senate Ballots

The 2013 Australian Federal Election took place on 7 September 2013. During the election, 1,370 completed Western Australian Senate ballot papers were misplaced prior to inclusion in the final tally, an error which was not discovered until a recount was undertaken due to the closeness of the initial result [54].

The discovery of lost ballots led to the results of the election being voided by the High Court of Australia (sitting as the Court of Disputed Returns), and a Special Election being held in Western Australia on 5 April 2014 in order obtain fresh Senate results [55, 56]. This situation was unprecedented for an Australian federal election.

After the election, the JSCEM produced a number of reports regarding the conduct of the election, with the final report delivered in April 2015, shortly after the 2015 New South Wales State Election (described in more detail in Section 2.7.5). In a foreword for the final report, the chair of the Committee, MP Tony Smith, summarised the Committee's position on the use of electronic voting as a solution to the issues encountered during the 2013 election [57]:

> The Committee analysed the benefits and risks associated with electronic electoral processes both in Australia and internationally. We concluded that to introduce large-scale electronic voting in the near future would dangerously compromise federal electoral integrity. Subsequent events at the 2015 New South Wales state election with the iVote system suggest that the Committee's cautious approach was warranted.

### 2.3.4 The 2016 Federal Election: Optical Scanning and Counting Delays

The 2016 Australian Federal Election was a double dissolution election, with both houses of parliament being dissolved and all members subject to re-election [58]. The election was also the first to use a new Senate voting process, requiring more complexity to the vote count and prompting the AEC to use optical character recognition for input of Senate ballots [59, 60].

The election resulted in an extremely close vote count, with eight days elapsing before the governing party in the House of Representatives could be determined with sufficient confidence. The final outcome in the Senate took over four weeks to be determined.

These unprecedented delays led to renewed calls for the widespread adoption of electronic voting in Australia, as a means to ensure that future elections would not be subject to the same issues [18, 17, 61, 62].

### 2.3.5 Electoral Integrity Project Study: The Australian Voter Experience

A study was conducted by Karp et al. in order to capture the views of a representative sample of Australian voters before and after the 2016 Federal Election [33]. The study was commissioned by the Australian Electoral Commission, and was conducted by members of the Electoral Integrity Project. The study involved a panel survey of voters, asking about their expeirence when voting, perceptions of the integrity and convenience of the registration and voting process, and confidence in the administration of elections.

The study highlighted three major findings. The first was that a majority of Australians expressed confidence in the electoral process, with two-thirds of respondents stating that they were 'very' or 'somewhat' satisfied with the fairness of Australian elections. A majority of respondents were satisfied with the AEC's ability to conduct elections, ensure accurate counting of ballots, and to maintain the privacy and anonyminity of voters [33].

The second finding was that a substantial minority of Australians expressed doubts about the integrity of the electoral process and outcome, with about a quarter of all respondents believing that fraud occurs 'usually' or 'always' during Australian elections. The proportion of respondents with this belief increased among supporters of minor parties, and voters who were less educated, female, or of younger generations [33].

As noted by Karp et al. , and as discussed in Section 2.2.5, this finding is significant because of the impact that voter perception of the voting process can have on the perceived legitimacy of election results [33]:

> These perceptions matter. Respondents who thought that electoral fraud was common were also significantly more likely to believe that Australian elections are conducted unfairly and that electoral laws were unfair, were less confident in the AEC and poll-workers, less trusting in the Australian parliament, political parties, and politicians, more politically cynical in their attitudes, and less satisfied with the overall performance of Australian democracy.
>
> Doubts about the integrity of elections, whether these perceptions are true or false, can undermine public faith in the legitimacy of the democratic process.

The third and final finding from the study related to specific registration and voting facilities, including potential reforms such as increased use of online voter registration and electronic voting systems [33]. On average, voters who responded to the survey did not find it difficult to vote at existing polling stations. A majority of respondents, however, believed that the existing Australian voting system is too complicated, and should be simplified. A similar proportion of respondents commented on the time taken to release election results.

Significantly, the survey identified a high degree of support for online voting, which was most prominent among younger voters and those familiar with technology. A majority of voters (61%) responded that they were confident that security and privacy of the vote can be maintained in online voting systems. There was a strong link between a respondent's familiarity with eCommerce and/or online banking systems, and subsequent confidence in online voting [33]:

> As one would expect, familiarity with online shopping and/or banking is strongly associated with higher confidence in online voting technologies.

> *As shown in Table 32 more than half of those never shopping or banking online exhibit no confidence at all in online voting, whereas only 10% show high confidence. The numbers are almost reversed for people who shop or bank online once a week.*

As discussed in Section 2.2.4, it is common for online voting systems to be seen as a comparable security challenge to internet-based eCommerce or banking systems.

## 2.4 Software Independence

The term "software independence" was introduced by Ronald Rivest with regard to voting systems, in one of the seminal works on electronic voting systems [19]. As defined by Rivest:

> *A voting system is software-independent if an (undetected) change or error its software cannot cause an undetectable change or error in an election outcome.*

It is generally accepted that in any large and complex system (a definition which describes most software) finding every error is almost impossible [19, 63]. Detecting malicious modification of software can be even more difficult, as noted by Ken Thompson in his seminal essay *Reflections on Trusting Trust* [64]. In this essay, Thompson provides an example of a maliciously modified source code compiler which will propagate a "back door" into any compiled software, making it invisible from even the most thorough source code review:

> You can't trust code that you did not totally create yourself. (Especially code from companies that employ people like me.) No amount of source-level verification or scrutiny will protect you from using untrusted code.

Because of these issues, it is essential that voting systems be designed in such a way that errors or malicious actions can be detected and corrected when they occur. Rivest summarises this with the maxim: *"Verify the outcome, not the equipment."* [63]

Rivest also introduced the concept of **weak** and **strong** software independence. Weak software independence describes systems where any change or error in the election outcome can be detected, but it cannot be corrected without repeating the voting process. Strong software independence provides for both detection and correction of errors, without having to re-run the election.

There are several examples of strongly software-independent voting systems, and these examples fall into two general categories: systems with a voter-verifiable paper record, and cryptographic voting systems.

Any system with a voter-verifiable paper record, which allows for the election result to be corrected through a recount or audit of the paper ballots, meets Rivest's definition for a strongly software-independent system [63].

Similarly, a number of modern cryptographic voting systems have been proposed which meet the requirements for strong software independence. Some such systems are known as *"End-to-End Verifiable Voting Systems"* or E2E systems, a definition which has the following requirements [63]:

- A voter may verify that their vote is *cast as intended,*

- Anyone may verify that a given vote is *collected as cast*, and

- Anyone may verify that the votes are *counted as collected.*

Some cryptographic voting systems only meet the definition for weak software independence. A voter using these systems has the ability to detect and prove that their vote was incorrectly recorded or counted, but recovery is not possible without re-running the election.

## 2.5   Evidence-Based Elections

In their 2012 paper, Stark and Wagner propose a framework for the design and use of voting systems known as *Evidence-Based Elections*, summarised by the simple equation: "Evidence = auditability + auditing" [20]

The framework proposed by Stark and Wagner consists of three parts:

1. Strongly software-independent voting systems (as discussed in Section 2.4);

2. Compliance audits to ensure that audit records are being correctly protected, ensuring the ability to conduct a full hand count of the audit trail if necessary; and

3. Risk-limiting audits to verify portions of the audit records and verify that any electronic result is sufficiently supported by the audit trail.

Stark and Wagner are critical of existing approaches in the United States for providing official certification of voting systems by election officials, and selecting voting systems based on these certifications,. They note that these approaches increase the cost of voting systems without meaningfully improving the ability of election results to be audited and verified [20].

These increased costs can manifest in jurisdictions being reluctant to purchase new voting equipment, resulting in obsolete equipment being used long past the point at which it should have been replaced. In addition, the high up-front costs of undergoing certification can limit competition and result in regulatory capture by existing providers [20].

As an alternative, Stark and Wagner suggest that requiring local election officials to conduct compliance and risk-limiting audits would be more effective in ensuring the accuracy of election outcomes, in addition to providing an incentive to use voting equipment which allows for compliance and risk-limiting audits to be conducted with a minimum of difficulty and cost [20].

## 2.6   Poll-Site Electronic Voting Systems

### 2.6.1   Definition and Considerations

Poll-site electronic voting systems maintain the use of government-controlled polling sites, but involve the use of electronic means for casting the ballot, counting votes, or both. These systems may or may not produce a 'paper trail' for later auditing of election results.

Poll-site electronic voting systems have been in use for many decades. Some systems simply provide an electronic interface for printing a paper ballot, which is then cast in the normal manner. Others systems rely on punch cards for casting a ballot, and then use electronic counting systems to speed up vote tallying [65].

Some poll-site electronic voting systems are regarded as end-to-end auditable or end-to-end voter verifiable (E2E), as described in Section 2.4. An E2E system requires strong levels of integrity, verifiability, and tamper-resistance [7]. Such systems often produce a paper ballot or receipt (a "paper trail") in combination with an electronic tally, which allows for a manual recount or audit to be conducted if required.

### 2.6.2   Direct Recording Electronic (DRE) Systems

Direct Recording Electronic (DRE) systems involve both electronic ballot casting and electronic vote counting, and early systems did not produce any form of paper trail for auditing purposes. A number of significant security flaws have been discovered in popular DRE systems [8, 66, 67, 14], and the use of such systems have been discontinued in a number of countries including the Netherlands, Ireland and Germany [7].

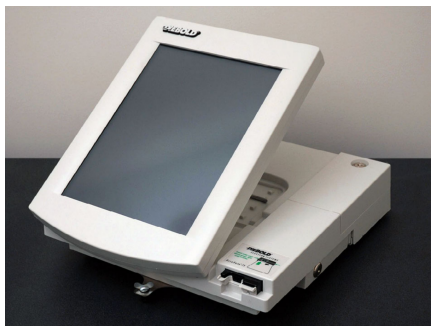

Fig. 1: The Diebold AccuVote TS Direct Recording Electronic voting machine. [68]

A major concern with the use of DRE systems is the inability to verify the correct recording and counting of votes, as there is often no "paper trail" generated in order to allow for auditing of the vote count. Because of the reliance on software to accurately record and tally ballots, and their inability to perform manual auditing, such systems do not meet the definition of software-independent systems as defined in Section 2.4.

DRE voting systems gained prominence in the United States following the 2000 Presidential Election and subsequent Bush v Gore lawsuit, which led to calls for electronic voting to be used in favour of the existing punch-card system. Diebold Election Systems was a major manufacturer of DRE voting systems during this period, with the Diebold AccuVote-TS voting machine being used by up to 10% of registered voters during the 2006 mid-term elections [68]. An example of this system can be seen in Figure 1.

Studies of the AccuVote-TS voting machine showed a number of significant security vulnerabilities, showing that an attacker could fully compromise the machine with brief physical access either to the system itself, or merely the memory cards used to transfer completed ballots [8]. Such a compromise could be crafted as a computer 'worm' to spread automatically between machines when memory cards were exchanged [68]. In addition, the design of the system meant that any malicious code had full access to audit logging, making detection impossible if the attacker was competent [8, 68].

### 2.6.3    ThreeBallot

The ThreeBallot voting system was designed by Ronald Rivest in order to illustrate the principles of E2E verifiable voting systems, without using cryptography. Instead, the ThreeBallot system relies upon voters casting **three** ballots: exactly two in favour of a candidate to support them, and exactly once in favour of a candidate to oppose them [69]. Each ballot is also marked with a unique ballot ID number, generated randomly.

After casting three ballots, the voter retains a copy of one of the cast ballots as a receipt, selected at random. This receipt can later be used to verify that the corresponding ballot was included in the count, by checking that the unique ballot ID number appears in the list of accepted ballots, and contains the correct vote. This provides a mechanism for detecting if votes are manipulated or deleted. This retained ballot cannot be used to prove how the voter voted (and therefore cannot be used to buy or coerce votes), because there is no guarantee as to which of the three votes the receipt corresponds to.

The ThreeBallot system was designed primarily for pedagogic purposes, and was not proposed as a practical voting system [69]. Indeed, the design of ThreeBallot was shown by Strauss to be vulnerable to correlation attacks if the election consists of more than a simple Yes or No question on the ballot [70, 71].

The ThreeBallot system would be difficult to apply to Australian preferential elections, as it is designed for use in first-past-the-post voting systems where a voter selects only one candidate.

### 2.6.4    Scantegrity

The Scantegrity voting system was proposed by Chaum et al. in 2008, and is designed to provide an E2E verifiable voting system without significant changes to the way voters in the United States traditionally completed ballots [46]. Scantegrity combines the cryptographic principles used by other E2E voting systems with the paper ballots and optical scan machines commonly in use in the United States, and was designed in such a way that no new equipment would be required at polling places [46].

The Scantegrity system is designed to provide an identical voter experience to existing optical scan systems, where a voter marks a specially designed ballot and feeds it into an optical scanner for central counting. The major departure from existing optical scan systems is the use of a receipt for voters to take home and verify that the vote was correctly recorded and tallied. This focus on simplicity and voter experi-ence was an intentional attempt to improve the acceptance of E2E voting systems, which had historically required expensive changes to existing voting processes [46].

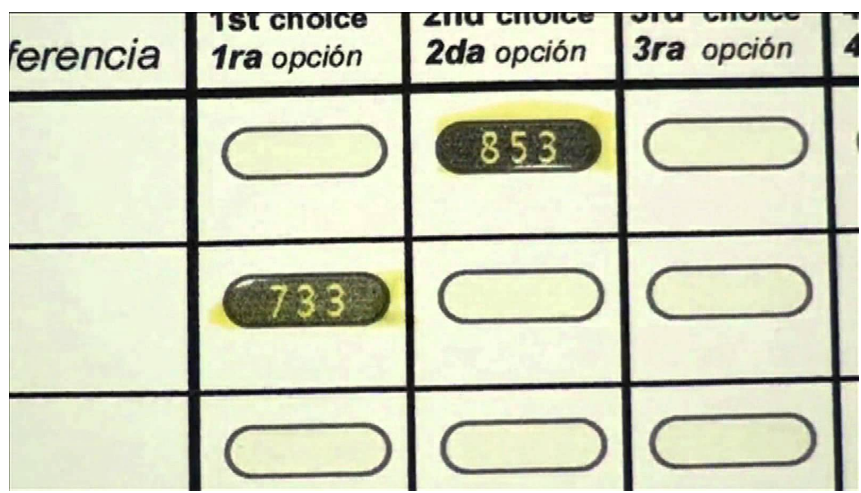

Fig. 2: An example of a Scantegrity ballot.

A later system named Scantegrity II improved on this design by using invisible ink markers, providing several benefits to voter security including immunity to subtle coercion methods such as the signature attack or "Italian attack" [46]. The Scantegrity II system was successfully used for a 2009 municipal election in Takoma Park, Maryland, marking the first time a poll site E2E verifiable voting system had been used in a binding government election [72, 73].

The Takoma Park municipal election uses a preferential Instant Runoff Voting system, allowing voters to rank candidates in order of preference. The Scantegrity implementation of this voting system involved each ballot containing a grid of cells to be marked, with each row denoting a specific candidate, and each column denoting a specific preference. Therefore, each ballot consisted of a grid of $N * N$ cells, where $N$ was the total number of candidates on the ballot. This implementation resulted in some confusion by voters, with an exit survey showing that some voters had difficulty understanding how to mark a preferential ballot [72].

This design has implications for the use of the Scantegrity design in Australian elections, where large numbers of candidates are common. As noted by Culnane et al., elections to the upper houses of Australian parliaments typically involve around 30 candidates, and a 30 by 30 grid of cells may be too complicated for an average voter [49]. Such complexity may thereby negatively impact the accessibility of the voting system.

### 2.6.5   STAR-Vote

STAR-Vote is a poll-site E2E electronic voting system which is designed provide a DRE-style interface to the voter, but is software independent, with a voter-verifiable audit trail and full end-to-end verifiability [74].

STAR-Vote was the result of collaboration between the Travis County (Austin), Texas elections office, and a number of high profile voting security experts. The STAR-Vote system is designed such that no part of the system is trusted. The system is intended to be robust to unintentional errors or intentional tampering by any party or device in the system, including voters, election officials, and the electronic devices used to enter, print, and scan ballots [74].

Within a polling place, STAR-Vote uses a local wired network which is not connected to the Internet or to any other external networks. The system uses distributed hash chains to provide redundancy between records produced by machines in the system, and uses homomorphic encryption in order to allow the tallying of votes without the counting system being able to see individual ballots, combined with an implementation of the "Benaloh challenge" which allows voters to test the system with a completed ballot prior to casting their final vote, in order to verify that the system is working correctly [74, 75].

The STAR-Vote system is designed to be used with risk-limiting audits (as described in Section 2.5), in order to ensure that any errors or tampering can be both detected and corrected. STAR-Vote therefore fulfils the requirements for Rivest's definition of strong software independence, as discussed in Section 2.4.

In 2017, Pereira and Wallach demonstrated improvements to the processes surrounding use of the STAR-Vote system, which would mitigate classes of attack known as "clash attacks" against the end-to-end verifiability of the system [76].

The design of STAR-Vote is not currently usable for Australian elections, as it is dependent on cryptographic techniques such as homomorphic tallying and zero knowledge proofs which are designed for ballots where only one selection is made—all candidates on a given ballot are either a one (selected) or a zero (not selected). This would not be appropriate for the preferential and STV voting systems used in Australian elections, as discussed in Section 2.3.1.

### 2.6.6   Prêt à Voter

Prêt à Voter is an E2E verifiable voting system which employs a split ballot paper and randomised candidate lists for each ballot. Only one half of the ballot contains the printed randomised candidate list, with this half being destroyed before a vote is cast [77, 78]. The remainder of the ballot contains only the voter markings for the specific candidate ordering they were provided, along with a cryptographic value which can be used to recreate the original candidate ordering and therefore determine which candidates were voted for.

Because of this cryptographic protection of candidate ordering, a voter can retain a copy of the marked portion of the ballot, which does not show candidate names and therefore cannot be trivially used in vote buying or coercion attacks. However, this method of masking candidates is susceptible to signature attacks or "Italian attacks" as described in Section 2.3.1 [46, 47, 48].

The Prêt à Voter system has been modified over the years to include support for additional election methods, including the preferential (IRV) and Single Transferable Vote (STV) systems used for Australian Elections [48, 79].

### 2.6.7 vVote: The 2014 Victorian State Election

For the 2014 Victorian State Election, a modified version of Prêt à Voter was employed for the *vVote* poll-site electronic voting system [80, 81, 82, 49, 79, 83]. The system accepted a total of 1121 votes, encompassing supervised polling places inside Victoria and at the Australian High Commission in London.

The vVote system employs a split ballot, an example of which can be seen in Figure 3. The example shows the two portions of the ballot: the left-hand side displays the randomly permuted candidate list, and the right-hand side displays the voter's preferences for those candidates, entered from 1 - 7. During the election, the voter obtains the candidate list from election officials, scans the QR code in a provided touch-screen system, and then enters their desired preferences. The voter then obtains a "preference receipt" which lists the voter's numbered preferences, but not the candidate names [82].

The security of the vVote system requires that the left-hand side of the ballot be removed and destroyed immediately after the vote has been cast, in order to ensure that no party can reconstruct the voter's completed vote using the candidate list [49]. Both sides contain a unique serial number, which allows the voter to later verify that the vote was cast and recorded correctly, by reproducing the specific permutation of candidates that the voter saw on the left-hand side of the ballot.

After receiving a blank ballot but before completing it, a voter has the ability to perform a "Benaloh challenge" to ensure that the ballot is correctly formed, *i.e.* that the randomly permuted candidate list is correctly stored in the public ledger which holds all of the encrypted candidate lists for later verification [49]. If the ballot is checked in this manner, it must then be destroyed and replaced with a new ballot.

Fig. 3: An example of a vVote ballot. [49]

Following the 2014 Victorian State Election, the Electoral Matters Committee conducted an *Inquiry into electronic voting* which produced a number of recommendations based on the experience using vVote and other electronic voting systems in other Australian jurisdictions, particularly the iVote remote online voting system used in New South Wales and Western Australia [83]. These recommendations included in-principle support for a remote online voting targeted at a limited category of electors, including electors with vision impairments and electors who are unable to vote in-person on election day and would otherwise have to use a postal vote.

The Committee also recommended the establishment of an Electronic Voting Board, including experts in electronic voting and representatives from registered political parties, to oversee security arrangements for any remote voting system used in subsequent elections [83].

In a foreword to the Committee's report, Chair Louise Asher MP made the following observation [83]:

> The Committee examined the different electronic options currently available across a number of Australian states and also examined a number of international case studies. With the exception of Estonia, many countries in the world are now moving away from electronic voting in the light of significant security concerns.

In a submission to the Electoral Matters Committee inquiry, Ronald Wen and Richard Buckland made a number of observations regarding perceived flaws in the vVote system as used for the 2014 election, including the complexity of the system, the risk of compromises to voter privacy due to modifications made to the original Prêt à Voter design, and an over-reliance on verifiability rather than prevention of failures [84].

## 2.7 Remote Online Voting Systems

### 2.7.1 Definition and Considerations

Remote online voting systems involve voting performed remotely via electronic means, typically using the internet as a transport mechanism. This process is similar to the 'postal vote' offered in many jurisdictions for voters who cannot attend a polling place on election day.

In most remote online voting systems, voters can use personal electronic devices such as desktop PCs or smartphones in order to cast a ballot. Some systems also require an application to be downloaded by the user for use when voting, while others are web-based and use standard web browsers to conduct the voting process.

**Remote Voting**

As a remote voting system, where voters cast a ballot in an uncontrolled location, this method of voting suffers from an inherent problem: it is difficult or impossible to prove that a completed ballot was cast without coercion [7]. This problem is inherent to any voting method where a voter can cast a ballot remotely, rather than at a polling place. For example, a voter casting a ballot at home can be threatened, bribed or otherwise coerced into voting for a particular candidate by an acquaintance or family member, without this coercion being easily detectable by authorities.

These security concerns were summarised by Ronald Rivest in 2001, in the context of elections in the United States [85]:

> Our largest security problem is likely to be absentee ballots. Absentee voting has increased dramatically over the past decade. Indeed, some states, such as Oregon, vote entirely by mail. Remote electronic voting can be viewed as a version of absentee voting.
>
> In my opinion, however, by allowing such an increase in absentee voting we have sacrificed too much security for the sake of voter convenience. While voters should certainly be allowed to vote by absentee ballot in cases of need, allowing voting by absentee ballot merely for convenience seems wrong-headed. I would prefer seeing "Voting Day" instituted as a national holiday to seeing the widespread adoption of unsupervised absentee or remote electronic voting.

As discussed in Section 2.3.1, an increasing number of voters in Australian elections are opting to vote using early voting methods, which encompasses remote postal voting. The number of voters using early voting has increased from 15% of ballots in 2007, to 27% in 2013.

**Security of Voter Devices**

Remote *electronic* voting raises additional issues above those posed by any remote voting system. These issues are due to the personal electronic devices owned by voters and used to cast ballots [86].

Modern cyber attacks have reinforced the difficulty of preventing general-purpose electronic devices from being compromised by sophisticated attackers, even for organisations with significant resources. In addition, the nature of elections and their impact on international politics suggests that any threat to a voting system is more likely to be a sophisticated attacker such as a nation state than would be true for other electronic systems [87, 26, 88].

This is a risk shared by all remote online voting systems: it is almost impossible to ensure the security and provenance of the devices used to cast ballots, and devices which have been compromised by an attacker may be used to view or modify ballots without any immediate indication to the voter [86]. In addition, voters may have no way to tell whether a vote was cast correctly if the systems they are using for voter verification have also been compromised.

**Voter Participation**

A common argument for the use of online voting systems is that it may increase voter participation, particularly among younger voters [89, 90, 91].

This perceived benefit of online voting has been disputed, with Oostveen and Besselaar concluding in 2004 that increased use of electronic voting systems will not influence turnout, and Bochsler finding in 2010 that the use of online voting in Estonia mostly substituted for existing votes at physical polling places, rather than increasing turnout [92, 93].

### 2.7.2   The Estonian I-Voting System

Estonia was the first country in the world to utilise widespread online voting for binding elections, in 2005. As of 2011, the number of voters utilising this voting method has increased 14-fold to reach 140,000 votes (a quarter of all votes cast), from less than 10,000 nationally in the 2005 local elections [94, 95].

There were three "Pillars of Success" for online voting in Estonia, according to Vinkel [94]:

Open Receptive Society   Estonia is a small country by both population and landmass, has widespread government use of technology services, and high levels of public trust in technology and government use of these services.

Secure Remote e-Authentication (eID)   The Estonian electronic identity card is mandatory for all citizens and permanent residents over the age of 15. The card includes two RSA public/private key pairs with corresponding user PINs, used for secure authentication with government services and signing digital documents.

Compliance with Electoral Principles   Legislation was designed to ensure that the principles of voting secrecy were met through the I-Voting system. This includes the feature that a voter may cast multiple ballots electronically, with only the most recent taking precedence, or a paper ballot if one is cast on voting day. This is intended as a safeguard against voter coercion.

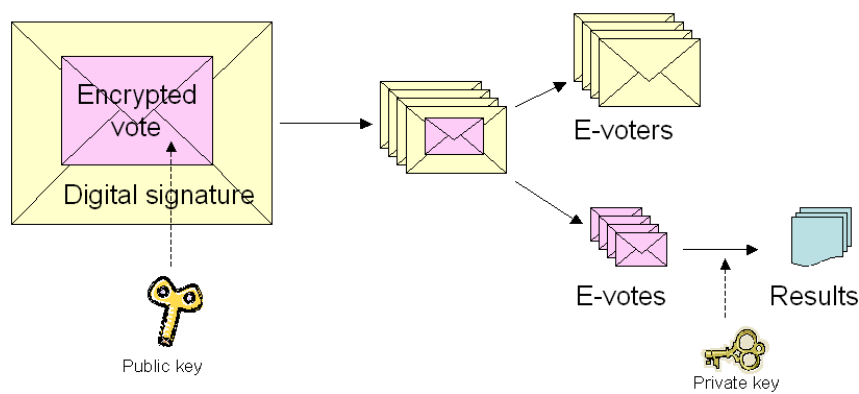

Fig. 4: The two-envelope method implemented using public key encryption. [94]

The Estonian I-Voting system can be most easily understood as the digital equivalent of the "two-envelope" method employed in many jurisdictions (including Australia) for postal voting [94]. An graphical description of this system can be seen in Figure 4.

When using this method for a postal vote, a voter inserts a completed ballot inside of an envelope free of any markings which may identify the voter. Next, the voter places this unmarked envelope inside of another envelope, this time including an identifier showing who they are and allowing the electoral authorities to check them against the voter roll. When the envelope is received, the outer envelope (containing the identification of the voter) is removed, and used to verify that the voter is eligible to vote. The second envelope remains sealed while being transferred to a separate location for counting. This intends to ensure that the voter's identity cannot be connected with the completed ballot at any point during and after the election.

The Estonian online voting system implements this process digitally, using asymmetric encryption [94]. The voter's filled ballot is first encrypted using the public key of the Estonian National Electoral Committee, then digitally signed using the private key stored on the voter's electronic identification card. Upon receipt, the signature is verified as belonging to an eligible voter, and the encrypted ballot is then sent to be counted.

Only during the counting process is the National Electoral Committee private key used to decrypt the ballots and allow for the vote to be tallied. By ensuring that ballots are kept encrypted until counting and with correct key management, this mechanism aims to ensure that a voter cannot be connected with a completed ballot.

During the Estonian local elections in 2013, a group of security researchers attended as observers, performing interviews with election staff, code reviews of the I-Voting software, and tests of the I-Voting software on a laboratory system [86].

The researchers identified a number of security flaws, including operational security issues which could allow for compromise of the counting servers used to tally votes, the risk of voter devices being compromised, and methods for circumventing the I-Voting voter verification system [86]. In addition, the researchers specifically identified the risk that a state actor such as Russia might have the ability to compromise the voting system [86]:

> We conclude that there are multiple ways that state-level attackers, sophisticated online criminals, or dishonest insiders could successfully attack the Estonian I-voting system. Such an attacker could plausibly change votes, compromise the secret ballot, disrupt elections, or cast doubt on the integrity of results. These problems are difficult to mitigate, because they stem from basic architectural choices and fundamental limitations on the security and transparency that can be provided by procedural controls. For these reasons, we recommend that Estonia discontinue the I-voting system.

In October 2017, it was reported that 750,000 Estonian digital ID cards contained a vulnerability allowing practical factorisation of the RSA private keys used to encrypt and sign votes for the Estonian online voting system [96, 97]. This vulnerability was due to a flaw in an RSA algorithm for generating RSA key pairs developed by Infineon, and discovered by Nemec et al. [98]. News of this vulnerability led Estonian officials to require all card holders to refresh their ID cards with new keys [99, 100].

### 2.7.3   The Helios Online Voting System

The Helios system was first described by Adida in 2008, and was the first publicly available implementation of an online open-audit voting system allowing for any observer to cryptographically verify that election results are sound [101]. The system is kept intentionally simpler than other cryptographic voting protocols, and provides two properties:

Ballot casting assurance. Each voter can obtain personal assurance that a vote was correctly captured.

Universal verifiability. Any observer can verify that all captured votes were properly tallied.

Adida stresses that Helios is designed for elections where trustworthy, secret ballot elections are required but coercion is not a serious concern, and that the system is not designed to be used for high-stakes elections due to the increased risk of coercion and voter devices being compromised [101].

The difficulty of conducting secure online voting for high-stakes elections is also noted on the frequently asked questions page for Helios itself [102]:

> Online elections are appropriate when one does not expect a large attempt at defrauding or coercing voters. For some elections, notably US Federal and State elections, the stakes are too high, and we recommend against capturing votes over the Internet. This has nothing to do with Helios itself: we just don't trust that people's home computers are secure enough to withstand significant attacks.

In 2016, Chang-Fong and Essex demonstrated several vulnerabilities in the design of Helios, including methods allowing a malicious election official to rig an election by providing intentionally weak cryptographic parameters during the configuration of the Helios server [103]. This allows the attacker to declare an arbitrary election result while still issuing a cryptographic proof showing that the results were correctly tallied.

Other vulnerabilities identified by Chang-Fong and Essex included the ability for a malicious voter to cast a "poisoned ballot" to preventing the voting tally from being decrypted at the completion of the election, and a cross-site scripting (XSS) vulnerability which would allow an attacker to cast a ballot on behalf of the voter through use of a dummy election [103].

In their discussion, Chang-Fong and Essex noted that existing approaches for E2E verifiable elections focus on verifying elections rather than the software or voting machines used to cast the ballots, and that errors in the verification design or implementation itself may result in a worse outcome, due to the false sense of security that this may introduce [103].

### 2.7.4   Pretty Good Democracy

Pretty Good Democracy (PGD) is a remote online voting system proposed by Ryan & Teague in 2009 [104].

The system is based on *code voting*, where voters are provided—by a separate channel such as the postal system—codes which map candidate choices for selection using an online voting platform. In such a system, the voter is in effect performing encryption of a ballot manually using a pre-shared code book, which makes it significantly more difficult for malicious software on a personal device to interfere with the voting process or change the voter's intention [105].

PGD adds a simple enhancement to existing code voting methods in order to provide a voter with a degree of verification that a vote was accepted, by matching an acknowledgement sent by the server to one provided by the code sheet [104]. The mechanism does not provide a coercer with information on how the voter actually voted, but the system remains vulnerable to a coercer who is present during the casting of the ballot itself.

In 2010, Heather, Ryan, & Teague proposed an enhanced version of PGD to encompass elections with more complex voting methods than the existing first-past-the-post design, where a voter selects only a single candidate [106]. The improved PGD design allows for voting schemes using preference ordering, including the preferential or Instant Runoff Voting (IRV) and Single Transferable Vote (STV) systems used in Australian elections.

### 2.7.5   iVote NSW: The 2015 New South Wales State Election

The 2015 New South Wales (NSW) State Election was the first widespread deployment of remote online voting in Australia. The NSW Electoral Commission (NSWEC) selected the Scytl iVote online voting system, and use of the system was restricted to voters who would be absent from the state on election day, or who suffered from a disability qualifying them to use iVote instead of attending a polling place [107].

Despite these restrictions limiting use of iVote to only 5% of the total votes cast, this still corresponded to over 280,000 completed ballots being delivered using iVote, making the iVote NSW the largest deployment of remote online voting in the world at the time [107, 108].

The iVote system is designed to perform encryption of ballots using JavaScript in the voter's browser, providing some measure of confidentiality between the voter and the iVote counting server. In contrast to remote online voting systems such as Pretty Good Democracy or Helios, the iVote system does not use code voting or provide cryptographic auditing of ballots [107].

#### Vulnerabilities in Transport Layer Security

During the election, security researchers Halderman & Teague determined that the iVote website included JavaScript for a third-party analytics tool known as Piwik [108]. This script was secured using Transport Layer Security (TLS), but the server serving the script was vulnerable to the FREAK and Logjam downgrade-to-export attacks. These attacks allow a man-in-the-middle attacker to force a connection to use obsolete "export-grade" cryptography and subsequently break the TLS encryption [108, 109]. If an attacker was able to exploit this vulnerability, it would have allowed the attacker to read or steal votes by injecting malicious JavaScript into a voter's iVote session.

In order to test the feasibility of such an attack, Halderman & Teague simulated a live FREAK attack on the iVote practice system, which was identical to the real voting system in its use of TLS and loading of third-party analytics scripts. This simulation was successful in forcing the TLS connection to the third-party analytics server to downgrade to a weaker encryption method, allowing the private key to be calculated in a matter of hours using an Amazon EC2 instance. With this key, it was then possible to impersonate the TLS server delivering the Piwik analytics scripts, and inject malicious JavaScript into the voter's session in order to read or modify votes [108].

**Telephone Verification Service**

The iVote system used for the 2015 NSW state election also incorporated a telephone verification service, which allowed for voters to confirm that votes cast using the iVote website had been correctly recorded [107, 108]. Use of this system required the voter to dial a phone number provided by the NSWEC and enter the voter ID, PIN, and the receipt number provided by the iVote website on submission of a vote. The telephone verification service would then read back the preferences submitted by the voter.

As discussed by Halderman and Teague, this process was without precedent in Australia, and introduced a number of opportunities for breaches of voter privacy or coercion which have not previously been available. For example, an attacker intending to sway the election result may offer money in return for iVote verification credentials which produce the desired preferences, or threaten retribution if such credentials are not provided [108].

In contrast to traditional weaknesses with remote voting methods such as postal voting (for example: requiring a photograph to be taken of a completed ballot), this form of attack could be undertaken from anywhere in the world, and would provide a greater degree of assurance for the attacker that voting instructions were followed.

**Response**

In response to the concerns raised by Halderman and Teague during the election, Scytl and members of the NSWEC claimed that the risks had been overstated, and that attack presented was not a significant threat to iVote due to the high level of technical expertise required. In addition, both Scytl and the NSWEC compared the likelihood of a successful attack using the methods described by Halderman and Teague to *"that of a postal worker manipulating a postal vote"* [110, 111].

This response appears to misunderstand the difference in scale and access to ballots provided by an internet-based vulnerability—even one requiring interception of traffic—compared to that of a physical postal worker. Techniques such as KARMA attacks, which trick user devices to connect to malicious wireless access points, can allow even unsophisticated attackers to perform man-in-the-middle attacks on voters [112]. More overt techniques for performing large-scale traffic interception, such as BGP Hijacking, are well within the capabilities of nation-state attackers [113, 114].

Following the completion of the election, and in response to the security issues identified with the iVote system, the NSW Joint Standing Committee on Electoral Matters (JSCEM) released a report on the conduct of the election and use of iVote, and provided the following recommendations relating to iVote [115]:

1. That the NSW Government not expand iVote beyond its existing role;

2. That the NSW Government establish an independent panel of experts to conduct a full enquiry into the iVote system prior to the 2019 state election, and that this panel contain members with expertise in online voting, privacy, security and cybercrime;

3. That the iVote system only be used for the 2019 state election if the security concerns highlighted by the Committee have been addressed; and

4. That the NSW Government make the iVote source code publicly available.

Since 2015, the iVote system has been used for a number of by-elections in NSW, including in April 2017. The use of iVote for these elections is discussed in greater detail in Section 3.7.

# 3   iVote WA: The 2017 Western Australian State Election

This section contains an description and analysis of the iVote remote online voting system used for the Western Australian (WA) State Election in February and March 2017, referred to hereafter as iVote WA or simply iVote. Section 3.1 provides an overview of the iVote system. Section 3.2 describes the use of a cloud-based service for proxying voter connections. Section 3.3 outlines the behaviour of this service and implications for the security of votes cast using iVote. Section 3.4 describes the impact of this service on the encryption of voter connections, and the implications for storage of encryption keys and voter privacy.

Section 3.5 describes the potential for attacks against the voter verification system. Section 3.6 describes the failure of the iVote system to correctly implement DDoS mitigation during the first several days of the election. Section 3.7 provides a follow-up analysis of the iVote system used for the New South Wales by-elections in April 2017, and Section 3.8 provides a summary.

## 3.1   Background and Overview

The iVote system is developed by Scytl Secure Electronic Voting, a company based in Barcelona, Spain. iVote has been used previously for the 2015 New South Wales State Election, described in more detail in Section 2.7.5.

For the 2017 Western Australian State Election, the Western Australian Electoral Commission (WAEC) approved the Scytl iVote online voting system for use by voters who could not vote without assistance because they had insufficient literacy skills, were sight impaired, or were otherwise incapacitated [116]. This was the first occasion that a remote online voting system had been used for a state election in WA.

Voters could register to use iVote from Monday 13 February 2017, with votes able to be cast from Monday 20 February 2017. Voting using iVote closed at 6:00pm WST on Saturday 11 March, the same time as the closure of all other polls. The iVote Receipt verification service, intended to allow voters to verify that a vote was processed in the final count, was available for three days from Monday 13 March until Wednesday 15 March [116].

A detailed analysis of the iVote system as used for this election was conducted by the author in conjunction with Dr Chris Culnane and Dr Vanessa Teague of the University of Melbourne, and Dr Aleksander Essex of the University of Western Ontario. The results of this analysis were used as the basis for a paper presented at the E-Vote-ID conference in Bregenz, Austria in 2017. [117]

### 3.1.1   Voting Process

Voting using the iVote WA system consisted of three main phases:

**Registration.**  During the registration period, a voter could visit the iVote registration website at `https://ivote-reg.elections.wa.gov.au` and enter personal details including name, address, date of birth, and passport number. During the process, the voter could select a 6-digit PIN which would be used to verify the voter's identity when voting. An 8-digit registration number known as an 'iVote ID' is then sent to the voter using a different method, such as post or SMS.

**Voting.**  During the voting period, a voter could visit the iVote Core Voting System (CVS) website at `https://ivote-cvs.elections.wa.gov.au`, and enter a registration number and PIN. This would allow them to select candidates and cast a ballot through the iVote system.  All votes are encrypted in the voter's browser using client-side JavaScript served by the iVote server.  Upon submission of a completed ballot, the voter receives a 12-digit receipt number, which can be used to verify that a vote was received correctly.

**Verification.**  After the close of polls, voters can call a phone-based verification service in order to confirm that a vote was received and tallied correctly. Upon calling the service, the voter enters a registration number, PIN, and receipt number, and is able to have their vote read back to them over the phone.

During the election, voter connections to the iVote voting and registration websites took place via a U.S. based Content Delivery Network known as Imperva Incapsula, which acted as a non-transparent proxy for all connections.

### 3.1.2   Security Assumptions

The security of ballots cast using the iVote voting system depends upon several assumptions.

Firstly, the design of the system is based upon encryption of ballots using client-side JavaScript, which assumes that the JavaScript is delivered unmodified to the voter from the iVote CVS. This connection is secured and authenticated using Transport Layer Security (TLS), but a third-party service is used as a non-transparent proxy. This has important implications for the security of the JavaScript served to voters, as will be discussed in later sections.

Secondly, the privacy of voters using iVote hinges on the effective separation of voter identity from the registration number for the voter (iVote IDs).  Voter identity is linked with a iVote ID during the registration phase, and iVote IDs are linked with completed ballots during the voting phase. If a link can be made between voter identity and completed ballots at any point, voter privacy would be lost and it would become possible for an attacker to coerce voters using the iVote system.

Thirdly, the iVote system is a remote online voting system, with voters able to cast a ballot from personal devices at any location with an internet connection. This presents a risk of malicious parties compromising voter devices directly, in order to modify ballots when they are cast. This problem is common to all remote online voting systems where voters use personal devices in order to cast a ballot, as discussed in Section 2.7.1.

Finally, the use of remote voting opens voters to the threat of coercion, because voting does not take place at a controlled location (such as a polling place), and voters are therefore susceptible to being threatened or bribed into voting a particular way. As discussed in Section 2.7.1, this risk is common to all remote voting systems which do not provide controlled locations for voting, including paper-based systems such as postal voting.

## 3.2   DDoS Mitigation: Imperva Incapsula

Imperva Incapsula is an cloud-based content delivery network (CDN) which provides services including mitigation of Distributed Denial of Service (DDoS) attacks. These attacks involve causing a large number of connections to flood a target website, overloading the website systems and preventing legitimate users from logging in. It was a DDoS attack which was blamed for the failure of the Australian Government online Census system in August 2016 [118, 119].

Incapsula's DDoS mitigation service operates by placing Incapsula servers between the user and the website - intercepting all communications to and from the website in order to filter malicious connections. For example, when connecting to the iVote Core Voting System (CVS) for the WA election (`https://ivote-cvs.elections.wa.gov.au`), the user's connection is encrypted using Transport Layer Security (TLS) and travels to a server owned by Incapsula (a "point-of-presence", or PoP), where it is decrypted, scanned, and then forwarded on to the actual WA Electoral Commission server [120].

This interception provides Incapsula with the ability to filter out malicious traffic (and thereby mitigate DDoS attacks), but also allows Incapsula to see all traffic travelling through the service. This is by design: modern DDoS mitigation methods rely on scanning the actual unencrypted, or plaintext, traffic being transmitted to the server they are protecting [120, 121]. Without this ability, they would have a much harder time determining legitimate connections from malicious ones.

```
<script type="text/javascript">
//<![CDATA[
(function() {
var _analytics_scr = document.createElement('script');
_analytics_scr.type = 'text/javascript'; _analytics_scr.async
    = true; _analytics_scr.src = '/_Incapsula_Resource?SWJIYLWA
    =2977d8d74f63d7f8fedbea018b7a1d05&ns=1';
var _analytics_elem = document.getElementsByTagName('script')
    [0]; _analytics_elem.parentNode.insertBefore(_analytics_scr
    , _analytics_elem);
})();
// ]]>
</script>
```

Fig. 5: HTML injected into iVote server responses

In effect, DDoS mitigation services are paid to perform the role of a benign man-in-the-middle, whose objective is only to filter malicious traffic.

## 3.3   Script Injection and User Fingerprinting

When a user connects to iVote, using the address `https://ivote-cvs.elections.wa.gov.au`, the connection is proxied through Incapsula using an Incapsula-controlled TLS certificate.

For the first connection, the response from the iVote server is modified by Incapsula to include JavaScript code at the end of the HTML response. This code is designed to perform fingerprinting of the user's system and aid in DDoS mitigation, as described by Incapsula in the blog post: *How Incapsula Protects Against Data Leaks* [122]. This behaviour is normal for connections intercepted by Incapsula.

The actual HTML code inserted into iVote server responses is shown in Figure 5. This code is inserted by Incapsula before the end of the web page returned by the iVote server, and results in the voter's web browser downloading a second script, which is heavily obfuscated. This obfuscated code, abbreviated for readability, can be seen in Figure 6.

### 3.3.1   Tracking Cookies

Once expanded into a readable form, the injected JavaScript is revealed as a tracking function designed to read various information from the user's system, including the

```
(function() { var z="";var b="2866756e6374...
//<9660 characters clipped for brevity>"
;for (var i=0;i<b.length;i+=2){z=z+parseInt(b.substring(i, i
   +2), 16)+",";}z = z.substring(0,z.length-1); eval(eval('
   String.fromCharCode('+z+')'));})();
```

Fig. 6: Obfuscated JavaScript code

browser being used, the CPU and operating system, what browser plugins they
have installed, and other information which is designed to fingerprint individual
users. This information is then written into a tracking cookie, stored on the user's
computer, and sent back to Incapsula in all other requests so that Incapsula can
better identify users sending the requests.

It is likely that the tracking cookie generated by this process is part of Incapsula's
DDoS mitigation system. Incapsula can determine which requests are likely to be
from legitimate users, based on the information obtained using the injected JavaS-
cript code.

### 3.3.2   Potential for Malicious Script Injection

In the context of iVote, the JavaScript injection performed by Incapsula's servers
poses a significant risk for voter security. As discussed in Section 3.1.2, the iVote
system is designed with the assumption that the TLS encryption and authentication
covering voter connections is secure between the voter and the iVote server itself,
preventing any third-party from reading or modifying the contents of the connection.
If a third party has the ability to intercept this communication and inject malicious
JavaScript into server responses, it would be possible to hijack the entire voting
process.

For example: with Incapsula being used to proxy iVote connections, non-malicious
JavaScript is being injected into server responses. If this was to change such that
malicious JavaScript was added to this code by an attacker—such as a rogue Incap-
sula employee, or a foreign intelligence service—it would be possible for the attacker
to steal the ID number and PIN code from the voter during the login stage. The
attacker could then subsequently modify the voter's ballot as it is transmitted to
the iVote server, with a very low chance of detection by either the voter or the iVote
server itself.

With Incapsula's cookies already being used to identify voters between both the
registration server and voting server, it would also be possible for such an attacker
to link voter identity (provided to the registration server) with a completed ballot

(provided to the voting server), impacting voter privacy and opening voters to the risk of vote-buying or coercion.

The fingerprinting behaviour may also allow these attacks to be performed in a selective fashion. Recent research by Cao et al has shown that the sort of JavaScript user fingerprinting performed by Incapsula can be used to identify users with a high degree of confidence even across different browsers [123]. This may provide an attacker with the ability to selectively target individual voters or electoral divisions, and to avoid targeting users who may notice changes to the injected JavaScript.

## 3.4   Delegation of Transport Layer Security

The correct use of Transport Layer Security (TLS) involves a trust relationship between the user (the client) and the service they are connecting to (the server). When first negotiating a TLS connection, the server presents a TLS certificate to the client which has been signed by a valid Certificate Authority: a third-party that the client trusts to verify that the server is legitimate.

When a DDoS mitigation service like Incapsula is involved, this process is slightly altered. As discussed in Section 3.2, TLS connections by a user to the iVote Core Voting System (CVS) initially terminate at an Incapsula server, before being re-encrypted and sent to the iVote server itself [120]. Incapsula's servers become an intermediary for all traffic to the iVote server.

As described by Incapsula, this process does **not** involve the surrender of the iVote TLS private key to Incapsula for use on Incapsula servers [120]. Instead, Incapsula outwardly presents their own validly signed TLS certificate to incoming user connections, and then uses the iVote server's TLS encryption for the second stage of the connection.

In order for the TLS certificate presented by Incapsula to be trusted by users, it must be a valid certificate for the domain the user is connecting to: `ivote-cvs.elections.wa.gov.au`. Incapsula achieved this by serving a TLS certificate to users containing the wildcard domain `*.elections.wa.gov.au`, which is accepted as a valid certificate by users connecting to any subdomain (including the iVote Core Voting System).

An example of this behaviour is shown in Figure 7. Here, a connection to `ivote-cvs.elections.wa.gov.au` presents a valid TLS certificate, which has been accepted by the web browser (shown by the green lock symbol next to the address bar). On further inspection, the actual certificate is shown to be an Incapsula certificate, rather than a certificate controlled by the WA Electoral Commission. This certificate contains a Subject Alternate Name (SAN) field which lists a number of domains,

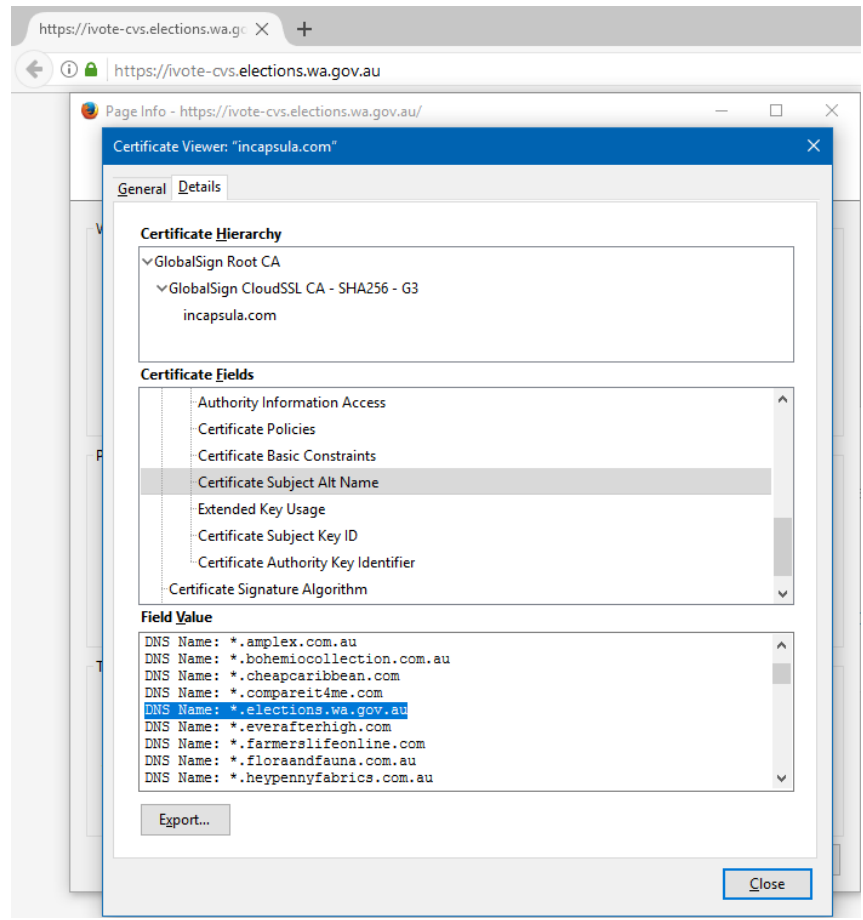

Fig. 7: The TLS certificate for `*.elections.wa.gov.au` provided by an Incapsula
    server in Melbourne

including the wildcard domain `*.elections.wa.gov.au`. The actual server host-
ing this certificate (`107.154.128.220`) is located in an Incapsula Point of Presence
(PoP) in Melbourne, Australia.

This behaviour is seen by any user connecting to the iVote CVS domain from within
Australia, due to the anycast routing design of Incapsula's systems: users who
perform a DNS lookup on any domain protected by Incapsula's services are directed
to the nearest PoP, rather than the actual server [124].

### 3.4.1 Foreign Hosting of TLS Private Keys

Incapsula's global network consists of 32 points-of-presence (PoPs), located across the Americas, Europe, the Middle East, and the Asia Pacific region[1]. Due to the design of Incapsula's network, certificates hosted in one PoP are propagated worldwide, so that users in any region served by Incapsula can have a connection proxied by the nearest PoP available.

As a consequence of this, the same wildcard certificate for `*.elections.wa.gov.au` is also stored on Incapsula servers in foreign countries. Figure 8 shows the certificate served if a user attempts to perform a TLS connection to the server at `45.64.65.220`. As shown in Figure 9, this IP is an Incapsula server located in Hong Kong.

The significance of this discovery is that there is a TLS private key located on a server in a foreign country which provides its owner with the ability to impersonate any domain matching the wildcard URL `*.elections.wa.gov.au`. This has implications for the security of the iVote Core Voting System and other systems reliant upon secure TLS connections to domains controlled by the WA Electoral Commission.

For example: a foreign government, as part of a legitimate domestic surveillance operation, may request that Incapsula provide access to the TLS private key for the domain `*.example.com` served by an Incapsula point-of-presence located in the foreign country. If this domain is listed in the same Incapsula TLS certificate as `*.elections.wa.gov.au`, obtaining this private key would also provide the foreign government with the ability to perform man-in-the-middle attacks on voters using iVote.

### 3.4.2 Implications for Voter Privacy

The privacy and anonymity of voters using iVote depends upon the assumption that the registration process (where the voter enters personal details) and voting process (where they cast completed ballot) cannot be linked. The design of the system aims to achieve this by ensuring that registration is conducted on a separate server from the voting process, and that the only information shared between the two is the Voter's ID number and PIN [107].

However, the use of Incapsula as a TLS proxy for both registration and voting introduces a single point where both registration and voting information can be combined. Indeed, the user fingerprinting behaviour of Incapsula—required for it to perform its intended function as a DDoS mitigation service—would likely provide for a high degree of confidence when linking the registration of individual voters with completed votes. This type of fingerprinting can be over 99% accurate for voters using the same web browser, and provides accurate identification even when a user uses different browsers on the same system [123].

---

[1] `https://www.incapsula.com/incapsula-global-network-map.html`

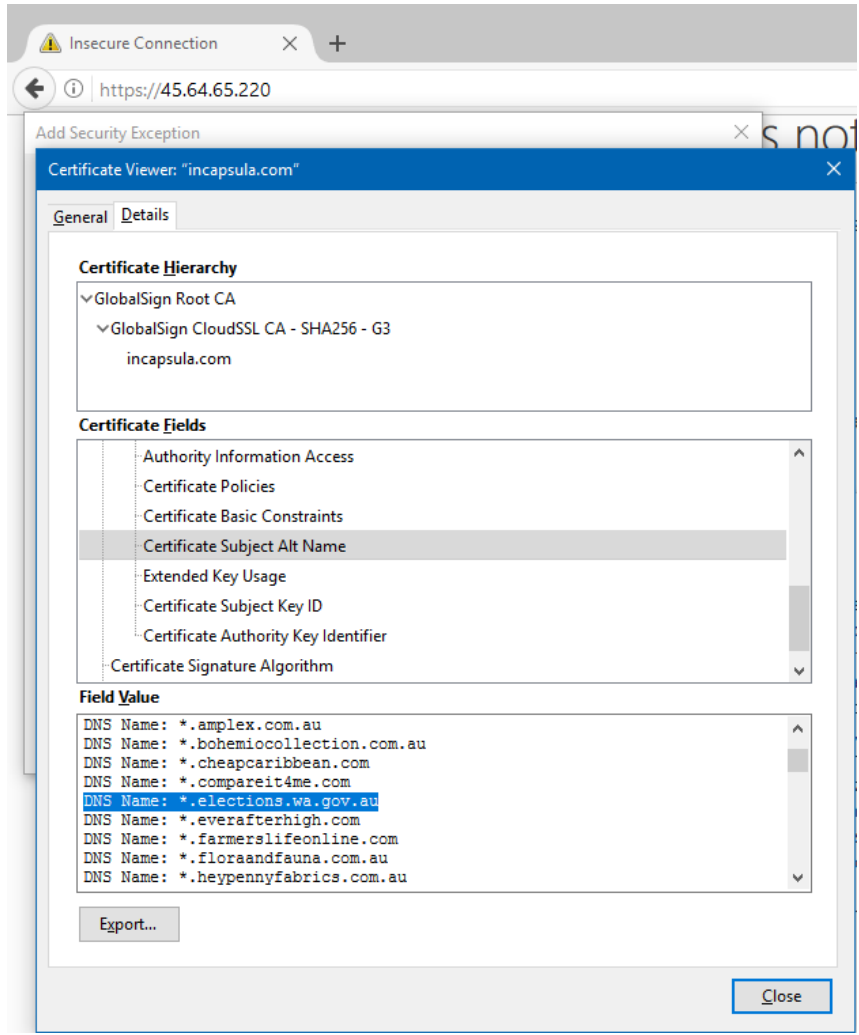

Fig. 8: The TLS certificate for `*.elections.wa.gov.au` provided by an Incapsula server with IP address `45.64.65.220`

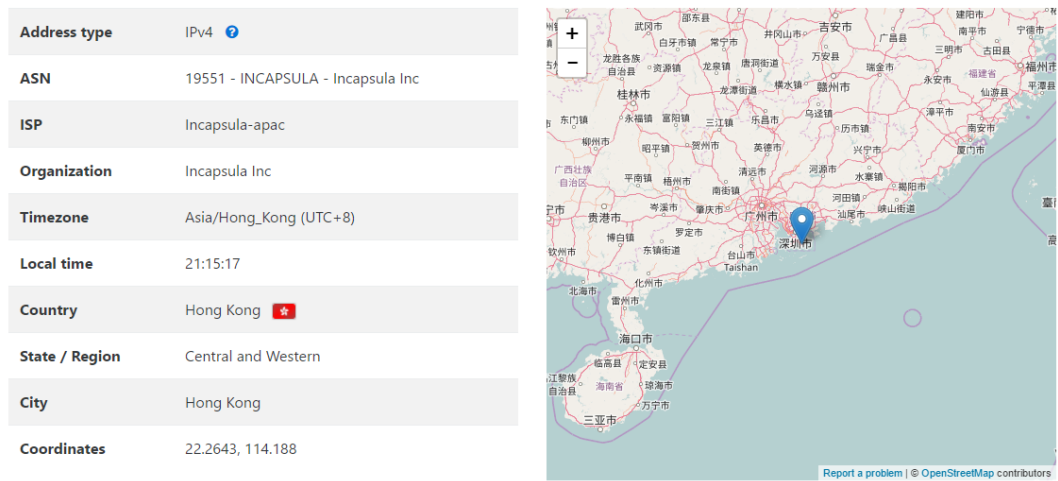

**Fig. 9:** The 'db-ip' geographical IP database showing the location of the IP address `45.64.65.220` in Hong Kong

## 3.5 Clash Attacks on Voter Verification

The iVote system incorporates a telephone verification service [125], which allows a voter to dial a provided number and connect with an interactive voice response (IVR) system.

The telephone verification service requires the voter's iVote ID number, PIN, and the receipt number provided by the iVote server after a vote has been successfully cast. After these three numbers have been provided, the telephone verification service reads back the list of candidates, in preference order, chosen by the voter in the completed ballot.

During the 2015 New South Wales State Election, which also used the iVote system, Halderman and Teague identified several potential attacks against this telephone verification system [108]. These attacks could allow an attacker, who had manipulated iVote ballots, to avoid detection from voters attempting to verify that their vote was cast as intended. One of these attacks is known as a "clash attack", and is designed to trick voters by manipulating the registration and vote confirmation pages to provide the ID number, PIN, and receipt number of a previous voter with the same candidate preferences.

The clash attack was first described by Küsters, Truderung and Vogt in 2012 [126]. When performing the attack, a previous voter's ballot (which has been allowed to be recorded unmodified) is then used as verification evidence of for multiple voters, whose actual votes can then be manipulated at-will with little chance of detection.

Crucially, the clash attack relies on accurate prediction of how a voter will vote prior to registration, so that they can be provided with the ID number and PIN of a like-minded voter who has submitted an unmodified ballot [108]. In addition, the attack relies upon providing voters with a PIN rather than allowing them to choose one. This may raise the suspicions of voters who are aware that the iVote system allows them to choose a PIN.

For the iVote WA system used in 2017, the effectiveness of the clash attack could be improved over prior iVote implementations, as consequence of Incapsula being used to proxy all voter connections to both the registration and voting servers. An attacker with access to Incapsula's systems would have the ability to link each voter's registration details with a completed ballot, provided the voter registers and votes using the same browser and therefore shares the Incapsula tracking cookie between connections.

From research by Cao, Li & Wijmans in 2017, it would also be possible to perform this linking of voters across browsers, through device fingerprinting methods similar to those employed by Incapsula for tracking cookies [123].

Due to Incapsula's position as a DDoS mitigation service for a number of other online services, an attacker with access to Incapsula's systems would have the ability to identify voter identity (and therefore likely voting preferences) with significantly more accuracy than if they only had access to the iVote system itself. This would allow for more accurate clash attacks to be performed.

## 3.6 Implementation Failure: The Importance of Hiding the Origin

As has been discussed in Section 3.2, DDoS mitigation services such as Incapsula operate by intercepting connections to a website, in order to filter out malicious connections and ensure only legitimate users can access the website. As noted by Vissers et al., this protection is heavily reliant upon the true IP address of the service being hidden, in order to avoid so-called direct-to-origin DDoS attacks [127]:

> The most popular Cloud-based Security Providers do not require the purchase of dedicated traffic-rerouting hardware, but rely solely on changing the DNS settings of a domain name to reroute a website's traffic through their security infrastructure. Consequently, this rerouting mechanism can be completely circumvented by directly attacking the website's hosting IP address. Therefore, it is crucial for the security and availability of these websites that their real IP address remains hidden from potential attackers.

During the first several days of voting in the 2017 WA state election, it was possible to identify the public IP address of the true server hosting the iVote WA Core Voting System (`https://ivote-cvs.elections.wa.gov.au`), through an analysis of HTTP responses and TLS certificates served by known iVote NSW infrastructure in Sydney. This same infrastructure was also used for iVote WA.

Th public IP address of this infrastructure could be publicly identified through DNS queries and other methods requiring little sophistication on the part of an attacker. With knowledge of this address, it would have been possible for an attacker to perform DDoS attacks directly against the public IP address for the iVote infrastructure, rendering Incapsula's protection ineffective.

Recommended practice for the use of DDoS mitigation services is to filter traffic from sources other than the DDoS mitigation service itself, in order to prevent attackers from identifying the true IP address of the website being protected [128, 129]. These protections were not correctly implemented at the opening of iVote for the WA state election, and the system remained vulnerable to direct-to-origin DDoS attacks for several days.

The WAEC was immediately notified as soon as this implementation failure was identified. Subsequently, all connections from non-Incapsula addresses were blocked from directly accessing the server hosting the iVote Core Voting System.

## 3.7   The 2017 New South Wales By-Elections

On the 8th of April 2017, three by-elections were held in the state of NSW for members of the Legislative Assembly which had resigned in the electorates of Gosford, Manly and North Shore.

For the by-elections, the New South Wales Electoral Commission (NSWEC) used the iVote online voting system in a similar manner to the 2015 NSW state election, with one major difference: Imperva Incapsula was employed to provide DDoS mitigation services for the iVote by-election registration domain (`registration.ivote.nsw.`

`gov.au`) and demonstration domains (`practise.ivote.nsw.gov.au` and `bypractise.ivote.nsw.gov.au`), in a similar manner to the iVote system used in Western Australia (iVote WA).

In contrast to iVote WA, however, Incapsula was **not** used to protect the iVote Core Voting System for the NSW by-elections, located at `cvs.ivote.nsw.gov.au`. This domain instead resolved directly to an IP address for a server located in Sydney and controlled by the NSWEC[2].

However, the use of Incapsula to protect the iVote NSW demonstration domains was implemented through the use of TLS certificates for the wildcard domain `*.ivote.nsw.gov.au`. These certificates were propagated to Incapsula servers globally, in a similar way to the certificates for `*.elections.wa.gov.au` described in Section 3.4.1. An example of one such certificate, served by an Incapsula server in Hong Kong (`45.64.64.88`), is shown in Figure 10.

As discussed in Section 3.4.1, the existence of this wildcard certificate on a foreign server has implications for the security of the iVote core voting system. An attacker with control of the TLS private key associated with this certificate would have the ability to impersonate any subdomain of the iVote NSW system.

Importantly, this also includes the iVote Core Voting System at `cvs.ivote.nsw.gov.au`. Although this domain was not protected by Incapsula during the election, it is still covered by the wildcard domain contained in Incapsula's TLS certificate. This demonstrates both the risk of using wildcard domains, and the risk of outsourcing TLS security to third parties.

## 3.8   iVote WA Summary: Security Trade-offs and Transparency

In this section we have examined the design of the iVote WA online voting system as implemented during the 2017 state election, and the security impact caused by use of a cloud-based DDoS mitigation service. We have also explored the use of iVote for by-elections held in NSW during April.

Use of Incapsula's content delivery network for iVote WA broke the security assumptions of iVote: namely that connections between voters and the voting server could not be intercepted by a third-party. Normal operation of services such as Incapsula requires the ability to view and modify the unencrypted traffic between voters and the voting server, and this operation also increased the scope for clash attacks to be performed on the iVote voter verification system.

---

[2] The same server was used to host the iVote Core Voting System for iVote WA.

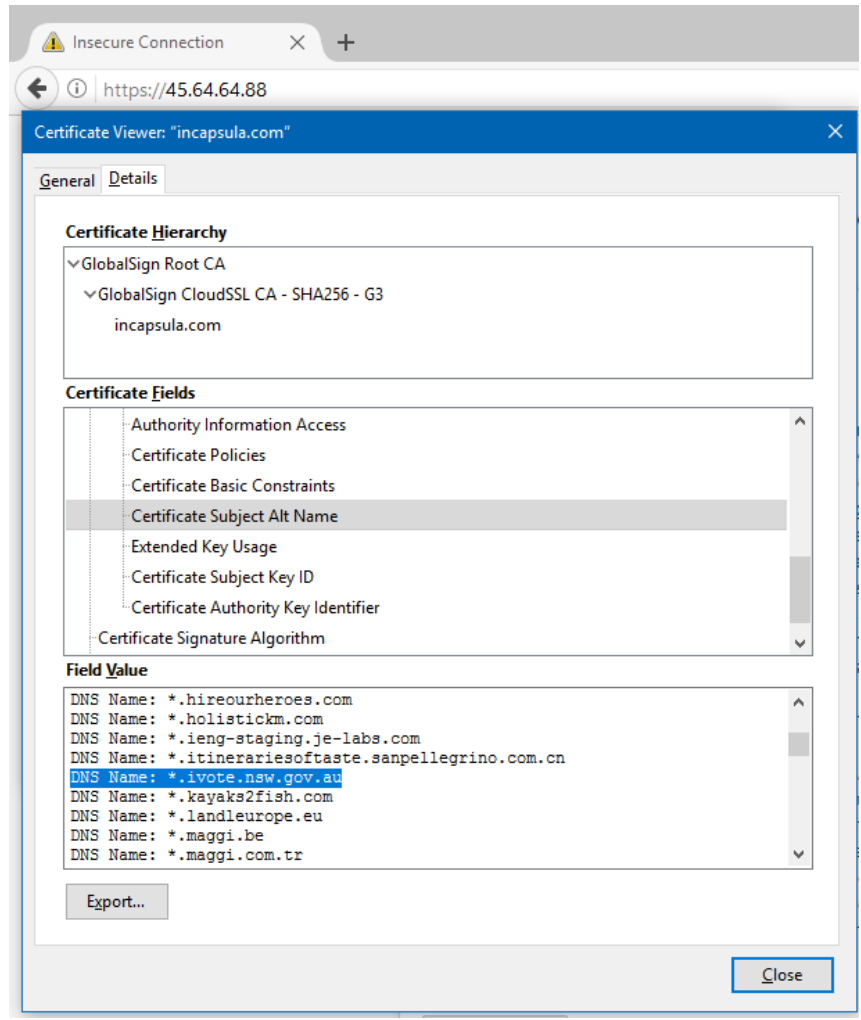

Fig. 10: The TLS certificate for `*.ivote.nsw.gov.au` provided by an Incapsula server in Hong Kong

In addition to this, use of Incapsula resulted in the TLS certificate for the iVote election infrastructure being shared with a number of unrelated websites, and the private key for this certificate was stored in Incapsula points-of-presence in foreign countries. This introduced a new risk: that a legal domestic law-enforcement request could compromise the TLS private key needed for an attacker to impersonate the WA or NSW election infrastructure.

Finally, the DDoS mitigation was not correctly implemented during the WA election, causing the iVote system to be vulnerable to direct-to-origin DDoS attacks and therefore defeating the purpose of using Incapsula.

There is a clear trade-off when using a content delivery network such as Incapsula to provide DDoS mitigation: these services ensure increased availability of the voting system during voting, but they can have significant impacts to the trust model of the system. It is important to keep these trade-offs in mind when implementing an online voting system, and it is also important that voters are made aware of the risks and benefits involved in voting using the system.

# 4 A Cross-Sectional Survey of Australian Voters

This section presents the results of a cross-sectional survey of Australian Voters regarding trust in voting systems. Section 4.1 provides background and outlines why a survey was required. Section 4.2 describes the purpose and scope of the survey. Section 4.3 outlines the methodology of the survey, including administration method, sample selection, questionnaire design, threats to validity, and the privacy, confidentiality, and ethical considerations of the survey.

Section 4.4 describes how results were analysed, including consideration of final sample size, demographic information, and the process used for qualitative analysis and coding. Section 4.5 provides the results of the survey, along with discussion of the results and their consequences for the design of electronic voting systems in Australia. Section 4.6 provides a summary of the survey.

## 4.1 Background

When conducting a democratic election the perceived fairness of the voting system, and its ability to accurately select the winning candidate based on the will of voters, can be just as important as the reality of the system itself [1, 21].

Future policy regarding the use of electronic voting systems in Australian federal elections is dependent on voter perception of those systems; both in directly influencing decisions regarding which systems to consider, and in order to determine additional education measures which may be needed to convince voters that a given system is trustworthy.

In order to determine how Australian voters perceive various voting systems, and a measure of the trust they have in each system, it is necessary to conduct a survey.

There are two major types of survey, each with its own typical administration methods [130]:

Self-administered questionnaire: where the respondent completes a form independently, either by hand ("paper and pencil") or electronically using a website or application; and

Interview: where the surveyor interacts with the respondent to elicit responses, either in person, over the telephone, or via an instant messaging application.

Selecting the type of survey and associated administration method requires balancing the desire for credible and accurate survey results with the available resources to complete a survey using the chosen administration method [130].

## 4.2 Purpose and Scope

A cross-sectional survey was performed using a sample of eligible Australian voters. The purpose of the survey was to obtain a measure of the perceived trustworthiness of three different voting systems, if applied for the next Australian federal election:

1. The existing paper-based voting system, as used in July 2016

2. The iVote remote online voting system, as used for the Western Australian State Election in February and March 2017

3. The vVote poll-site electronic voting system, as used for the Victorian State Election in November 2014

For each system, survey respondents were provided with a short video describing how each voting system functions, from the perspective of the voter's interaction with the system.

The survey then asked respondents to rate their level of trust in each system on various measures, their own perceived level of understanding of how the system works, and whether or not they would expect accidental errors or intentional tampering to be detected by election officials.

## 4.3 Methodology

### 4.3.1 Survey Type and Administration Method

The survey was conducted using a self-administered questionnaire, delivered electronically using the SurveyMonkey online survey platform.

The primary delivery method was via social media (Facebook), through initial sharing of the survey link and requests for respondents to voluntarily complete the survey and share it with their online connections. The secondary delivery method was via direct email and word-of-mouth, in order to effectively reach demographics which were less likely to use Facebook.

Collection of results occurred over a one-month period from 10 September 2017 through to 10 October 2017.

### 4.3.2 Survey Questionnaire

The survey questionnaire consisted of five pages. The full survey questionnaire, containing all questions and links to explanatory videos, can be found in Appendix A.

## Introduction

The first page of the questionnaire provided an introduction to the survey, containing a brief written description of the survey and its purpose, along with a short video describing the survey.

## Demographic Information

The second page contained demographic questions about the participant, including:

- Whether or not the participant was an Australian citizen;

- The age of the participant; and

- The highest level of formal education the participant had completed.

## Voting System Descriptions and Trust Questions

The third, fourth and fifth pages of the survey contained questions regarding each of the three voting systems. Each page was devoted to one of the three systems, and contained a thirty-second video description of the system, followed by a series of questions about the participant's level of trust and understanding of the system.

The design of each video description was closely controlled in order to provide, to the extent possible, an objective overview of the operation of the system from the perspective of a voter using the system to cast a ballot. The script for each video was designed to contain only factual information about the operation of each system, without any subjective opinion or comment.

The focus of each video description was to describe the following stages of casting a ballot, from the perspective of a typical voter:

- Visiting a polling site or voting website

- Obtaining a ballot paper or logging into the voting website

- Completing a ballot

- Casting a completed ballot

- Counting of ballot papers

- Scrutineering of counting (if applicable)

- Verifying receipt of the completed ballot (if applicable)

A strict time limit on each video made it difficult to include all relevant information about each system. The potential effect of this on survey results is discussed in more detail in Section 4.3.5.

In order to reduce possible bias caused by the ordering of the voting systems within the questionnaire, the three voting system pages were presented in a random order to each participant. This is described in further detail in Section 4.3.5.

The questionnaire page for each voting system contained the following questions:

1. How the participant would rate their level of trust in the system to:

   - Accurately collect votes
   - Accurately count votes
   - Resist attempts to tamper with the election
   - Protect the privacy of voters
   - Ensure that all votes are kept anonymous
   - Be accessible to all voters
   - Be usable by all voters
   - Return election results in a timely manner

2. An open-ended question asking the participant to briefly explain why they gave the ratings that they did

3. How the participant would rate their level of understanding of how each part of the voting system works:

   - Voter registration and obtaining ballot papers
   - Completing and casting a ballot
   - Counting the votes and producing a result
   - Supervision of counting by candidates (scrutineering)
   - Verification of election results by voters

4. To what extent the participant thinks that their understanding of the system impacts their level of trust in the system

5. Whether the participant believes that errors or attempted tampering with the voting system would be detected by election officials

6. An open-ended question asking the participant to briefly explain why responded the way they did to the previous question

For Question 1, the aspects of the voting system to be queried were selected based on the characteristics of a well designed voting system, as discussed in Section 2.2.1. These four characteristics were expanded to 8 distinct aspects of a voting system, in order to capture as much information as possible from participants who may draw distinctions between concepts such as 'anonymity' and 'privacy', or 'accessible' and 'usable' when considering particular systems. These concepts have a signficant degree of overlap, but were asked separately in order to assist in obtaining responses from participants who did not engage with a particular choice of term.

An additional question for "return election results in a timely manner" was added in order to capture expected responses from voters based on the experiences of the 2016 Australian Federal Election (discussed in Section 2.3.4).

For Question 3, the components of the voting system asked of each participant were the same components described in each video description.

### 4.3.3 Pilot Testing

Pilot testing is essential for any well-conducted survey [130]. As the survey was be administered electronically using a self-administered questionnaire, it was essential that the language used in the questionnaire be thoroughly tested for clarity.

This was particularly important given the technical nature of the subject, and the fact that respondents were unlikely to be well-versed in information security or the design of voting systems.

Pilot testing was conducted over a one-week period between 27 August and 3 October 2017, and involved a small number of testers from different demographics. Feedback from pilot testing was used to modify the language used in the system description videos and survey questions, the format and style of questions, and to improve the introductory page of the survey.

### 4.3.4 Sample Selection

The purpose of the survey was to obtain a measure of the perceived trustworthiness of different voting systems from a representative sample of eligible Australian voters.

Given a population size $N$, margin of error $E$, confidence level $c$, and response distribution $r$, the required sample size $n$ can be calculated using the following equations: [131, 132]

$$x = (Z_{\alpha/2})^2 r(100 - r) \tag{1}$$

$$n = \left( \frac{Nx}{(N-1)E^2 + x} \right) \tag{2}$$

Where $Z_{\alpha/2}$ is the *critical value*. For responses which are normally distributed, the critical value for a confidence level of 95% is 1.96 ($\alpha = 0.05$)

$$Z_{\alpha/2} = 1.96$$

According to the Australian Electoral Commission [133], the total number of eligible Australian voters as of 31 December 2016 was 16,589,009.

Using this information, the required sample size for the survey was calculated, using the following parameters:

- Population size $N$: 16,589,009

- Desired margin of error $E$: 5%

- Confidence level $c$: 95%

With these parameters, the required sample size for the survey was $n = 385$.

### 4.3.5 Threats to Validity

**Collection method**

A likely source of bias in the survey results is associated with the collection method: an online self-administered questionnaire manually targeted using social and family connections of the author.

Due to resource constraints, the primary method for soliciting survey responses was via social media platforms (primarily LinkedIn and Facebook) and email or word-of-mouth to existing contacts. This resulted in survey participants being highly likely to have an existing connection to the author, and therefore biases responses towards the employment background, socio-economic status, age, and education level of the author and the author's close relations.

An effort was made to counteract this bias through solicitation of friends and family members to share the questionnaire with their own connections, and therefore dilute the effects of demographic bias. The recorded demographic information for survey respondents, including the effects of this bias, is discussed in greater detail in Section 4.4.2.

Despire attempts to reduce demographic bias, there is a high likelihood that South Australian voters are disproprortionately represented in the survey responses. It is not clear whether state of residence would have an impact on survey results, and this was not a question asked during in the survey questionnaire.

## Voting system descriptions

The video description summarising each voting system introduced a significant potential source of bias. If these video descriptions were to contain a point-of-view regarding the relative merits of each system, it is likely that they would influence responses. This is especially likely for systems where the respondent has no existing familiarity.

As a mitigation for this, the video descriptions were carefully scripted to ensure that they contained only the process of using the system as would be experienced by a voter. The descriptions did not include any point-of-view information about the relative merits of each system.

The restriction on permitted time for each description video may have also introduced bias. Each video was not identical in length, with the existing system, iVote, and vVote, totalling 44, 47, and 42 seconds respectively. In addition, mandating an equal time limit for each system is likely to disproportionately affect more complex systems, or systems where the respondent does not have an existing familiarity with the system.

This component of the survey was the subject of significant review during the pilot testing process, as described in Section 4.3.3.

## Participant familiarity with systems

Each participant in the survey will have had a different familiarity with the three voting systems presented. All participants are likely to be familiar with the existing paper-based voting system, unless they have not voted in a previous Australian election. This familiarity may result in a bias towards trusting the existing paper-based system at the expense of the iVote and vVote systems.

A small number of participants may have also been familiar with the iVote or vVote systems, if they have used these systems previously or if they have followed media reports about the systems.

The description video for each system is intended to help normalise responses by ensuring that all participants have a base level of understanding of how each system works from the perspective of the voter. However, it is difficult to avoid the effects of existing familiarity with the existing paper-based system.

**Page Ordering**

The order in which each system is presented to the survey participant may have introduced bias to the results, e.g. through so-called 'priming' effects and fatigue effects on questions asked at the end of a questionnaire [134, 135].

In order to avoid bias introduced by the ordering of the three systems, the order these pages were presented was randomised. Each participant received a random permutation of the three pages containing voting system questions.

### 4.3.6 Privacy and Confidentiality

Results collected during the survey were anonymised to remove references to individual respondents, and respondent IP addresses were not collected.

Results were only analysed in aggregate, and free-form responses are only repeated in this thesis if there is no identifying information contained in the response.

### 4.3.7 Ethical Review

As a human research project being undertaken by a student of the University of Adelaide, this survey required an appropriate level of ethical approval by the University Human Research Ethics Committee (HREC).

In accordance with the *Ethical Issues Checklist for Human Research* provided by the HREC[3], this survey meets the requirements for "negligible risk research" as defined by the HREC, and was therefore exempted from HREC review.

## 4.4 Analysis

### 4.4.1 Sample Size Adjustment

During the collection period for the survey, a total of 178 completed responses were obtained from a total of 247 responses. A total of four respondents answered 'No' to the question asking whether or not they are Australian citizens, and were removed from consideration.

---

[3] `https://www.adelaide.edu.au/research-services/oreci/human/docs/hrec-ethics-checklist.doc`

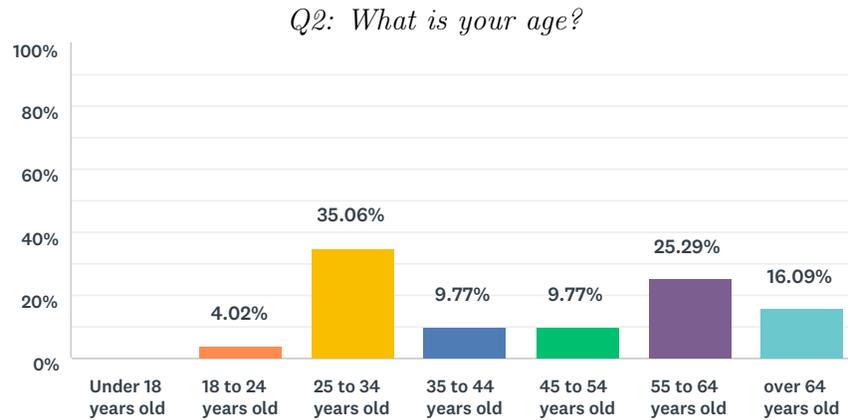

Fig. 11: Demographic results for respondent age range

Therefore, the final sample size was 174, less than the 385 required for a 5% margin of error as discussed in Section 4.3.4.

With a sample size $n$ of 174, the revised margin of error can be calculated using the following equation (with all other parameters kept at the same values):

$$E = \sqrt{\left(\frac{x(N-n)}{n(N-1)}\right)} \tag{3}$$

Where $x$ is given by Equation 1, as before.

With a sample size of 174, the revised margin of error is $E = 0.075$, or 7.5%. This error margin was deemed acceptable given the distribution of responses, as the responses for each voting system differed by greater than 7.5%. This is discussed in further detail in Section 4.4.3.

### 4.4.2 Demographic information

**Respondent age group**

Results showing respondent age grouping are shown in Figure 11. The results show peaks for the age groups 25-34 and 55-64, and a low response rate for the age group 18-24.

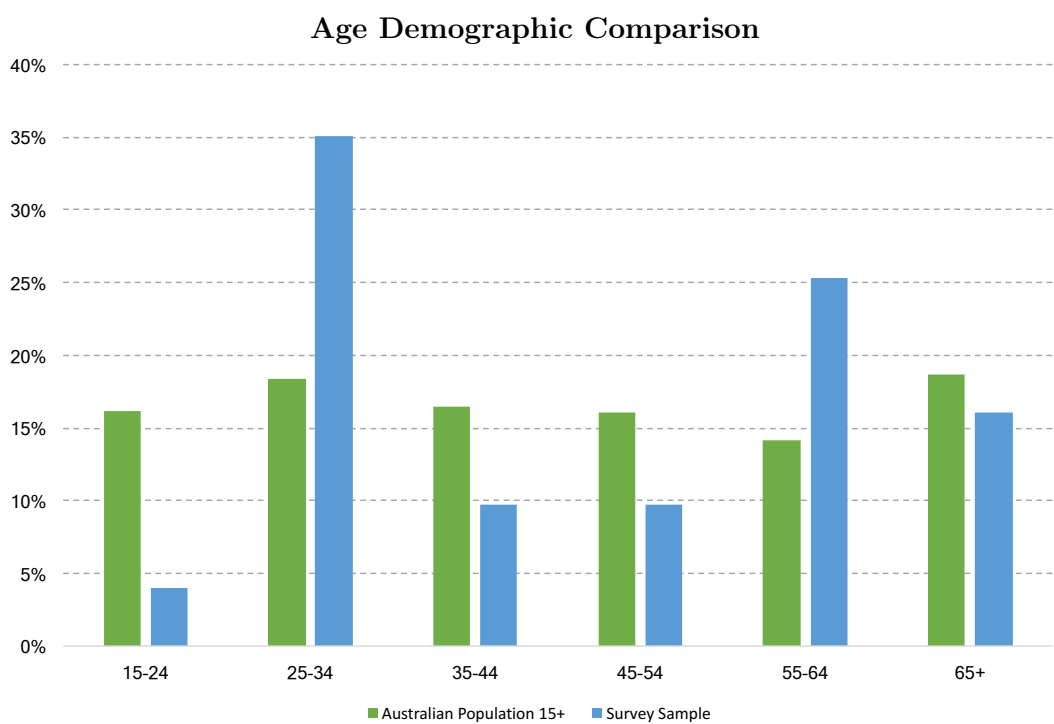

**Fig. 12:** Comparison of sample ages with Australian age distribution (ABS, 2017)

As discussed in Section 4.3.5, the peaks for ages of 25-34 and 55-64 are likely to be caused by the survey collection method. The 25-34 year age group matches the author and the author's immediate social circle, and so increased responses from this demographic are to be expected.

The 55-64 age group correspondents to members of the author's immediate family. Additional effort was made to reach this demographic in order to offset biases caused by disproportionate repsentation of younger age groups, and this manifests as a spike in responses from this age group.

A comparison was made between the age distribution of survey respondents and the estimated Australian age distribution at 30 June 2016. These figures were obtained from the Australian Bureau of Statististics, and are based on the results of the 2016 Australian Census [136].

The results of this comparison can be seen in Figure 12. All percentages are as a percentage of relevant population: in the case of the survey respondents, as a percentage the final sample size; in the case of the Australian population, as a percentage of all census respondents over the age of 15.

*Q3: What is the highest level of formal education you have completed?*

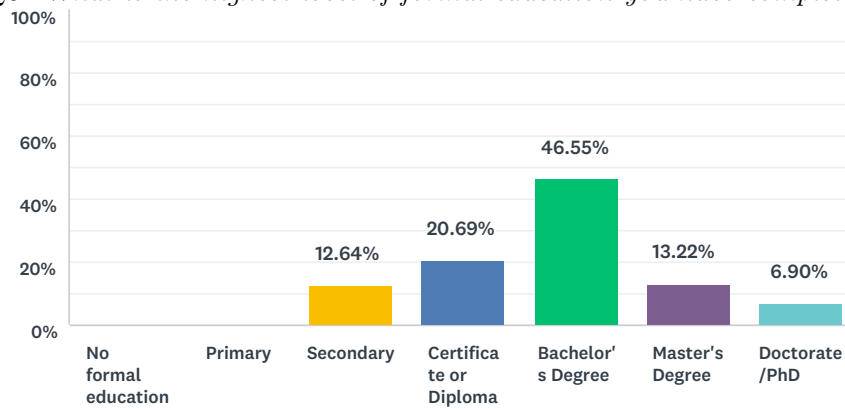

Fig. 13: Demographic results for respondent education level

It is important to note that the statistics reported by the ABS includes respondents who are not Australian citizens, and would therefore be ineligible to vote. It is also important to note that the 15-24 age group, which is specified in ABS statistics, is underrepresented in the survey sample due to the minimum age being 18 years of age for eligible Australian voters.

With these caveats, it can be seen that the 25-34 and 55-64 age groups are likely to be overrepresented in the survey sample.

## Highest level of education attainment

Results showing respondent level of educational attainment are shown in Figure 13. Similarly to the age distribution results, and as discussed in Section 4.3.5, the peak for "Bachelor's Degree" is likely to be caused by the survey collection method causing a bias towards the author's family and social connections.

Notably, 100% of respondents included in this analysis listed their completed education level as Secondary or above. This is significant as the proportion of Australian residents aged 20 to 64 with this level of education is estimated by the Australian Bureau of Statistics (ABS) to be 66% as of May 2016: [137]

> *The proportion of people aged 20 to 64 years with Year 12 or equivalent has increased from 53% in 2006 to 66% in 2016.*

Therefore, the results of the survey questionnaire do not include any responses from the 34% of Australian residents aged 20 to 64 who have no formal education, or only completed a primary school education.

As with the age comparison, it is important to note that the statistics reported by the ABS includes respondents who are not Australian citizens, and would therefore be ineligible to vote. In addition, the statistics exclude several demographics which were not explicitly excluded from the voting system survey, including respondents aged 65 or above.

Despite these differences in survey scope, it remains notable that the survey questionnaire does not include any responses from such a significant portion of Australians.

### Comparative demographic analysis of results

As shown in Figure 11, there were two peaks in the age groups of respondents to the survey questionnaire. The largest was the 25 to 34 age group, with 35.06% of respondents. This was followed by the 55 to 64 age group, with 25.29% of respondents.

Because these two age groups are generationally distinct and comprise a large proportion of respondents, additional analysis was performed to determine any statistically significant differences in the responses between the age groups.

For these comparisons, a standard definition for statistical significance (95% confidence, or $p = 0.05$) was used. The results of this comparison are shown in Section 4.5.5.

### 4.4.3   Comparison of Results

Comparison of results is made difficult due to the impact of subjective interpretation of each question by participants. For example, two participants may have the same level of trust in a particular system, but each may interpret the wording of individual options (such as 'strongly trust') in a different manner.

Despite these limitations, it is useful to provide some measure of the statistical significance of results between the various systems. In order to compare the results for reported trust levels in each voting system, a basic Z-test will be used. Note that use of a Z-test is optimal for responses which fit a normal distribution, which by definition is not true for a Likert scale (where responses fit an ordinal distribuition).

For each system, options are ranked from 1 to 5, with 1 being 'strongly distrust' and 5 being 'strongly trust'. A weighted average is produced based on the responses, with this weighted average being used as the mean for the Z-test

The Z-value can then be determined from the following equation: [138]

$$Z = \frac{(\bar{X}_1 - \bar{X}_2)}{\sqrt{\sigma_{x_1}^2 + \sigma_{x_2}^2}}$$

Where $\bar{X}_i$ is the weighted average of the responses to be compared, and $\sigma_x$ is the standard deviation of the responses divided by the square root of the number of data points.

In general, the following values for $Z$ were used to guide statements regarding statistical significance: [138]

- If $Z < 2$, there is no difference between the two sets of responses

- If $2.0 < Z < 2.5$, there is a marginal difference between the two sets of responses

- If $2.5 < Z < 3.0$, there is a significant difference between the two sets of responses

- If $Z > 3.0$, there is a highly significant difference between the two sets of responses

This approach will be used during the analysis of results when comparing the results of the primary Likert scale for each voting system (Questions 4, 10, and 16, respectively).

### 4.4.4 Qualititative Analysis and Coding

For each of the three pages in the survey questionnaire devoted to the voting systems, there were two questions allowing for open-ended text responses from responses. Properly analysing these responses required a coding approach, as discussed by Saldaña [139].

The approach used involved a collection of pre-set codes informed by the characteristics of voting systems as discussed in Section 2.2.1, such as "anonymity is good", or "resistent to tampering". The pre-set codes also included aspects of voting systems described in the system description videos, such as "fast counting speed".

As discussed in Section 4.3.2, additional pre-set codes were included to reflect the additional aspects of a voting system listed in the survey questionnaire, in order to capture as much information as possible from participants who may draw a distinction between concepts such as 'anonyminity' and 'privacy', or 'accessible' and 'usable' when considering particular systems. The full list of pre-set codes can be found in Appendix B.

In addition to the pre-set codes, a number of emergent codes were added during the coding process, in order to better fit the provided responses. These included codes such as "mature system": for example, this code was used to categorise responses such as *"if it 'aint broke, don't fix it"* regarding the existing paper-based voting system.

During code consolidation, several pre-set codes were combined with other coded responses to improve the readability of results. For example, responses coded as "checks and balances are effective" were combined with responses coded as "scrutineering is effective", when the latter only resulted in a small number of coded responses.

The former code was an emergent code used for responses which commented on general checks and balances within a system, without specifically mentioning scrutineering by representatives of candidates. Because scrutineering is a form of check and balance within a voting system, it was deemed acceptable to combine these codes.

**Issues with page randomisation**

As outlined in Section 4.3.5, each of the pages devoted to one of the three voting systems was presented in a random order to each respondent, in order to control for biases introduced by ordering the voting systems in a specific way.

The fact that each page would be presented in a random order was not explained to participants during the introduction to the survey questionnaire. This led to issues with some open-ended responses, which contained comments such as "same as previous system" or similar. These respondents likely assumed that the ordering they received each page was common to all participants. These responses have not been included in coding, as it was not possible to determine which order they received their survey pages.

## 4.5 Results and Discussion

### 4.5.1 Results Summary: Comparison of Voting Systems

The first question posed to participants for each voting system asked for a rating of trust or distrust in the system to meet various requirements of an election. This

Tab. 1: Trust in voting systems: Likert scale response statistics

*How much would you trust or distrust this system to meet the following requirements for an election?*

|  | Existing System | | iVote | | vVote | |
|---|---|---|---|---|---|---|
|  | Weighted Average | Std. Dev. | Weighted Average | Std. Dev. | Weighted Average | Std. Dev. |
| Accurately collect votes? | 4.06 | 0.89 | 3.65 | 1.18 | 4.01 | 0.92 |
| Accurately count votes? | 3.71 | 0.96 | 4.04 | 1.04 | 4.26 | 0.87 |
| Resist attempts to tamper with the election? | 3.76 | 1.02 | 2.58 | 1.15 | 3.25 | 1.18 |
| Protect the privacy of voters? | 4.40 | 0.77 | 2.85 | 1.23 | 3.40 | 1.17 |
| Ensure that all votes are kept anonymous? | 4.45 | 0.72 | 2.74 | 1.23 | 3.36 | 1.19 |
| Be accessible to all voters? | 4.02 | 0.84 | 2.71 | 1.20 | 3.43 | 1.09 |
| Be usable by all voters? | 4.03 | 0.82 | 2.63 | 1.11 | 3.25 | 1.12 |
| Return election results in a timely manner? | 3.16 | 1.26 | 4.29 | 0.81 | 4.35 | 0.82 |

Tab. 2: Z-test values for significance of voting system comparisons

| Z-test (Existing/iVote) | Z-test (Existing/vVote) | Z-test (iVote/vVote) |
|---|---|---|
| 3.65 | 0.51 | 3.16 |
| 3.07 | 5.58 | 2.13 |
| 10.10 | 4.30 | 5.35 |
| 14.05 | 9.39 | 4.26 |
| 15.78 | 10.31 | 4.76 |
| 11.76 | 5.64 | 5.84 |
| 13.34 | 7.39 | 5.17 |
| 9.92 | 10.41 | 0.68 |

question was posed in the form of a Likert scale, with valid responses from one (strongly distrust) through to five (strongly trust).

A simplified analysis of these responses used a weighted average as a measure for comparison between each voting system, with results on a scale from 1-5. An answer of 'don't know' was ignored for this weighting.

The weighted average of each response category for each voting system can be seen in Figure 14, and the complete weighted averages and standard deviations for each question across the three voting systems can be seen in Table 1.

As discussed in Section 4.4.3, a Z-test was used to determine a measure of statistical signficance between the results for each system, with Z-values over 2.5 denoting a significant difference. A summary chart shoting these Z-values for each question across the three voting systems can be seen in Table 2.

Overall, these results show a higher level of voter trust in the existing paper-based system across five of the seven categories measured, when compared to both the iVote

**Trust in Voting Systems**

*Q4/10/16: How much would you trust or distrust this system to meet the following requirements for an election?*

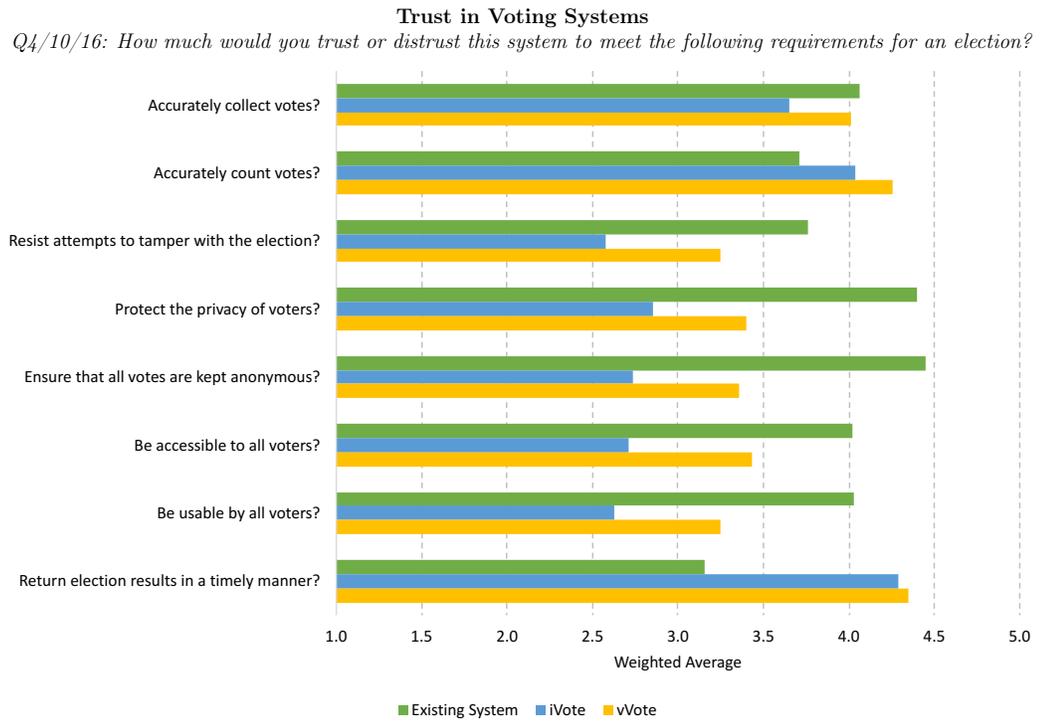

Fig. 14: Respondent trust in voting system components: weighted averages

and vVote systems. These differences were highly significant ($Z > 3$). The categories in which voters showed a higher level of trust in the existing system included: "resist attempts to tamper with the election", "protect the privacy of voters", "ensure that all votes are kept anonymous", "be accessible to all voters", and "be usable by all voters".

In contrast, on the question of whether or not the system would "accurately count votes", the iVote and vVote systems rated higher than the existing system. There was only a marginal difference between the iVote and vVote systems on this question, with vVote returning higher levels of trust on average.

On the question of whether the system would "return election results in a timely manner", both the iVote and vVote systems rated higher than the existing system, with a high degree of significance ($Z > 3$). There was no significant difference betweeen the iVote and vVote systems on this measure.

On the question of whether or not the system would "accurately collect votes", the iVote system scored lower than both the existing and vVote systems, with a high degree of significance ($Z > 3$). There was no significant difference between the existing and vVote systems on this measure.

### 4.5.2   The Existing Paper-Based Voting System

**Voter Trust**

Question 4 of the survey questionnaire asked participants to rate their level of trust or distrust in the existing paper-based voting system to meet various requirements of an election. The results of this question are shown in Figure 15 and Table 3.

Figure 15 presents the results in the form of a stacked bar chart, in order to assist in visualisation of the overall responses.

The results for the existing paper-based voting system show an overall high degree of trust in the system, with a majority of respondents recording either "trust" or "strongly trust" for 8 of the 9 characteristics listed.

The only characteristic not to achieve a majority was whether or not the respondent trusted the system to *"Return election results in a timely manner"*. This result is somewhat unsurprising given recent high-profile issues with the speed of election results returned by this system, as discussed in Section 2.3.4.

The highest level of trust in the system was recorded for *"Protect the privacy of voters"* and *"Ensure that all votes are kept anonymous"*. Almost 90% of respondents trusted the existing paper-based system to meet both of these requirements (89.6%

**The Existing Paper-Based Voting System**

*Q4: How much would you trust or distrust this system to meet the following requirements for an election?*

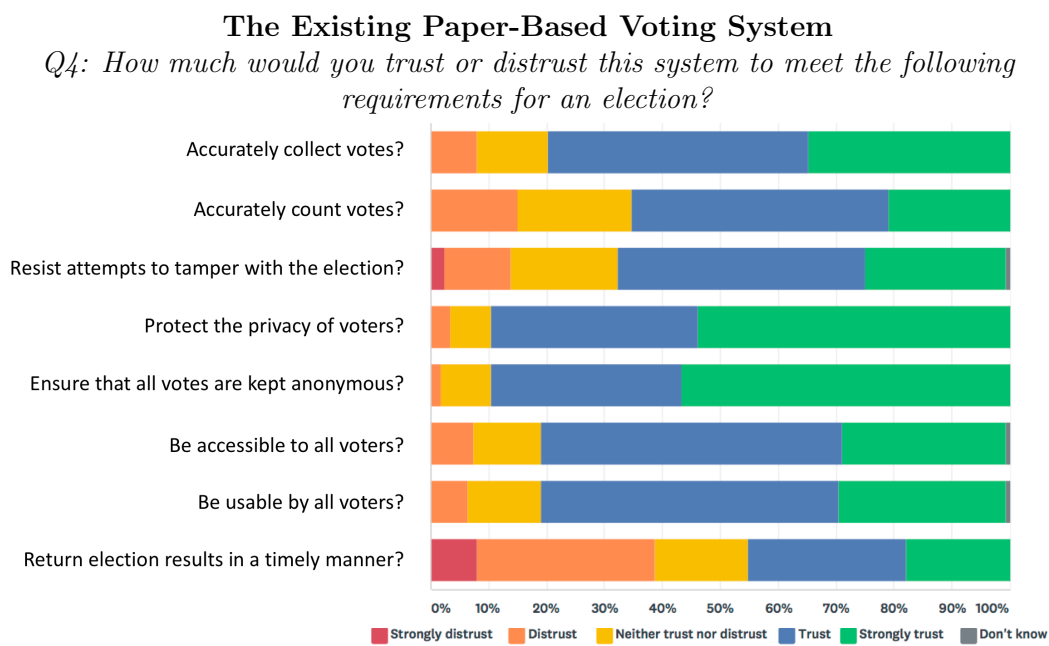

Fig. 15: Question 4 results

of responses), with a majority reporting strong trust in both cases (53.8% and 56.7%, respectively).

Reported levels of distrust in the existing paper-based system were low, with combined 'distrust' and 'strongly distrust' responses remaining below 40% across all characteristics. Only *"Resist attempts to tamper with the election"* and *"Return election results in a timely manner"* contained any responses with strong levels of distrust.

Question 5 of the survey collected open-ended text responses from participants regarding their trust in the existing paper-based system, by asking the question *"Briefly explain why you gave these ratings."*

The results of qualitative coding of these free-form responses is shown in Figure 16. Percentages are provided as a proportion of the total number of participants who provided a response to this question.

A substantial number of respondents provided comments to the effect that the existing paper-based system was a "mature system" (29.8%), had effective checks and balances in place (20.5%), maintained voter anonymity (19.2%), and was generally accessible to voters (15.2%), which positively affected their trust in the system. A representative response was provided by respondent #139:

**The Existing Paper-Based Voting System**
*Q5: Briefly explain why you gave these ratings.*

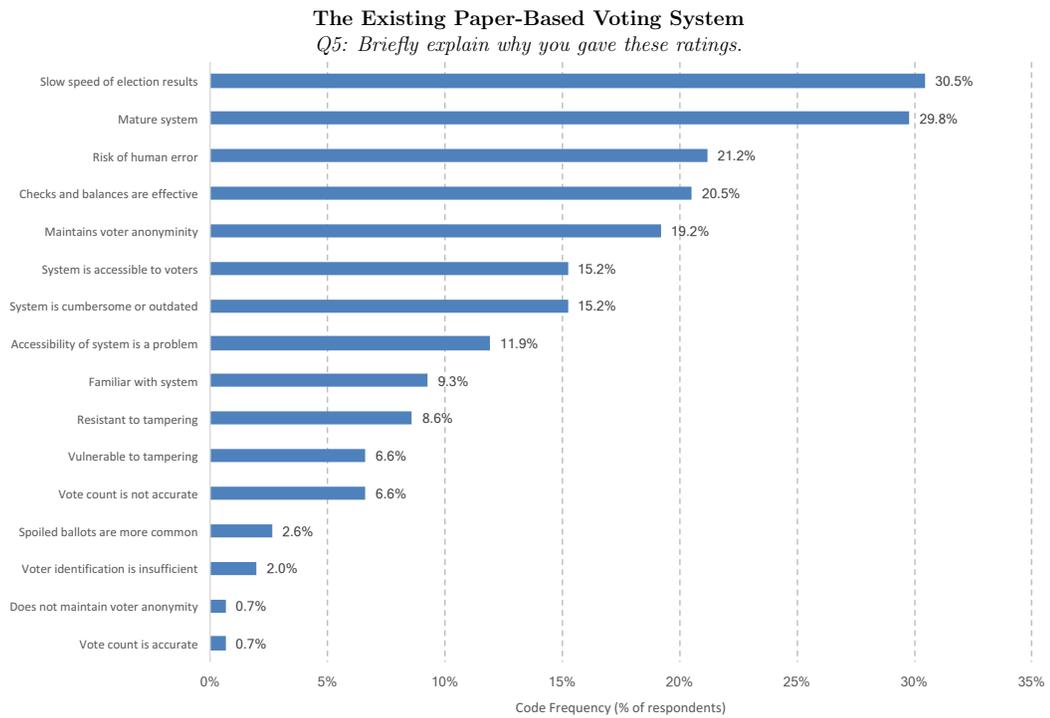

Fig. 16: Frequency of codes for Question 5

Tab. 3: Question 4 detailed response data

**The Existing Paper-Based Voting System**

*Q4: How much would you trust or distrust this system to meet the following requirements for an election?*

| | STRONGLY DISTRUST | DISTRUST | NEITHER TRUST NOR DISTRUST | TRUST | STRONGLY TRUST | DON'T KNOW | TOTAL | WEIGHTED AVERAGE |
|---|---|---|---|---|---|---|---|---|
| Accurately collect votes? | 0.00% 0 | 8.09% 14 | 12.14% 21 | 45.09% 78 | 34.68% 60 | 0.00% 0 | 173 | 4.06 |
| Accurately count votes? | 0.00% 0 | 15.03% 26 | 19.65% 34 | 44.51% 77 | 20.81% 36 | 0.00% 0 | 173 | 3.71 |
| Resist attempts to tamper with the election? | 2.31% 4 | 11.56% 20 | 18.50% 32 | 42.77% 74 | 24.28% 42 | 0.58% 1 | 173 | 3.76 |
| Protect the privacy of voters? | 0.00% 0 | 3.47% 6 | 6.94% 12 | 35.84% 62 | 53.76% 93 | 0.00% 0 | 173 | 4.40 |
| Ensure that all votes are kept anonymous? | 0.00% 0 | 1.73% 3 | 8.67% 15 | 32.95% 57 | 56.65% 98 | 0.00% 0 | 173 | 4.45 |
| Be accessible to all voters? | 0.00% 0 | 7.51% 13 | 11.56% 20 | 52.02% 90 | 28.32% 49 | 0.58% 1 | 173 | 4.02 |
| Be usable by all voters? | 0.00% 0 | 6.36% 11 | 12.72% 22 | 51.45% 89 | 28.90% 50 | 0.58% 1 | 173 | 4.03 |
| Return election results in a timely manner? | 8.09% 14 | 30.64% 53 | 16.18% 28 | 27.17% 47 | 17.92% 31 | 0.00% 0 | 173 | 3.16 |

*I think the system has served us well to date.*

A number of respondents (9.3%) specifically cited their level of familiarity with the system as a reason for their trust in the system:

*It's a system I'm familiar with and I trust the AEC and its processes.*

Conversely, a large number of respondents commented on the slow speed of results produced by the system (30.5%), the potential for human error during counting (21.2%), and that the system was generally cumbersome or outdated (15.2%). A representative response was provided by respondent #121:

*This method appears to be the most inefficient in returning results due to the high level of manual labour involved. It is also the most accessible as no electronics technology is required. This system has been in place for a long time and have been proven to work, however a more efficient system would be beneficial.*

In addition, a number of respondents expressed concern that the existing system was prone to inaccuracies when counting ballots (6.6%), vulnerable to tampering (6.6%), or did not provide sufficient identification of voters before casting a ballot (2%).

A representative response was provided by respondent #107:

> *A motivated individual could quite easily tamper with the election by providing false votes or in the central counting stage. Privacy and anonymity appears to protected most effectively via this method, but this reduces checks and balances on accuracy.*

## Voter Understanding

Questions 6 of the survey questionnaire asked participants to self-rate their understanding of how each component of the existing paper-based voting system functions during an election. As discussed in Section 4.3.5, it is to be expected that results for these questions will include a level of positive bias due to respondent familiarity with the existing system, when compared to the iVote and vVote systems.

The majority of respondents rated their understanding as 'good' for four of the five system components: voter registration; completing and casting ballots; counting of votes; and supervision of counting. Only a small minority of respondents reported having a 'poor' understanding of these four components, with this option chosen by less than 10% of respondents.

For the fifth voting system component, "verification of election results by voters", a large minority of 21.51% reported having a 'poor' understanding of how this component functioned. This is notable, as the existing paper-based voting system does not have any method for verification of election results by voters. These responses therefore indicate that a large amount of voters do not properly understand what "voter verification" means in this context.

Results showing self-reported voter understanding of each component of the existing system are shown in Figure 17, with full results shown in Table 4.

Question 7 asked respondents to what extent their understanding of the existing system impacted their level of trust in the system. 85% of respondents selected 'somewhat' or 'a great deal' for this question.

Results for this question can be seen in Figure 18 and Table 5.

## Detection of errors or tampering by election officials

Question 8 asked respondents whether they believed that unintentional errors or deliberate tampering with the existing system would be detected by election officials. Results of this question can be seen in Figure 19 and Table 6.

**The Existing Paper-Based Voting System**

*Q6: How would you rate your level of understanding of how each part of the voting system works?*

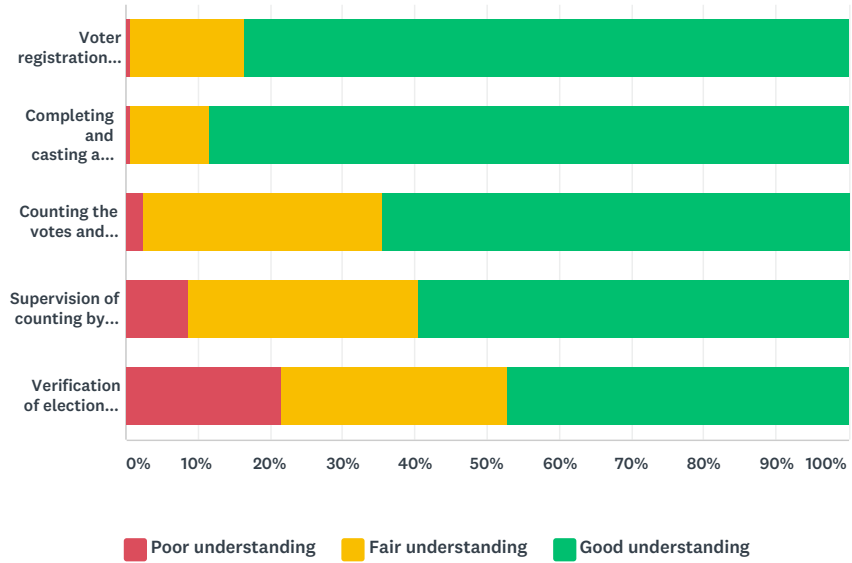

Fig. 17: Question 6 results

Tab. 4: Question 6 detailed response data

**The Existing Paper-Based Voting System**

*Q6: How would you rate your level of understanding of how each part of the voting system works?*

| | POOR UNDERSTANDING | FAIR UNDERSTANDING | GOOD UNDERSTANDING | TOTAL | WEIGHTED AVERAGE |
|---|---|---|---|---|---|
| Voter registration and obtaining ballot papers | 0.58%<br>1 | 15.79%<br>27 | 83.63%<br>143 | 171 | 2.83 |
| Completing and casting a ballot | 0.58%<br>1 | 10.98%<br>19 | 88.44%<br>153 | 173 | 2.88 |
| Counting the votes and producing a result | 2.33%<br>4 | 33.14%<br>57 | 64.53%<br>111 | 172 | 2.62 |
| Supervision of counting by candidates (scrutineering) | 8.67%<br>15 | 31.79%<br>55 | 59.54%<br>103 | 173 | 2.51 |
| Verification of election results by voters | 21.51%<br>37 | 31.40%<br>54 | 47.09%<br>81 | 172 | 2.26 |

**The Existing Paper-Based Voting System**

*Q7: To what extent do you think that your understanding of the system impacts your level of trust in the system?*

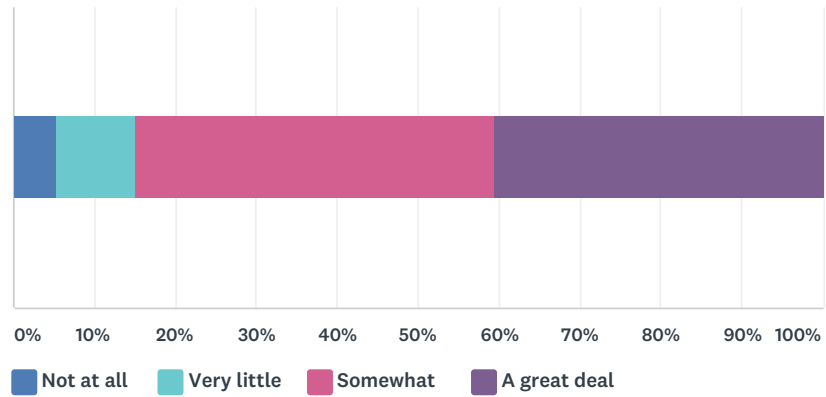

Fig. 18: Question 7 results

Tab. 5: Question 7 detailed response data

**The Existing Paper-Based Voting System**

*Q7: To what extent do you think that your understanding of the system impacts your level of trust in the system?*

| ANSWER CHOICES | RESPONSES | |
|---|---|---|
| Not at all | 5.20% | 9 |
| Very little | 9.83% | 17 |
| Somewhat | 44.51% | 77 |
| A great deal | 40.46% | 70 |
| TOTAL | | 173 |

**The Existing Paper-Based Voting System**

*Q8: If there were errors in counting votes or someone attempted to tamper with the system, do you believe that it would be detected by election officials?*

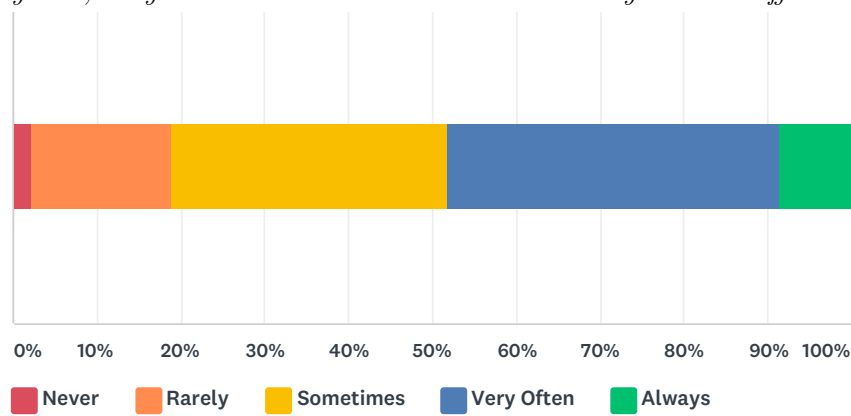

Fig. 19: Question 8 results - voter belief in detection of errors or tampering

Responses to this question were mixed, with a minority of respondents selecting the options 'never' or 'always' (2.30% and 8.62%, respectively). A relative majority of respondents selected 'very often' (39.66%), closely followed by 'sometimes' (32.76%).

Question 9 was an open-ended question which asked respondents to provide free-form responses explaining their answers to Question 8.

The results of qualitative coding of these free-form responses is shown in Figure 20. Percentages are provided as a proportion of the total number of participants who provided a response to this question.

A relative majority of respondents (45.93%) provided answers to the effect that "checks and balances are effective" when explaining their answer to Question 8. Note that this code also included responses citing the use of candidate-appointed scrutineers to observe the counting.

A representative response was provided by Respondent #51:

> *The number of checks and balances present, as well as the presence of officials from the major parties, with conflicting but strong interests make me confident any tampering would be picked up.*

A significant minority of respondents (17.78%) stated that their response to Question 8 was due to the existing system being reliant on humans, or not having any technology involved. This was seen as a negative by these respondents. This reflects the voters citing "human error" in responses to Question 5.

Tab. 6: Question 8 detailed response data

**The Existing Paper-Based Voting System**

*Q8: If there were errors in counting votes or someone attempted to tamper with the system, do you believe that it would be detected by election officials?*

| ANSWER CHOICES | RESPONSES | |
| --- | --- | --- |
| Never | 2.30% | 4 |
| Rarely | 16.67% | 29 |
| Sometimes | 32.76% | 57 |
| Very Often | 39.66% | 69 |
| Always | 8.62% | 15 |
| TOTAL | | 174 |

A representative response was provided by Respondent #19:

> *I think humans are more likely to make mistakes than machines and paper votes are more easily tampered with.*

### 4.5.3 The iVote Online Voting System

**Voter Trust**

As with the previous system, the first question for the iVote online voting system asked participants to rate their level of trust or distrust in the system to meet various requirements of an election. The results for this question can be seen in Figure 21 and Table 7. Figure 21 once again presents the results in the form of a stacked bar chart for ease of visualisation.

The results for the iVote system are extremely mixed, with three of the listed requirements—*"accurately collect votes", "accurately count votes"* and *"return election results in a timely manner"*—showing high levels of trust, while the remaining five showed substantial levels of distrust. In addition, a larger proportion of respondents entered a response of 'don't know' than for the existing paper-based voting system.

The highest level of trust was in the iVote system's ability to return election results in a timely manner, with 85.0% of respondents reporting that they 'trust' (39.9%) or 'strongly trust' (45.1%) the system to meet this requirement. A large number of respondents also reported that they 'trust' or 'strongly trust' the iVote system to *accurately count votes* (77.5%), and *accurately collect votes* (64.7%).

On the other side of the scale, only a minority of respondents reported that they trusted the iVote system to resist tampering attempts, maintain voter privacy and

**The Existing Paper-Based Voting System**

*Q9: In a few words, please explain your answer to the previous question.*

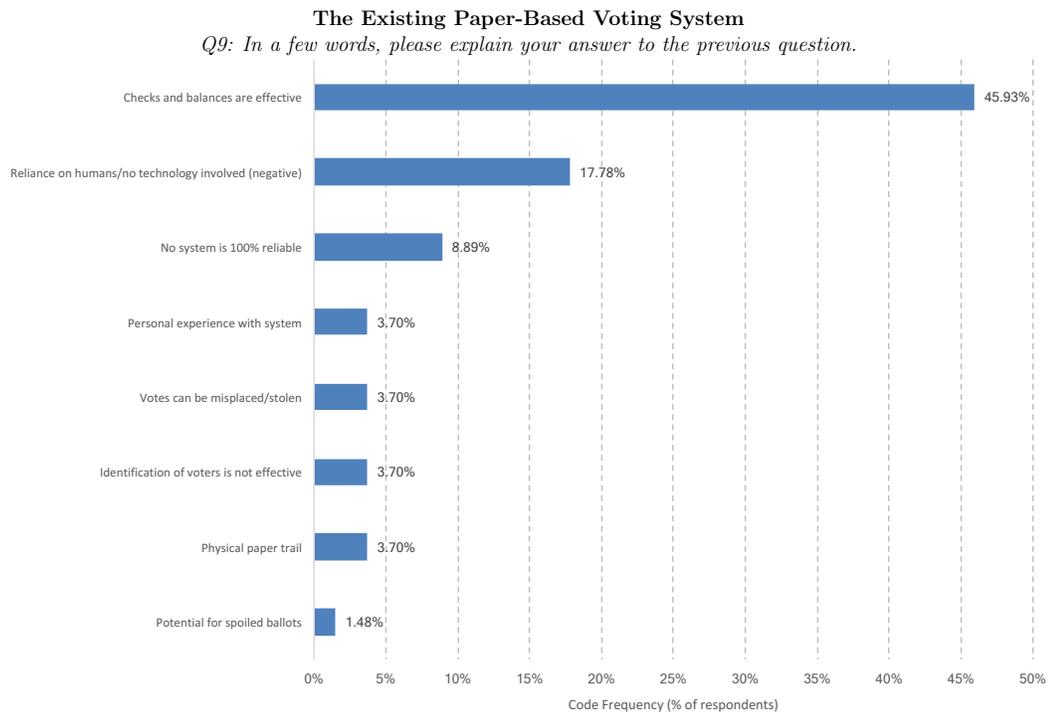

Fig. 20: Frequency of codes for Question 9

**The iVote Online Voting System**

*Q10: How much would you trust or distrust this system to meet the following requirements for an election?*

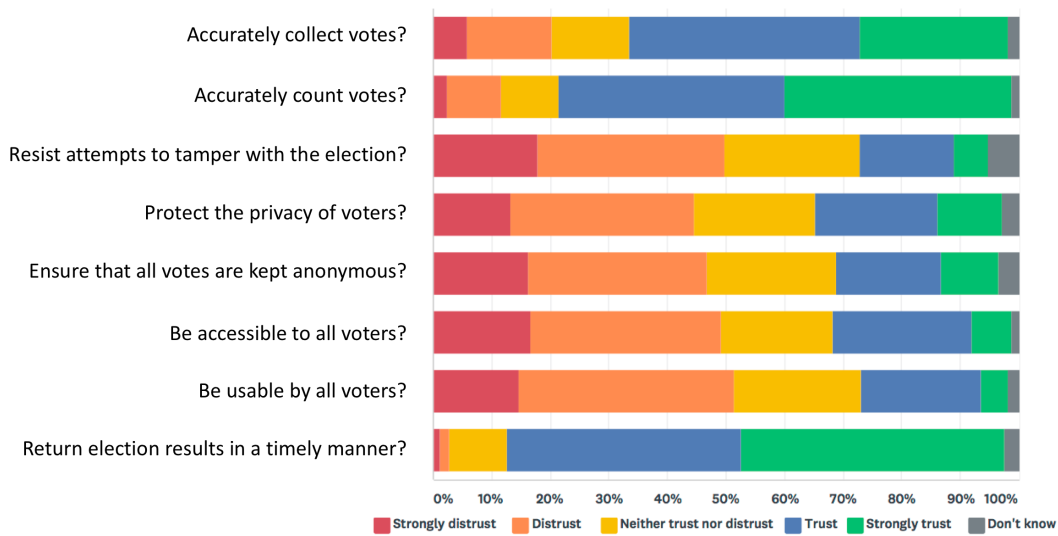

Fig. 21: Question 10 results

anonymity, and be accessible and usable by all voters, with approximately half of all respondents reporting that they would 'distrust' or 'strongly distrust' the system to meet each of these requirements.

Question 11 of the survey collected open-ended text responses from participants regarding their trust in the iVote online voting system, by asking the question *"Briefly explain why you gave these ratings."*

The results of qualitative coding of these free-form responses is shown in Figure 22 as a proportion of the total number of participants who provided a response to this question.

A relative majority of respondents (41.8%) expressed concern that the iVote system was vulnerable to cyber attack, through either remote hacking or methods such as denial of service. This total also includes responses which expressed generic concerns regarding possible comprosmise of internet-based systems.

A representative response was provided by respondent #47:

> *I do not trust digital storage. We so often see hacks on data no matter how high the level of protect (sic) and encryption. I believe something as important as an election would be a large target for foreign countries or hacking groups.*

Tab. 7: Question 10 detailed response data

**The iVote Online Voting System**

*Q10: How much would you trust or distrust this system to meet the following requirements for an election?*

| | STRONGLY DISTRUST | DISTRUST | NEITHER TRUST NOR DISTRUST | TRUST | STRONGLY TRUST | DON'T KNOW | TOTAL | WEIGHTED AVERAGE |
|---|---|---|---|---|---|---|---|---|
| Accurately collect votes? | 5.78% 10 | 14.45% 25 | 13.29% 23 | 39.31% 68 | 25.43% 44 | 1.73% 3 | 173 | 3.65 |
| Accurately count votes? | 2.31% 4 | 9.25% 16 | 9.83% 17 | 38.73% 67 | 38.73% 67 | 1.16% 2 | 173 | 4.04 |
| Resist attempts to tamper with the election? | 17.92% 31 | 31.79% 55 | 23.12% 40 | 16.18% 28 | 5.78% 10 | 5.20% 9 | 173 | 2.58 |
| Protect the privacy of voters? | 13.29% 23 | 31.21% 54 | 20.81% 36 | 20.81% 36 | 10.98% 19 | 2.89% 5 | 173 | 2.85 |
| Ensure that all votes are kept anonymous? | 16.18% 28 | 30.64% 53 | 21.97% 38 | 17.92% 31 | 9.83% 17 | 3.47% 6 | 173 | 2.74 |
| Be accessible to all voters? | 16.76% 29 | 32.37% 56 | 19.08% 33 | 23.70% 41 | 6.94% 12 | 1.16% 2 | 173 | 2.71 |
| Be usable by all voters? | 14.62% 25 | 36.84% 63 | 21.64% 37 | 20.47% 35 | 4.68% 8 | 1.75% 3 | 171 | 2.63 |
| Return election results in a timely manner? | 1.16% 2 | 1.73% 3 | 9.83% 17 | 39.88% 69 | 45.09% 78 | 2.31% 4 | 173 | 4.29 |

**The iVote Online Voting System**
*Q11: Briefly explain why you gave these ratings.*

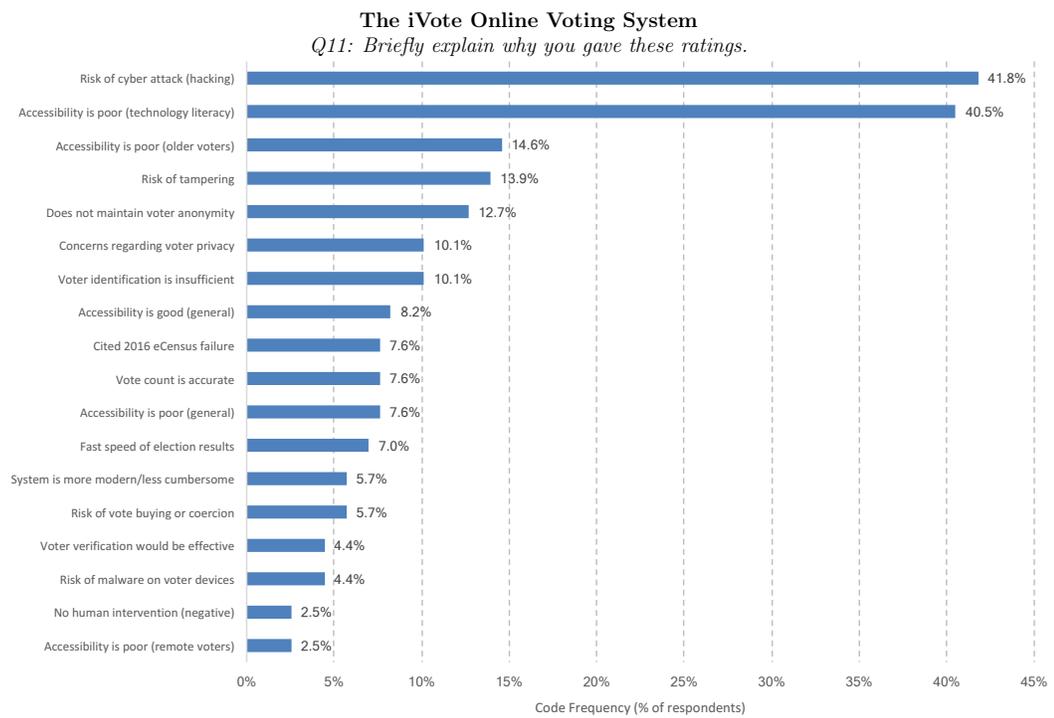

Fig. 22: Frequency of codes for Question 11

A smaller number of respondents (7.6%) specifically cited the issues experienced by the 2016 Australian eCensus as a contributor to their concerns about the risk of attacks against the iVote system. The 2016 eCensus suffered from a distributed denial of service attack which prevented participants from submitting census forms, and these issues were widely publicised during the census [118, 119].

Another group of respondents (13.9%) mentioned a risk of vote tampering, but did not specify that their concern was related to the internet-based nature of the iVote system. A number of respondents expressed concern about the identification of voters (10.1%), and the potential for voters to be bribed or coerced into voting in a particular way (5.7%). A small group of respondents specifically mentioned the potential for voter's devices to be compromised and thereby manipulate their vote (4.4%).

When it came to accessibility, a substantial number of respondents expressed concern that the iVote system would not be easily usable by certain voting demographics, such as voters who were not comfortable with or did not have easy access to technology (40.5%) or older voters (14.6%). Note that these two responses had a large degree of overlap, with many respondents mentioning both.

A representative response was provided by respondent #85:

> I think this system could easily be hacked or compromised. Many people, some older do not have a computer, tablet or smart phone or would have no idea how to use one. It would be very intimidating. My parents in their 80's would be horrified at being asked to vote electronically and would not be capable.

A smaller proportion of respondents commented that the iVote system had poor accessibility in general (7.6%), however this was narrowly exceeded by the proportion of respondents who commented the opposite: that general accessibilitty was good (8.2%).

Interestingly, a small number of respondents expressed concern that the iVote system would not easily be accessible to remote voters (2.5%) or voters with disabilities (1.3%). These numbers are not statistically significant, but the existance of these responses is notable given that the stated purpose of the iVote system is to improve accessability of voting to both demographics, as discussed in Sections 2.7.5 and 3.1.

A proportion of voters specifically mentioned concerns regarding the privacy of voters and the ability of the iVote system to keep votes anonymous (10.1% and 12.7%, respectively).

On a more positive note, 7.6% of voters expressed confidence in the ability of the iVote system to provide accurate results, and 7% commented that counting of ballots would be fast. As described by respondent #126:

> *Online voting would be far more convenient - for people who have access to a device and internet, and are computer literate. Automated counting would be immediate and accurate (not subject to human error). However, online systems are more easily tampered with - and the risk is centralised rather than dispersed (many human eyes counting paper votes). I don't believe that anonymity could possibly be guaranteed. Also why would votes be verified by phone? That seems very strange. I wonder also if staggered voting but immediate counting would sway people's votes - particularly if results were leaked or otherwise released prior to ballots closing.*

A small number of respondents commented on the telephone verification component of the iVote system, with 4.4% positive about the ability of the system to verify votes, while 1.3% claimed that the system would not provide them with any confidence about their vote.

One respondent (#137) commented that large-scale movement to an online voting system would negatively impact the political culture of Australia, by removing a community sense of acting together on polling day:

> *I've been a OIC for a Federal Election Polling Place, and I also develop database driven web application and iPhone/iPad apps for a living. In my opinion, the current online security levels are insufficient for online voting, and the technological expertise in the Australian community has insufficient spread to making voting accessible. (I base this assertion on user support experiences.)*
>
> *But most importantly, voting is not a mechanical process, it is a cultural and political activity The political culture of Australia would be severely damaged by online voting. It is a regressive proposal that would move political debate towards tweet psychology. It would remove the community's sense of acting together. I live in the online world and eagerly approve of most online developments, but this is not an activity that would benefit from online automation at the individual level. (It would be more effective to [automate] Polling Place activities with online software.)*

## Voter Understanding

Question 12 of the questionnaire asked participants to rate their understanding of how different components of the iVote online voting system functioned during an election.

**The iVote Online Voting System**

*Q12: How would you rate your level of understanding of how each part of the voting system works?*

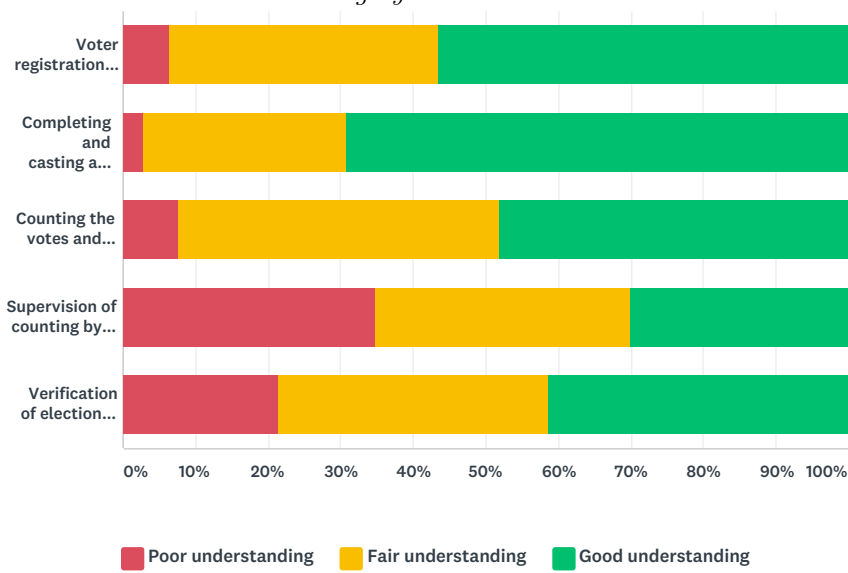

Fig. 23: Question 12 results

A majority of respondents rated their understanding as 'good' for two of the system components: voter registration (56.65%) and casting a ballot (69.36%). Compared with the existing paper-based voting system, significantly more respondents rated their understanding as 'poor' for "supervision of counting by candidates" (34.68%).

For the fifth component, "verification of election results by voters", the number of respondents rating their understanding as 'poor' was the same as that for the existing system, within margin of error (21.51%). This result is notable because the iVote system incorporates a phone-based vote verification method, while the existing paper-based system does not have any voter verification method at all.

This again indicates that a large amount of voters do not properly understand what "voter verification" means in this context.

Results showing self-reported voter understanding of each component of the iVote system are shown in Figure 23, with full results shown in Table 8.

Question 13 asked respondents to what extent their understanding of the iVote system components impacted their level of trust in the system. 81% of respondents selected 'somewhat' or 'a great deal' for this question.

Results for this question can be seen in Figure 24 and Table 9.

Tab. 8: Question 12 detailed response data

**The iVote Online Voting System**

*Q12: How would you rate your level of understanding of how each part of the voting system works?*

|  | POOR UNDERSTANDING | FAIR UNDERSTANDING | GOOD UNDERSTANDING | TOTAL | WEIGHTED AVERAGE |
|---|---|---|---|---|---|
| Voter registration and obtaining ballot papers | 6.36%  11 | 36.99%  64 | 56.65%  98 | 173 | 2.50 |
| Completing and casting a ballot | 2.89%  5 | 27.75%  48 | 69.36%  120 | 173 | 2.66 |
| Counting the votes and producing a result | 7.56%  13 | 44.19%  76 | 48.26%  83 | 172 | 2.41 |
| Supervision of counting by candidates (scrutineering) | 34.68%  60 | 35.26%  61 | 30.06%  52 | 173 | 1.95 |
| Verification of election results by voters | 21.51%  37 | 37.21%  64 | 41.28%  71 | 172 | 2.20 |

**The iVote Online Voting System**

*Q13: To what extent do you think that your understanding of the system impacts your level of trust in the system?*

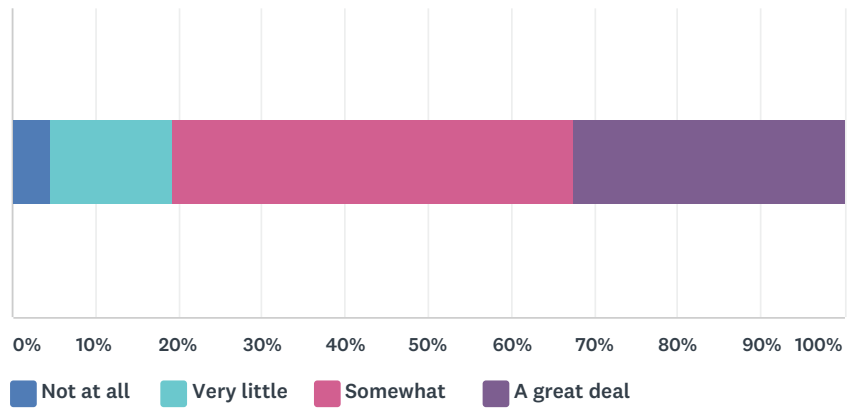

Fig. 24: Question 13 results

Tab. 9: Question 13 detailed response data

**The iVote Online Voting System**

*Q13: To what extent do you think that your understanding of the system impacts your level of trust in the system?*

| ANSWER CHOICES | RESPONSES | |
|---|---|---|
| Not at all | 4.65% | 8 |
| Very little | 14.53% | 25 |
| Somewhat | 48.26% | 83 |
| A great deal | 32.56% | 56 |
| TOTAL | | 172 |

Tab. 10: Question 14 detailed response data

**The iVote Online Voting System**

*Q14: If there were errors in counting votes or someone attempted to tamper with the system, do you believe that it would be detected by election officials?*

| ANSWER CHOICES | RESPONSES | |
|---|---|---|
| Never | 2.31% | 4 |
| Rarely | 27.17% | 47 |
| Sometimes | 46.24% | 80 |
| Very Often | 22.54% | 39 |
| Always | 1.73% | 3 |
| TOTAL | | 173 |

**Detection of errors or tampering by election officials**

Question 14 asked respondents whether they believed that unintentional errors or deliberate tampering with the existing system would be detected by election officials. Results of this question can be seen in Figure 25 and Table 10.

Responses to this question were substantially more negative than for the existing paper-based system, with almost 30% of respondents selecting 'never' or 'rarely', and a relative majority (46.24%) responding with 'sometimes'.

Only a minority of respondents were confident enough in the system to detect errors or deliberate tampering 'very often' (22.54%) or 'always' (1.73%).

Question 15 was an open-ended question which asked respondents to provide free-form responses explaining their answers to Question 14.

The results of qualitative coding of these free-form responses is shown in Figure 26. Percentages are provided as a proportion of the total number of participants who

**The iVote Online Voting System**

*Q14: If there were errors in counting votes or someone attempted to tamper with the system, do you believe that it would be detected by election officials?*

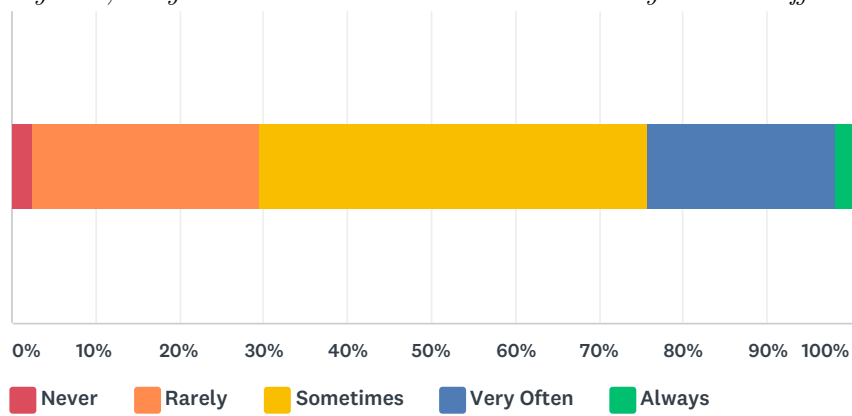

Fig. 25: Question 14 results - voter belief in detection of errors or tampering

provided a response to this question.

A relative majority of respondents (30%) provided answers which reflected concern about the risk of hacking or other cyber attack, and 6.43% of respondents specifically cited the problems with the 2016 Australian eCensus as a contributor to their response. In addition, a significant minority (17.14%) commented that they did not believe checks and balances were sufficient. Note that this code also incuded responses citing the lack of candidate-appointed scrutineers to observe counting of ballots.

A representative response was provided by Respondent #146:

> *The intelligence and military services of foreign governments will beat the AEC every day of the week. Sure, AEC will employ large external consulting groups to help them as they will not have the required expertise to run such a system internally. But, that did not prevent the stuff up on the last Census night. And that was just amateurs interfering. Wait till we encounter the real experts!!* (sic)

Other common responses included comments to the effect that their trust in the system to detect errors or tampering was "dependent on the implementation and technical skills of election officials" (16.43%).

Positive responses included comments to the effect that errors or tampering are easier to detect in electronic systems (15.71%) and that the checks and balances built into the iVote system were effective (4.29%).

A representative response was provided by Respondent #29:

> *There would be computer systems in charge of finding errors so more likely they would be found.*

### 4.5.4   The vVote Electronic Voting System

#### Voter Trust

As with the previous system, the first question for the vVote electronic voting system asked participants to rate their level of trust or distrust in the system to meet various requirements of an election. The results for this question can be seen in Figure 27 and Table 11. Figure 27 once again presents the results in the form of a stacked bar chart for ease of visualisation.

The results for the vVote system are similarly mixed as those for the iVote system. As with iVote, three of the listed requirements—*"accurately collect votes"*, *"accurately count votes"* and *"return election results in a timely manner"*—showed high levels of trust from respondents. The remaining five requirements showed higher levels of distrust, but to a lesser extent than for iVote. Once again, a larger proportion of respondents entered a response of 'don't know' than for the existing paper-based voting system.

Respondents reported the greatest level of trust in the vVote system's ability to return election results in a timely manner, with with 87.9% of respondents reporting that they 'trust' (37.0%) or 'strongly trust' (50.9%) the system to meet this requirement. Interestingly, the number of responses reporting strong trust in the vVote system to meet this requirement exceeded that for the iVote system (50.9% versus 45.1%, respectively).

A substantial majority of respondents trusted the vVote system to accurately collect and count votes, with over 80% of respondents reporting that they 'trust' or 'strongly trust' the system to meet both of these requirements.

The remaining five requirements showed higher levels of distrust among respondents, with approximately 30% of respondents reporting 'distrust' or 'strongly distrust' in the vVote system's ability to resist tampering attempts, maintain voter privacy and anonymity, and be accessible and usable by all voters. These five requirements also showed the highest proportion of responses in the 'neither trust nor distrust' and 'don't know' category.

Question 17 of the survey collected open-ended text responses from participants regarding their trust in the vVote electronic voting system, by asking the question *"Briefly explain why you gave these ratings."*

**The iVote Online Voting System**

*Q15: In a few words, please explain your answer to the previous question.*

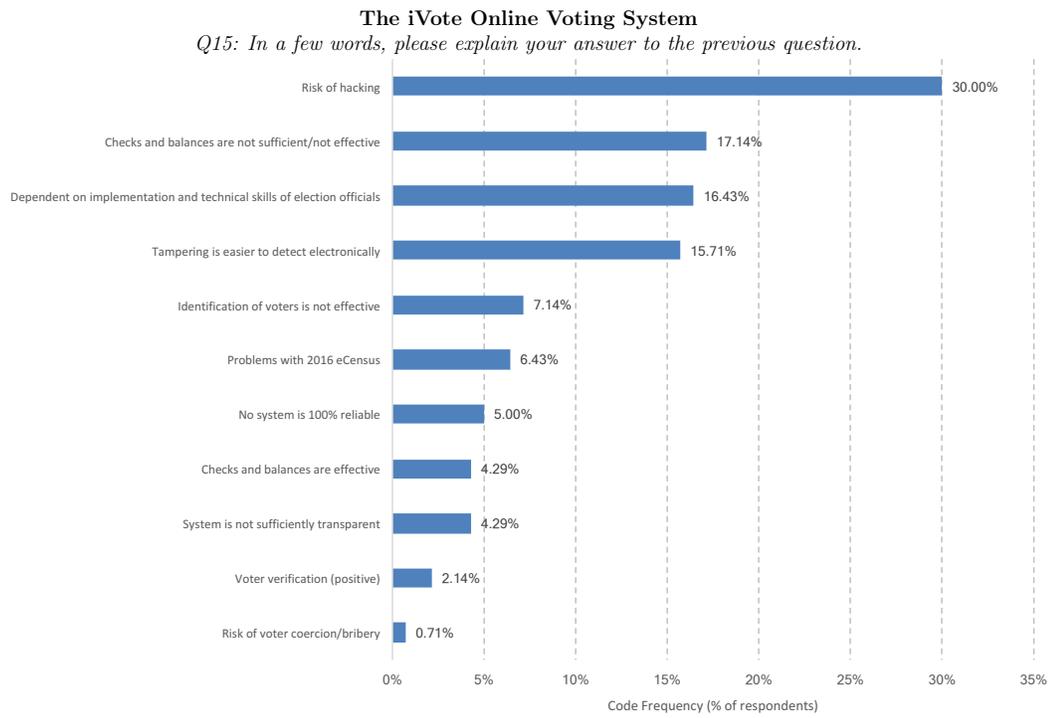

Fig. 26: Frequency of codes for Question 15

Tab. 11: Question 16 detailed response data

### The vVote Electronic Voting System

*Q16: How much would you trust or distrust this system to meet the following requirements for an election?*

| | STRONGLY DISTRUST | DISTRUST | NEITHER TRUST NOR DISTRUST | TRUST | STRONGLY TRUST | DON'T KNOW | TOTAL | WEIGHTED AVERAGE |
|---|---|---|---|---|---|---|---|---|
| Accurately collect votes? | 1.16% 2 | 9.25% 16 | 6.36% 11 | 52.02% 90 | 29.48% 51 | 1.73% 3 | 173 | 4.01 |
| Accurately count votes? | 0.58% 1 | 6.36% 11 | 5.20% 9 | 41.62% 72 | 45.09% 78 | 1.16% 2 | 173 | 4.26 |
| Resist attempts to tamper with the election? | 7.51% 13 | 20.23% 35 | 24.86% 43 | 28.90% 50 | 15.03% 26 | 3.47% 6 | 173 | 3.25 |
| Protect the privacy of voters? | 5.20% 9 | 20.23% 35 | 21.39% 37 | 31.79% 55 | 19.08% 33 | 2.31% 4 | 173 | 3.40 |
| Ensure that all votes are kept anonymous? | 6.43% 11 | 20.47% 35 | 19.30% 33 | 33.33% 57 | 17.54% 30 | 2.92% 5 | 171 | 3.36 |
| Be accessible to all voters? | 2.89% 5 | 21.39% 37 | 18.50% 32 | 37.57% 65 | 15.61% 27 | 4.05% 7 | 173 | 3.43 |
| Be usable by all voters? | 4.68% 8 | 23.98% 41 | 23.39% 40 | 30.99% 53 | 13.45% 23 | 3.51% 6 | 171 | 3.25 |
| Return election results in a timely manner? | 1.16% 2 | 2.31% 4 | 8.09% 14 | 36.99% 64 | 50.87% 88 | 0.58% 1 | 173 | 4.35 |

**The vVote Electronic Voting System**

*Q16: How much would you trust or distrust this system to meet the following requirements for an election?*

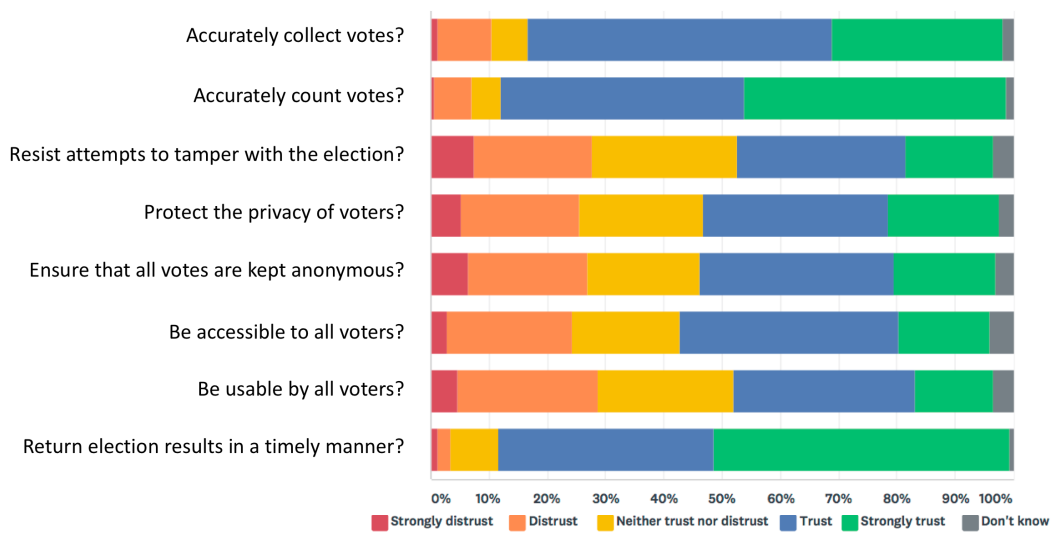

Fig. 27: Question 16 results

The results of qualitative coding of these free-form responses is shown in Figure 28 as a proportion of the total number of participants who provided a response to this question.

Among the open-ended responses, the most common comment was regarding the vVote system's vulnerability to cyber attacks such as hacking or denial of service, with 24.7% of respondents listing this as a concern. Related to this were responses commenting that the system would be vulnerable to tampering (10.7%), had privacy concerns (9.3%), and would not sufficiently keep votes anonymous (7.3%). Only a small number of respondents (4.0%) commented that they would trust the vVote system's voter verification feature.

A representative response was provided by respondent #47:

> *Once again, I found this system dangerous because of its leave of reliance on digital system that are open for hacking. The system could crash, the data could be stolen or altered. This would also open up the issue of people changing [their] mind and calling up claiming the system has the wrong vote recorded. The system then has no way to know if the person has just changed their mind and lying or there was a hack.* (sic)

And another by respondent #40:

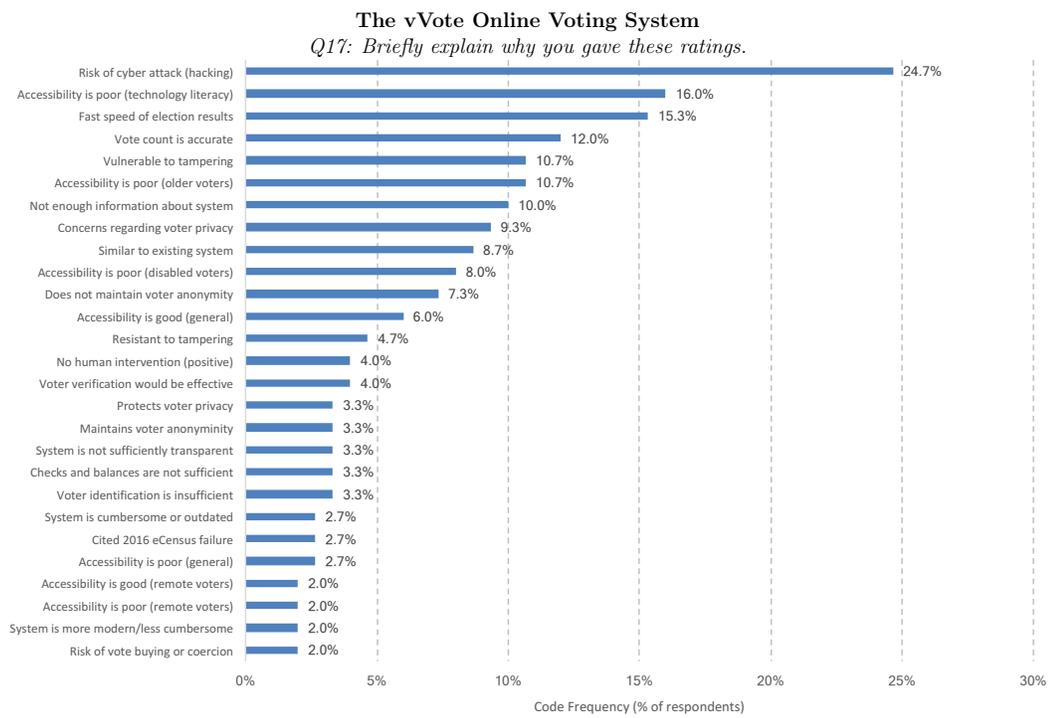

Fig. 28: Frequency of codes for Question 17

> *As one of the younger generations, I suppose I am skeptical of just how easy it is to track data. Numerous companies have faced litigation due to it, and I don't know if I would trust the integrity of an electronic system.*

A substantial number of respondents raised concerns regarding the accessibility and usability of the vVote system, with 16.0% of responses mentioning a potential difficulty for voters who were not comfortable with or did not have easy access to technology. A number of responses additionally raised concerns regarding accessibility for older voters (10.7%) and voters with disabilities (8.0%). A representative response was provided by respondent #110:

> *Seems like a straightforward process but would require the technology to be sound and working completely on the day and ensure there is redundancy should failure occur. Not sure how elderly people would cope.*

One respondent (#44) pointed out logistical and cost concerns when using a fully electronic voting system with existing polling places:

> *Most buildings where polling booths are located do not have sufficient power points/supply/access to be able to power numerous touch screens and other computers etc. A large portion of the population would require assistance to cast a vote electronically. The overall numbers of computer required is mammoth and so unsustainable, specially as they would need to be held in storage until the next election - this is also cost prohibitive and one of the reasons why timber voting screens were replaced by cardboard.*

In contrast to the above responses, 6.0% of respondents provided comments which were positive about the general level of accessibility of the system.

Positive responses were also provided regarding the vVote system's ability to provide fast election results (15.3%) and an accurate count (12.0%). A representative response was provided by respondent #37:

> *This system would make real time tracking of the results more accurate and quicker.*

10% of all respondents commented that they did not have sufficient information about the system. As discussed in Section 4.3.5, this is somewhat unsurprising given the lack of familiarity that respondents would have with the vVote system, and the relatively complex nature of the system's counting and voter verification features. A representative response was provided by respondent #142:

*I don't know enough about the security of the system to know if, once the electronic data is collected, it can't be tampered with, and linked back to the person who submitted it.*

## Voter Understanding

Question 18 of the questionnaire asked participants to rate their understanding of how different components of the vVote online voting system functioned during an election.

A majority of respondents rated their understanding as 'good' for three of the system components: voter registration (57.80%), casting a ballot (65.90%), and counting of votes (52.91%). As with the iVote system, significantly more respondents rated their understanding as 'poor' for "supervision of counting by candidates" (31.21%), when compared to the existing paper-based voting system.

For the fifth component, "verification of election results by voters", the number of respondents rating their understanding as 'poor' was the same as that for both the existing and iVote systems, within margin of error (21.97%). This result is again notable because the vVote system incorporates end-to-end voter verification, allowing voters to cryptographically confirm that their ballot was included in the final tally.

This again confirms that there is a lack of understanding among voters of what "voter verification" entails.

Results showing self-reported voter understanding of each component of the vVote system are shown in Figure 29, with full results shown in Table 12.

Question 13 asked respondents to what extent their understanding of the vVote system components impacted their level of trust in the system. 87.3% of respondents selected 'somewhat' or 'a great deal' for this question.

Results for this question can be seen in Figure 30 and Table 13.

## Detection of errors or tampering by election officials

Question 20 asked respondents whether they believed that unintentional errors or deliberate tampering with the existing system would be detected by election officials. Results of this question can be seen in Figure 31 and Table 14.

Negative responses to this question were on-par with those for the existing paper-based system, with approximately 20% of respondents selecting 'never' or 'rarely'.

**The vVote Electronic Voting System**

*Q18: How would you rate your level of understanding of how each part of the voting system works?*

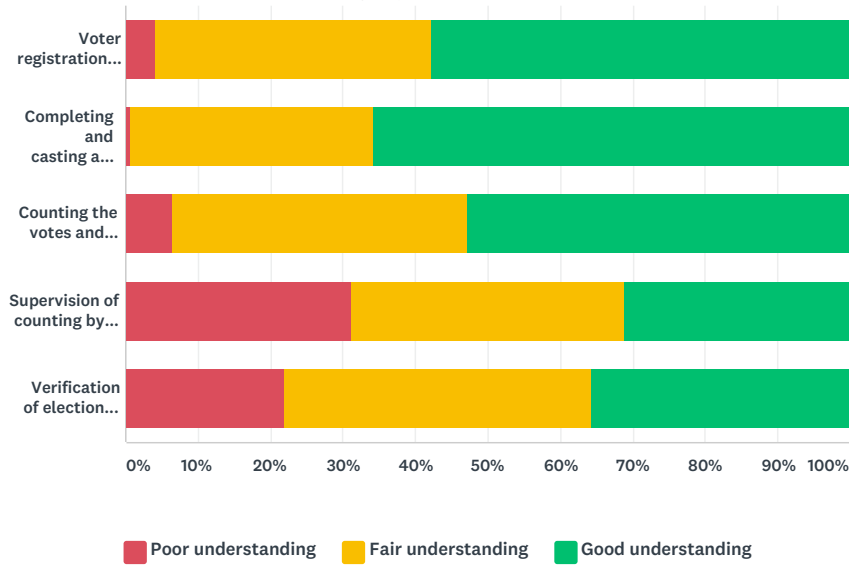

Fig. 29: Question 18 results

Tab. 12: Question 18 detailed response data

**The vVote Electronic Voting System**

*Q18: How would you rate your level of understanding of how each part of the voting system works?*

| | POOR UNDERSTANDING | FAIR UNDERSTANDING | GOOD UNDERSTANDING | TOTAL | WEIGHTED AVERAGE |
|---|---|---|---|---|---|
| Voter registration and obtaining ballot papers | 4.05%<br>7 | 38.15%<br>66 | 57.80%<br>100 | 173 | 2.54 |
| Completing and casting a ballot | 0.58%<br>1 | 33.53%<br>58 | 65.90%<br>114 | 173 | 2.65 |
| Counting the votes and producing a result | 6.40%<br>11 | 40.70%<br>70 | 52.91%<br>91 | 172 | 2.47 |
| Supervision of counting by candidates (scrutineering) | 31.21%<br>54 | 37.57%<br>65 | 31.21%<br>54 | 173 | 2.00 |
| Verification of election results by voters | 21.97%<br>38 | 42.20%<br>73 | 35.84%<br>62 | 173 | 2.14 |

**The vVote Electronic Voting System**

*Q19: To what extent do you think that your understanding of the system impacts your level of trust in the system?*

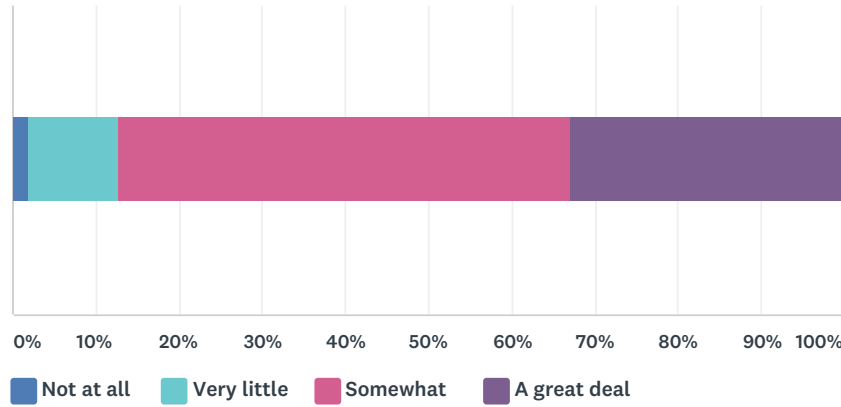

Fig. 30: Question 19 results

Tab. 13: Question 19 detailed response data

**The vVote Electronic Voting System**

*Q19: To what extent do you think that your understanding of the system impacts your level of trust in the system?*

| ANSWER CHOICES | RESPONSES | |
|---|---|---|
| Not at all | 1.73% | 3 |
| Very little | 10.98% | 19 |
| Somewhat | 54.34% | 94 |
| A great deal | 32.95% | 57 |
| TOTAL | | 173 |

**The vVote Electronic Voting System**

*Q20: If there were errors in counting votes or someone attempted to tamper with the system, do you believe that it would be detected by election officials?*

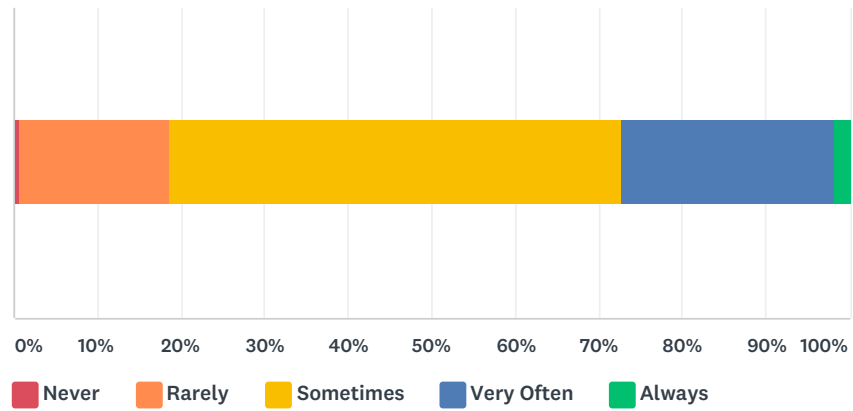

Fig. 31: Question 20 results - voter belief in detection of errors or tampering

A majority of respondents (54.07%) selected 'sometimes', a significantly greater amount than for the existing paper-based system.

As with the iVote system, only a minority of respondents were confident enough in the system to detect errors or deliberate tampering 'very often' (25.58%) or 'always' (1.74%).

Question 21 was an open-ended question which asked respondents to provide free-form responses explaining their answers to Question 20.

The results of qualitative coding of these free-form responses is shown in Figure 32. Percentages are again provided as a proportion of the total number of participants who provided a response to this question.

As with the iVote system, a relative majority of respondents (20.16%) provided answers reflecting concern about the risk of hacking or other cyber attack, and that the checks and balances were not sufficient (15.5%). Only a small number (1.55%) cited the 2016 Australian eCensus as a contributor to their response.

A representative response was provided by Respondent #40:

> *It can be hard to trace electronic fraud. No matter how good the anti-fraud process is, there's always someone ahead of the game. If people can hack pacemakers or insulin pumps, surely they'll find a way to hack voting machines.*

Tab. 14: Question 20 detailed response data

**The vVote Electronic Voting System**

*Q20: If there were errors in counting votes or someone attempted to tamper with the system, do you believe that it would be detected by election officials?*

| ANSWER CHOICES | RESPONSES | |
| --- | --- | --- |
| Never | 0.58% | 1 |
| Rarely | 18.02% | 31 |
| Sometimes | 54.07% | 93 |
| Very Often | 25.58% | 44 |
| Always | 1.74% | 3 |
| TOTAL | | 172 |

A smaller number of respondents commented that they did not have sufficient information about the system (9.30%) or that the system was not sufficiently transparent (3.88%).

As with the iVote system, another common response was that their belief in the likelihood of errors or tampering being detected was dependent on the implementation of the system and the technical skills of election officials (18.60%).

Positive responses included comments that errors or tampering are easier to detect in electronic systems (11.63%) and that the checks and balances built into the iVote system were effective (12.40%).

A representative response was provided by Respondent #129:

> *I think it would be relatively easy to build checks and balances into this system that would show up any errors or deliberate false responses.*

### 4.5.5 Demographic Comparison

As discussed in Section 4.4.2, a comparison was performed between two respondent age groups: respondents between 25 and 34, and respondents between 55 and 64. For the purposes of this comparison, the two groups will be referred to as "younger voters" and "older voters", respectively.

Tests for statistical significance were made when comparing responses, to a 95% confidence level ($p = 0.05$). Statistical significance was measured using the in-built comparison functionality of the SurveyMonkey results analysis tool.

**The vVote Electronic Voting System**

*Q21: In a few words, please explain your answer to the previous question.*

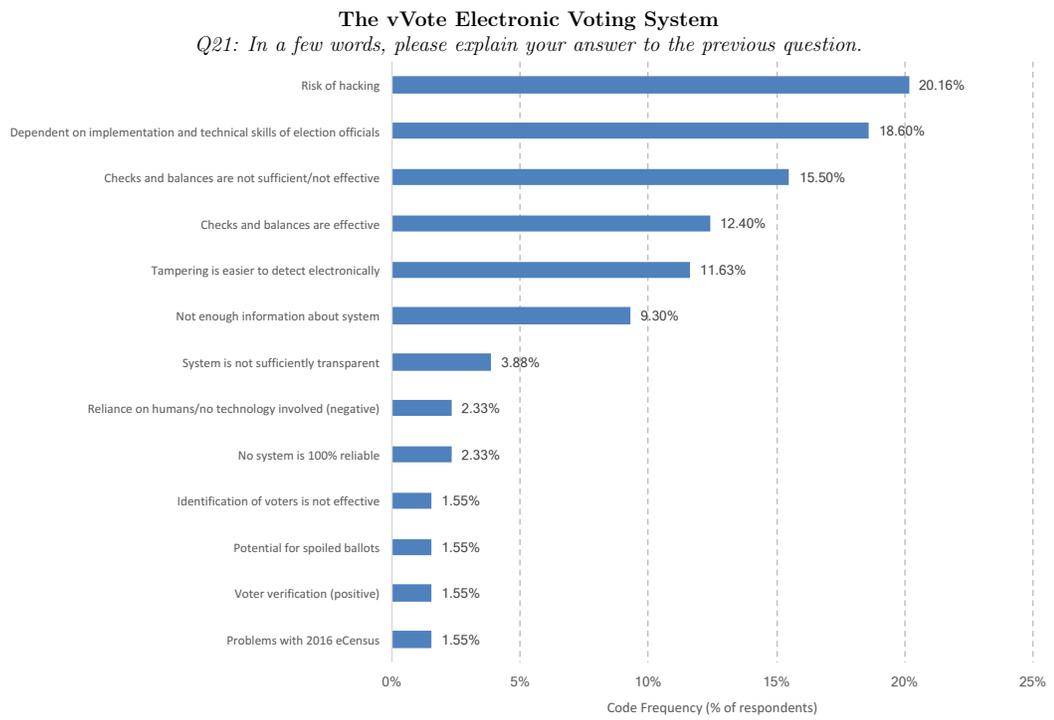

Fig. 32: Frequency of codes for Question 21

*Q3: What is the highest level of formal education you have completed?*

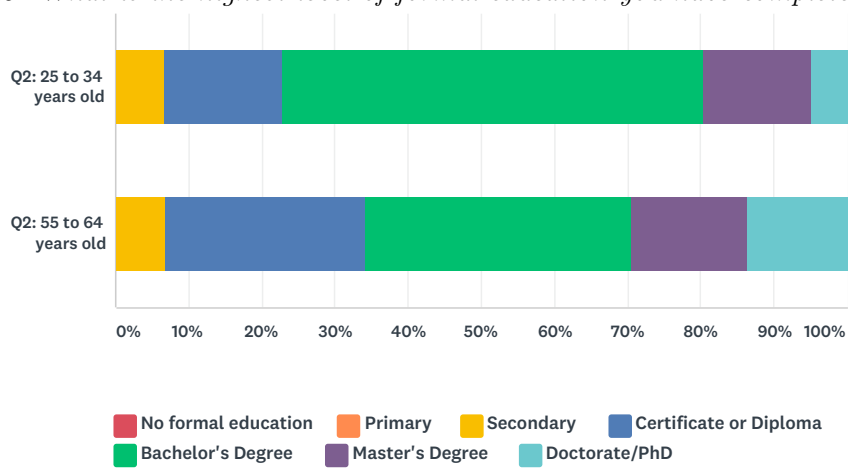

Fig. 33: Comparison of responses to Question 3, for age groups 25-34 and 55-64

## Education level

There was a statistically significant difference in the number of respondents who reported a Bachelor's Degree as their highest level of educational attainment between the younger voters (57.33%) and older voters (36.36%). A chart showing the comparison of these two age groups can be found in Figure 33.

## The Existing Paper-Based Voting System

For the existing paper-based system, differences in responses between the two age groups were not statistically significant for the majority of questions.

For the question of whether the voting system would "be accessible to all voters", both demographics responded with high levels of trust, but older voters were significantly more likely to respond that they had 'strong trust' in the system to meet this requirement, while younger voters were significantly more likely to respond with 'trust'.

When asked whether the system would "return election results in a timely manner", younger voters responded 'distrust' at a rate of 47.54%, significantly different to the 25.58% for the older demographic. Conversely, older voters responded with 'strongly trust' at a rate of 32.65%, compared with 4.92% for the younger demographic.

Finally, for the question regarding whether the participant believed that an error or attempt to tamper with the voting system would be detected by election officials,

**The Existing Paper-Based Voting System**

*Q8: If there were errors in counting votes or someone attempted to tamper with the system, do you believe that it would be detected by election officials?*

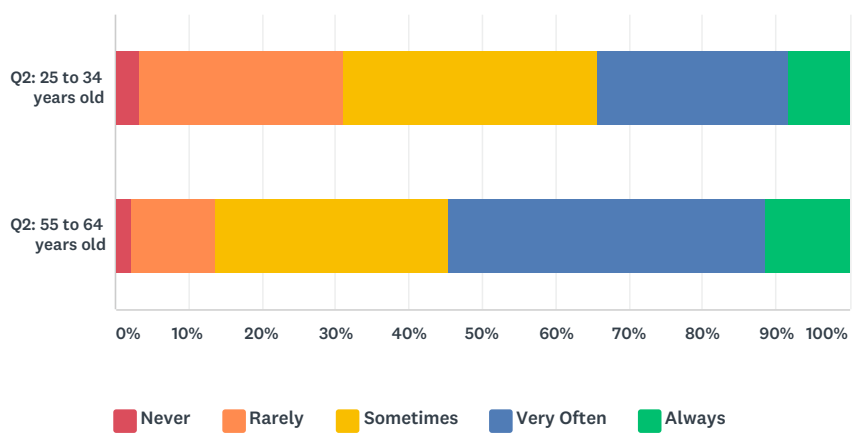

Fig. 34: Comparison of responses to Question 8, for age groups 25-34 and 55-64

younger voters responded with 'rarely' at a significantly higher rate than that for older voters (27.87% versus 11.36%).

These results can be seen in Figure 34.

**The iVote Online Voting System**

For the iVote system, a number of significant differences were observed.

For the question of whether the iVote system could be trusted to "accurately collect votes", older voters were significantly more likely to respond with 'neither trust not distrust' (11.60% versus 4.92%). This which may reflect a lower level of comfort and/or familiarity with online technologies among older respondents.

For the question of whether the iVote system could be trusted to "resist attempts to tamper with the election", younger voters were significantly more likely to respond with 'strongly distrust' (24.59% versus 9.30%). Younger voters also showed a higher rate of responses for 'distrust', though this difference was not statistically significant. For the question of whether the iVote system could be trusted to "ensure that all votes are kept anonymous", younger voters were significantly more likely to respond with 'strong distrust' (26.23% versus 6.98%).

As discussed in Section 2.7.1, a common argument for the use of online voting systems is that it may increase engagement among younger voters. On the contrary,

**The iVote Online Voting System**

*Q14: If there were errors in counting votes or someone attempted to tamper with the system, do you believe that it would be detected by election officials?*

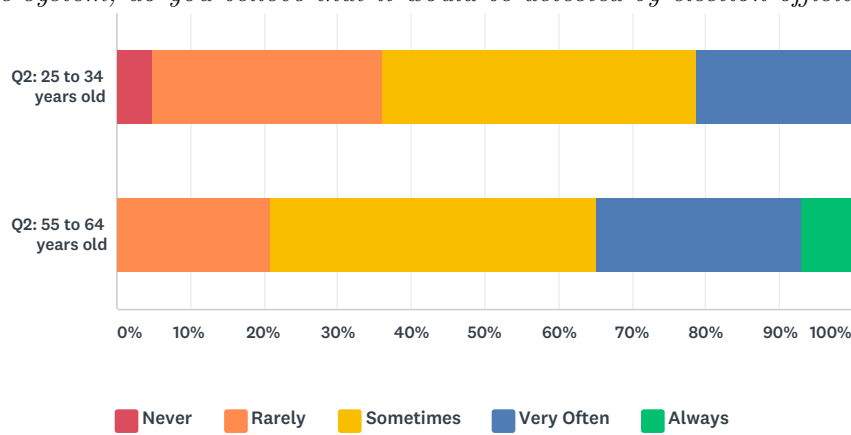

Fig. 35: Comparison of responses to Question 14, for age groups 25-34 and 55-64

these results indicate that younger voters are more likely to distrust online voting systems than their parents' generation.

Finally, for the question regarding whether the participant believed that an error or attempt to tamper with the iVote system would be detected by election officials, older voters responded with 'always' at a significantly higher rate than that for younger voters (6.98% versus 0%).

These results can be seen in Figure 35.

**The vVote Electronic Voting System**

The vVote system also showed several significant differences between the two demographics.

For the question of whether the vVote system could be trusted to "resist attempts to tamper with the election", older voters were significantly more likely to respond with 'strongly trust' (23.26% versus 8.20%). For the question of whether the vVote system could be trusted to "protect the privacy of voters", younger voters were significantly more likely to respond with 'distrust' (24.59% versus 9.30%). For the question of whether the vVote system could be trusted to "ensure that all votes are kept anonymous", older voters were significantly more likely to respond with 'neither trust nor distrust' (27.91% versus 11.48%).

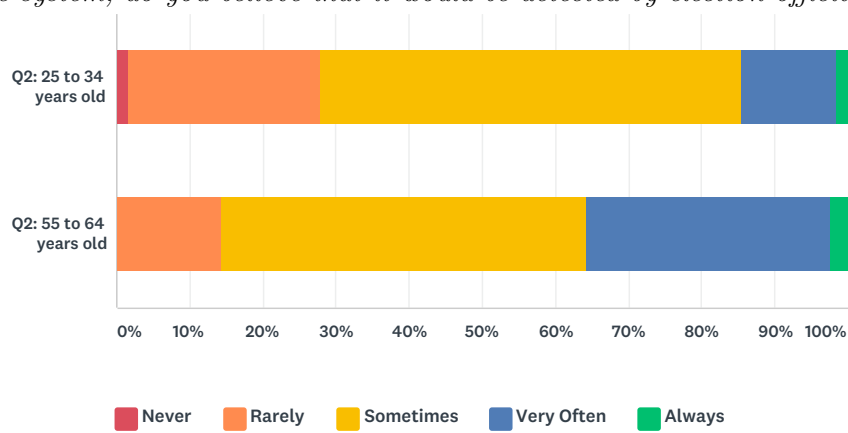

**The vVote Electronic Voting System**

*Q20: If there were errors in counting votes or someone attempted to tamper with the system, do you believe that it would be detected by election officials?*

Fig. 36: Comparison of responses to Question 20, for age groups 25-34 and 55-64

As with the iVote system, these results suggest that younger voters are more likely to distrust electronic voting systems than their parents' generation.

Finally, for the question regarding whether the participant believed that an error or attempt to tamper with the vVote system would be detected by election officials, older voters responded with 'very often' at a significantly higher rate than that for younger voters (33.33% versus 13.11%).

These results can be seen in Figure 36.

### 4.5.6 Discussion

#### iVote Accessibility

A substantial number of respondents reported concerns about the accessibility of the iVote system. This is an notable result, given that one of the core motivations for the introduction of the iVote system was to provide accessibility to voters who have difficulty voting using the existing paper-based system [140, 141].

As noted in Section 2.3.1, the existing paper-based voting system used for Australian federal elections does not provide a secret ballot for blind or visually impaired voters. The introduction of the iVote system was intended to address this issue, and properly afford a secret ballot to these voters.

It is important to note that blind or visually impaired persons are unlikely to have been represented in the survey responses, and would be much more likely to respond positively towards a system which is targeted towards their use. It would be useful to conduct a follow-up survey targeting this demographic, in order to obtain a measure of trust in the iVote system from the section of the community who the system is intended for.

### Voter Verification - iVote and vVote

From the results of the survey, it was apparent that voters do not sufficiently understand the meaning of "voter verification" as it is commonly applied to election systems.

In particular, answers to Question 6, 12 and 18 (the existing, iVote, and vVote systems, respectively) showed that respondents claimed the same level of understanding of how each system provided "verification of election results by voters", within the margin of error for the survey. This is notable, as each system has a very different capacity for voters to verify election results: the vVote system provides cryptographic verifiability for voters to ensure their vote was included in the count; the iVote system provides a telephone-based system to read back votes after they are cast; and the existing system has no voter verifiability at all.

One reason for these results could be that "verifiability" is poorly understood by voters, and the use of systems such as iVote and vVote (which claim voter-verifiable results) is not widespread. Therefore, it is likely that respondents are not familiar with the meaning of this term as it applies to election systems.

### Demographic Comparison

As discussed in Section 4.5.5, significant differences were seen regarding the trust in all three voting systems between younger voters and older voters. Overall, younger voters were less likely than older voters to trust the ability of iVote and vVote to resist tampering and protect the privacy and anonymity of voters. This is notable given that a common justification for the use of electronic voting systems is to appeal to younger voters and improve voter turnout, as discussed in Section 2.7.1.

It is important to note that younger voters were also less likely to respond with high levels of trust in the validity of the existing paper-based voting system, although these differences were not statistically significant.

A lack of trust in the election system among younger voters when compared to their parents generation was noted by Foa and Mounk in 2016 [142, 143]:

> *Across numerous countries, including Australia, Britain, the Netherlands, New Zealand, Sweden and the United States, the percentage of*

> *people who say it is "essential" to live in a democracy has plummeted,*
> *and it is especially low among younger generations.*

This finding was mirrored by the study conducted by Karp et al. discussed in Section 2.3.5. The use of voting systems which are trusted by voters will be an important component in reversing this trend, but is likely to require a focus on issues external to the election system itself.

**Voting System Selection**

The obvious question to be posed following these survey results is: "which voting system should we use?" The simple and unsatsifying answer is: "it depends". No voting system can encompass all use cases or requirements for an election, and selecting a system requires careful analysis of the risks and benefits of that system.

An argued benefit of online voting systems is the ability to enfranchise voters who cannot reach a polling booth on or prior to election day, and to extend a secret ballot to voters who are currently unable to vote without assistance. This is a minority of voters, but provides a worthy justification for the use of a well-designed online voting solution.

In addition, there are compelling arguments for the use of a well-designed and implemented poll-site electronic voting system during regular Australian federal elections, following the approaches of E2E voting systems such as STAR-Vote and vVote.

Note, however, that the results of this survey indicate that Australian voters do not currently trust electronic voting systems to the extent required for these systems to collect a large proportion of results in an Australian federal election, and this lack of trust is increased among younger voters.

Given the ongoing decline in trust towards the Australian electoral system demonstrated by Karp et al. [33], any increased use of electronic voting systems needs to be carefully considered if this trend is to be reversed. It is essential that any such systems be properly designed and implemented, and also that voters be properly convinced as to their benefits.

## 4.6    Survey Summary: The Importance of Trust and Voter Education

The results of this survey pose challenges for the design and implementation of future electronic voting systems to be used in Australian elections. For the most part, survey respondents are happy with the existing paper-based system, and are unlikely to trust electronic systems unless they are satisfied that concerns regarding security, privacy, and accessibility have been met.

Any future electronic voting system must be designed in such a way that a majority of voters will trust the result of an election conducted using the system. This will require both a focus on the technical security of the system, as well as the accessibility of this technical design to voters.

# 5   Future Work

**Analysis of Future Australian Electronic Voting Systems**

The increased usage of electronic voting systems in Australian elections remains a consideration at both the commonwealth and state level, with a number of state electoral commissions intending to either continue use of existing systems, or to conduct trials of systems in the future. There remains a large amount of security analysis which can be undertaken on existing systems or systems undergoing trials prior to full implementation.

The use of electronic voting systems is of particular interest as a method for replacing current remote voting methods such as the postal vote. This was a primary driver for the use of the iVote system in New South Wales in 2015. A compelling area of future research is to examine online voting options for Australian elections in greater detail, and explore the increased use of remote and absentee voting (both paper and electronic) and their impact on election security and voter trust in election outcomes.

The Electoral Commission of South Australia (ECSA), as part of its Disability Access and Inclusion Plan, intends to conduct a trial of "alternate electronic methods of voting" during the 2018 SA State Election [144]. This election is scheduled to take place on 17 March 2018. If possible, it would be beneficial for any such system to undergo analysis by qualified security researchers in partnership with the ECSA.

The New South Wales Electoral Commission (NSWEC) has planned an open tender process to update the iVote online voting system for the 2019 New South Wales state election, with a comprehensive Request for Proposals (RFP) scheduled to be released in November 2017 [145]. If possible, it would be beneficial to perform an analysis of the iVote system update in partnership with the NSWEC before it is deployed for the 2019 state election.

**Improvements to Survey Sample**

There are a number of improvements which could be made to the survey of Australian voters in any follow-up study, both in terms of the sample selection and survey questionnaire.

As noted in Section 4.4.1, the final sample size was lower than desired due to resource constraints, and resulted in a margin of error of 7.5%. Any follow-up survey would be well served by increasing this sample size to provide for a margin of error of 5% or lower.

In addition, as noted in Section 4.4.2, the final sample did not include any responses from participants with no formal education, or who had only completed a primary school education. This excludes a significant portion of eligible Australian voters from consideration in the survey results [137]. For future surveys, it would be beneficial to better target collection in order to include a representative sample from this demographic.

Finally, the survey questionnaire did not ask any demographic questions regarding the respondent's level of disability or whether or not they would be considered a remote voter. As noted in Sections 2.7.5, 3.1, and 4.5, improving accessibility for remote voters and voters with vision impairment is a primary design goal for a number of electronic voting systems., notably the iVote online voting system.

There is an opportunity for a follow-up survey to target these demographics in order to test the intended accessibility benefits of these systems, and to determine whether any such benefits come at the expense of decreased voter trust in those systems. This question was partially addressed in this survey, but did not specifically target the intended users of these electronic voting systems and did not collect any information to identify respondents who were members of these demographics.

**Improvements to Survey Questionnaire and System Descriptions**

The survey questionnaire could also be improved in several ways. As discussed in Section 4.4.4, a number of open-ended responses were unable to be used, due to the respondent not being aware that the ordering of each voting system page was randomised. If participants were informed of this behaviour prior to beginning the survey, these responses would be able to be incorporated.

As noted in Section 4.5.4, a number of respondents commented that they did not have enough information about the design and operation of the vVote system in order to properly answer questions regarding trust in the system. This issue was expected prior to collection, and was discussed more detail in Section 4.3.5. A trade-off was present between the ability to provide more information about each system to participants, without causing the voting system descriptions (and the survey overall) to become too long and onerous for participants.

One method to mitigate this problem would be for a future survey to specifically target each voting system with a single questionnaire, such that each participant receives only questions about one system. This would allow for more explanation to be provided regarding the design and operation of each system, at the expense of tripling the number of survey responses required for the same level of statistical accuracy.

# 6   Conclusion

In this thesis we have explored background and existing work in the field of electronic voting systems, including the unique security requirements of elections, the specific characteristics of the Australian federal election system, and case studies of electronic voting systems used in jurisdictions around the world.

We have presented an analysis of the iVote online voting system used for the Western Australian State Election in February and March 2017, including: the system design and security model; security issues associated with the use of a third-party Distributed Denial of Service mitigation service; and a discussion of risks introduced to the voting system as a result of this design. This analysis was presented to the International Joint Conference on Electronic Voting in October 2017.

We have designed and conducted a cross-sectional survey of Australian voters with the aim of determining the level of voter trust in three voting systems: the existing paper-based voting system, the iVote online voting system, and the vVote electronic voting system. We have presented the results of this survey, including a comparison of the trust held by voters towards components of each system, and have provided recommendations for improving voter trust in future electronic voting systems.

Finally, we have outlined future work which can be undertaken in this area, with the aim of improving both the security and voter trust of electronic voting systems used for Australian federal elections.

Voters require the ability to make informed decisions when using an election system, both regarding their vote and also the system they are using to cast it. The election is the foundation on which democratic government is built, and it is essential that this foundation be based on trustworthy election systems.

# References

[1] Bruce Schneier. American Elections Will Be Hacked. *The New York Times*, November 2016. URL: `http://www.nytimes.com/2016/11/09/opinion/american-elections-will-be-hacked.html`.

[2] Marshall D. Abrams and Michael V. Joyce. Trusted system concepts. *Computers & Security*, 14(1):45–56, January 1995. URL: `http://www.sciencedirect.com/science/article/pii/0167404895970256`, `doi:10.1016/0167-4048(95)97025-6`.

[3] Malcolm Stevens and Michael Pope. Data diodes. Technical Report DSTO-TR-0209, Defence Science and Technology Organisation, July 1995.

[4] Malcolm W Stevens. An Implementation of an Optical Data Diode. Technical Report DSTO-TR-0785, Defence Science and Technology Organisation, 1999.

[5] W. Hasselbring and R. Reussner. Toward trustworthy software systems. *Computer*, 39(4):91–92, April 2006. `doi:10.1109/MC.2006.142`.

[6] J. Bowen and V. Stavridou. Safety-critical systems, formal methods and standards. *Software Engineering Journal*, 8(4):189–209, July 1993. `doi:10.1049/sej.1993.0025`.

[7] Feng Hao and Peter Y. A. Ryan, editors. *Real-world Electronic Voting: Design, Analysis and Deployment*. Taylor & Francis, January 2016.

[8] T. Kohno, A. Stubblefield, A. D. Rubin, and D. S. Wallach. Analysis of an electronic voting system. In *2004 IEEE Symposium on Security and Privacy, 2004. Proceedings*, pages 27–40, May 2004. `doi:10.1109/SECPRI.2004.1301313`.

[9] National Museum of Australia. Defining Moments in Australian History: Secret Ballot Introduced, May 2017. URL: `http://www.nma.gov.au/online_features/defining_moments/featured/secret_ballot_introduced`.

[10] The Editors of Encyclopædia Britannica. The Australian Ballot, May 2017. URL: `https://www.britannica.com/topic/Australian-ballot`.

[11] Lorrie Faith Cranor and Simson Garfinkel. *Security and Usability: Designing Secure Systems that People Can Use*. "O'Reilly Media, Inc.", August 2005.

[12] Bruce Schneier. Security Design: Stop Trying to Fix the User, October 2016. URL: `https://www.schneier.com/blog/archives/2016/10/security_design.html`.

[13] Brenton Holmes. e-Voting: The Promise and the Practice. text, Parliament of Australia. Department of Parliamentary Services, Parliamentary Library, October 2012. URL: `http://www.aph.gov.au/About_Parliament/Parliamentary_Departments/Parliamentary_Library/pubs/BN/2012-2013/EVoting`.

[14] Matthew Bernhard, Josh Benaloh, J. Alex Halderman, Ronald L. Rivest, Peter Y. A. Ryan, Philip B. Stark, Vanessa Teague, Poorvi L. Vora, and Dan S. Wallach. Public Evidence from Secret Ballots. In *Electronic Voting*, Lecture Notes in Computer Science, pages 84–109. Springer, Cham, October 2017. URL: `https://link.springer.com/chapter/10.1007/978-3-319-68687-5_6`, doi:10.1007/978-3-319-68687-5_6.

[15] Australian Electoral Commission. Scrutineers, December 2010. URL: `http://www.aec.gov.au/Voting/scrutineers.htm`.

[16] Australian Electoral Commission. Overview of the AEC, August 2015. URL: `http://www.aec.gov.au/About_AEC/index.htm`.

[17] ABC/AAC. 'It's the 21st century': Turnbull, Shorten back move to e-voting, July 2016. URL: `http://www.abc.net.au/news/2016-07-10/election-2016-turnbull-shorten-back-electronic-voting/7584594`.

[18] Jason Cartwright. AIIA calls for electronic voting in Australia. *techAU*, July 2016. URL: `http://techau.com.au/aiia-calls-for-electronic-voting-in-australia/`.

[19] Ronald L. Rivest. On the Notion of 'Software Independence' in Voting Systems. *Philosophical Transactions of the Royal Society of London A: Mathematical, Physical and Engineering Sciences*, 366(1881):3759–3767, October 2008. URL: `http://rsta.royalsocietypublishing.org/content/366/1881/3759`, doi:10.1098/rsta.2008.0149.

[20] P. B. Stark and D. Wagner. Evidence-Based Elections. *IEEE Security Privacy*, 10(5):33–41, September 2012. doi:10.1109/MSP.2012.62.

[21] Scott Shackelford, Bruce Schneier, Michael Sulmeyer, Anne E. Boustead, Ben Buchanan, Amanda Craig, Trey Herr, Malekos Smith, and Jessica Zhanna. Making Democracy Harder to Hack: Should Elections Be Classified as 'Critical Infrastructure?'. SSRN Scholarly Paper ID 2852461, Social Science Research Network, Rochester, NY, October 2016. URL: `https://papers.ssrn.com/abstract=2870599`.

[22] If I Can Shop and Bank Online, Why Can't I Vote Online?, November 2011. URL: `https://www.verifiedvoting.org/resources/internet-voting/vote-online/`.

[23] G. Schryen and E. Rich. Security in Large-Scale Internet Elections: A Retrospective Analysis of Elections in Estonia, The Netherlands, and Switzerland. *IEEE Transactions on Information Forensics and Security*, 4(4):729–744, December 2009. URL: `http://ieeexplore.ieee.org/document/5272405/`, `doi:10.1109/TIFS.2009.2033230`.

[24] Chris Culnane, Rajeev Gore, and Vanessa Teague. Election explainer: why can't Australians vote online? URL: `http://theconversation.com/election-explainer-why-cant-australians-vote-online-57738`.

[25] Joint Standing Committee on Electoral Matters. Second interim report on the inquiry into the conduct of the 2013 federal election: An assessment of electronic voting options. Interim Report, The Parliament of the Commonwealth of Australia, Canberra, November 2014. URL: `http://www.aph.gov.au/Parliamentary_Business/Committees/Joint/Electoral_Matters/2013_General_Election/Second_Interim_Report/Preliminary_pages`.

[26] Greg Miller and Adam Entous. Declassified report says Putin 'ordered' effort to undermine faith in U.S. election and help Trump. *Washington Post*, January 2017. URL: `https://www.washingtonpost.com/world/national-security/intelligence-chiefs-expected-in-new-york-to-brief-trump-on-russian-hacking/2017/01/06/5f591416-d41a-11e6-9cb0-54ab630851e8_story.html`.

[27] Latanya Sweeney, Ji Su Yoo, and and Jinyan Zang. Voter Identity Theft: Submitting Changes to Voter Registrations Online to Disrupt Elections. *Technology Science*, September 2017. URL: `https://techscience.org/a/2017090601/`.

[28] Dan Goodin. Georgia's lax voting security exposed just in time for crucial special election, June 2017. URL: `https://arstechnica.com/tech-policy/2017/06/georgias-voting-system-is-uniquely-vulnerable-to-election-tampering-hackers/`.

[29] Jason Burke. Kenya election monitors urge losing candidates to accept poll results. *The Guardian*, August 2017. URL: `http://www.theguardian.com/world/2017/aug/10/kenya-election-observers-urge-defeated-candidates-to-accept-result`.

6 Conclusion

[30] Court annuls Kenya presidential election. *BBC News*, September 2017. URL: `http://www.bbc.com/news/world-africa-41123329`.

[31] Marine Le Pennetier, Leigh Thomas, and John Stonestreet. France drops electronic voting for citizens abroad over cybersecurity fears. *Reuters*, June 2017. URL: `http://www.reuters.com/article/us-france-election-cyber-idUSKBN16D233`.

[32] Dutch will hand count ballots due to hacking fears. *Reuters*, February 2017. URL: `https://www.reuters.com/article/us-netherlands-election-cyber/dutch-to-hand-count-ballots-due-to-hacking-fears-rlt-idUSKBN15G55A`.

[33] Jeffrey Karp, Alessandro Nai, Ferran Martinez i Coma, Max Grömping, and Pippa Norris. The Australian Voter Experience: Trust and confidence in the 2016 federal election. Technical report, Department of Government and International Relations, University of Sydney, January 2017. URL: `https://www.electoralintegrityproject.com/the-australian-voter-experience/`.

[34] Office of the Press Secretary. Statement by Secretary Johnson on the Designation of Election Infrastructure as a Critical Infrastructure Subsector, January 2017. URL: `https://www.dhs.gov/news/2017/01/06/statement-secretary-johnson-designation-election-infrastructure-critical`.

[35] Department of Homeland Security. Critical Infrastructure Sectors. URL: `https://www.dhs.gov/critical-infrastructure-sectors`.

[36] Kaveh Waddell. Why Elections Are Now Classified as 'Critical Infrastructure'. *The Atlantic*, January 2017. URL: `https://www.theatlantic.com/technology/archive/2017/01/why-the-government-classified-elections-as-critical-infrastructure/513122/`.

[37] Roland Wen, Vanessa Teague, and Richard Buckland. Best Practices for E-election Systems. *Computing Research and Education Association of Australasia (CORE) Supplementary Submission to the Inquiry into the 2010 Federal Election*, 2010.

[38] Scott Bennett and Rob Lundie. Australian electoral systems. Technical report, Parliament of Australia. Department of Parliamentary Services, 2007. URL: `http://www.aph.gov.au/About_Parliament/Parliamentary_Departments/Parliamentary_Library/pubs/rp/RP0708/08rp05`.

[39] The World Factbook — Central Intelligence Agency. URL: `https://www.cia.gov/library/publications/the-world-factbook/fields/2123.html`.

[40] Lisa Hill. On the Reasonableness of Compelling Citizens to 'Vote': The Australian Case. *Political Studies*, 50(1):80–101, March 2002. URL: `http://journals.sagepub.com/doi/abs/10.1111/1467-9248.00360`, `doi:10.1111/1467-9248.00360`.

[41] Tim Colebatch. Metre-long ballot paper means voters will need to read the fine print. *The Sydney Morning Herald*, August 2013. URL: `http://bit.ly/2rQ4H5S`.

[42] Gary Kemble. Size matters: voters grapple with mammoth ballot papers. *ABC News*, August 2013. URL: `http://www.abc.net.au/news/2013-08-22/size-matters3a-voters-grapple-with-mammoth-ballot-papers/4905092`.

[43] Roberto Di Cosmo. On privacy and anonymity in electronic and non electronic voting: the ballot-as-signature attack. April 2007. URL: `https://hal.archives-ouvertes.fr/hal-00142440`.

[44] J. Benaloh, T. Moran, L. Naish, K. Ramchen, and V. Teague. Shuffle-Sum: Coercion-Resistant Verifiable Tallying for STV Voting. *IEEE Transactions on Information Forensics and Security*, 4(4):685–698, December 2009. `doi:10.1109/TIFS.2009.2033757`.

[45] John Kelsey, Andrew Regenscheid, Tal Moran, and David Chaum. Attacking Paper-Based E2e Voting Systems. *Lecture Notes in Computer Science*, 6000:370, 2010. URL: `http://adsabs.harvard.edu/abs/2010LNCS.6000.370K`, `doi:10.1007/978-3-642-12980-3_23`.

[46] D. Chaum, A. Essex, R. Carback, J. Clark, S. Popoveniuc, A. Sherman, and P. Vora. Scantegrity: End-to-End Voter-Verifiable Optical- Scan Voting. *IEEE Security Privacy*, 6(3):40–46, May 2008. `doi:10.1109/MSP.2008.70`.

[47] Kun Peng and Feng Bao. A Design of Secure Preferential E-Voting. In *E-Voting and Identity*, pages 141–156. Springer, Berlin, Heidelberg, September 2009. URL: `https://link.springer.com/chapter/10.1007/978-3-642-04135-8_9`, `doi:10.1007/978-3-642-04135-8_9`.

[48] Zhe Xia, Chris Culnane, James Heather, Hugo Jonker, Peter YA Ryan, Steve Schneider, and Sriramkrishnan Srinivasan. Versatile Prêt à Voter: Handling multiple election methods with a unified interface. In *International Conference on Cryptology in India*, pages 98–114. Springer, 2010.

[49] Chris Culnane, Peter Y. A. Ryan, Steve Schneider, and Vanessa Teague. vVote: A Verifiable Voting System. *ACM Trans. Inf. Syst. Secur.*, 18(1):3:1–3:30, June 2015. URL: `http://doi.acm.org/10.1145/2746338`, `doi:10.1145/2746338`.

[50] Vision Australia calls for remote electronic voting system for blind and vision impaired voters, June 2016. URL: `http://www.abc.net.au/radionational/programs/breakfast/vision-australia-calls-for-remote-electronic-voting/7504870`.

[51] Australian Electoral Commission. Voting Options, November 2016. URL: `http://www.aec.gov.au/Voting/ways_to_vote/#early`.

[52] Australian Electoral Commission. 2016 Federal Election Key Facts and Figures, August 2016. URL: `http://www.aec.gov.au/Elections/Federal_Elections/2016/key-facts.htm`.

[53] Joint Standing Committee on Electoral Matters. Report on the 2007 federal election electronic voting trials: Interim report of the inquiry into the conduct of the 2007 election and matters related thereto. text, March 2009. URL: `http://www.aph.gov.au/Parliamentary_Business/Committees/House_of_Representatives_Committees?url=em/elect07/report.htm`.

[54] M J Keelty AO. Inquiry into the 2013 WA Senate Election. Technical report, December 2013. URL: `http://www.aec.gov.au/About_AEC/Publications/Reports_On_Federal_Electoral_Events/2013/keelty-report.htm`.

[55] Rob Lundie. The disputed 2013 WA Senate election, November 2013. URL: `http://www.aph.gov.au/About_Parliament/Parliamentary_Departments/Parliamentary_Library/FlagPost/2013/November/The_disputed_2013_WA_Senate_election`.

[56] Australian Associated Press. Fresh WA poll throws Canberra into spin. *News*. URL: `http://www.sbs.com.au/news/article/2014/02/20/fresh-wa-poll-throws-canberra-spin`.

[57] Joint Standing Committee on Electoral Matters. The 2013 Federal Election: Report on the conduct of the 2013 election and matters related thereto. text, The Parliament of the Commonwealth of Australia, Canberra, April 2015. URL: `http://www.aph.gov.au/Parliamentary_Business/Committees/Joint/Electoral_Matters/2013_General_Election/Final_Report`.

[58] Parliament of Australia. Double dissolution. URL: `http://www.aph.gov.au/Parliamentary_Business/Double_dissolution`.

[59] Australian Electoral Commission. Federal Election 2016 Central Senate Scrutiny: Frequently Asked Questions, July 2016.

[60] Berj Chilingirian, Zara Perumal, Ronald L. Rivest, Grahame Bowland, Andrew Conway, Philip B. Stark, Michelle Blom, Chris Culnane, and Vanessa Teague. Auditing Australian Senate Ballots. *arXiv:1610.00127 [cs]*, October 2016. arXiv: 1610.00127. URL: `http://arxiv.org/abs/1610.00127`.

[61] Benedict Brook. ABC pundit Antony Green says Australia needs to consider electronic voting to avoid long wait for results. *NewsComAu*, July 2016. URL: `http://bit.ly/2rQj0aq`.

[62] Benedict Brook. Call for Australia to follow Estonia's lead and embrace electronic voting. *NewsComAu*, July 2016. URL: `http://bit.ly/2rvvbGE`.

[63] Ronald L. Rivest and Madars Virza. Software Independence Revisited. In *Real-world Electronic Voting: Design, Analysis and Deployment*. Taylor & Francis, January 2016.

[64] Ken Thompson. Reflections on Trusting Trust. *Commun. ACM*, 27(8):761–763, August 1984. URL: `http://doi.acm.org/10.1145/358198.358210`, `doi:10.1145/358198.358210`.

[65] Roy Saltman. Effective Use of Computing Technology in Vote-Tallying. Final Project Report NBSIR 75-687, National Institute of Standards and Technology, Office of Federal Elections, March 1975. URL: `http://csrc.nist.gov/publications/nistpubs/NBS_SP_500-30.pdf`.

[66] Adam Aviv, Pavol Černy, Sandy Clark, Eric Cronin, Gaurav Shah, Micah Sherr, and Matt Blaze. Security Evaluation of ES&S Voting Machines and Election Management System. In *Proceedings of the Conference on Electronic Voting Technology*, EVT'08, pages 11:1–11:13, Berkeley, CA, USA, 2008. USENIX Association. URL: `http://dl.acm.org/citation.cfm?id=1496739.1496750`.

[67] Matt Blaze, Arel Cordero, Sophie Engle, Chris Karlof, Naveen Sastry, Micah Sherr, Till Stegers, and Ka-Ping Yee. Source code review of the Sequoia voting system. *State of California's Top to Bottom Review*, 2, 2007. URL: `http://www.chriskarlof.com/papers/sequoia-source-public-jul26.pdf`.

[68] Ariel J. Feldman, J. Alex Halderman, and Edward W. Felten. Security analysis of the Diebold AccuVote-TS voting machine. In *Proceedings of the 2007 USENIX/ACCURATE Electronic Voting Technology Workshop,*

2006.   URL: `https://www.usenix.org/event/evt07/tech/full_papers/` `feldman/feldman_html/`.

[69]  Ronald L. Rivest.  The ThreeBallot Voting System.  Working Paper, Caltech/MIT Voting Technology Project, October 2006. URL: `http://dspace.` `mit.edu/handle/1721.1/96593`.

[70]  Charlie Strauss. The trouble with triples: A critical review of the triple ballot (3ballot) scheme, part 1. *Verified Voting New Mexico*, 2006.

[71]  Andrew W Appel.   How to defeat rivest's threeballot voting system. *Manuskrypt, pazdziernik*, 2006.

[72]  Richard Carback, David Chaum, Jeremy Clark, John Conway, Aleksander Essex, Paul S. Herrnson, Travis Mayberry, Stefan Popoveniuc, Ronald L. Rivest, Emily Shen, Alan T. Sherman, and Poorvi L. Vora. Scantegrity II Municipal Election at Takoma Park: The First E2e Binding Governmental Election with Ballot Privacy. In *Proceedings of the 19th USENIX Conference on Security*, USENIX Security'10, pages 19–19, Berkeley, CA, USA, 2010. USENIX Association. URL: `http://dl.acm.org/citation.cfm?id=1929820.1929846`.

[73]  Richard T Carback, David Chaum, Jeremy Clark, Aleksander Essex, Travis Mayberry, Stefan Popoveniuc, and Ronald L Rivest. The Scantegrity Voting System and Its Use in the Takoma Park Elections. In *Real-World Electronic Voting: Design, Analysis and Deployment*, page 237. CRC Press, 2016.

[74]  Susan Bell, Josh Benaloh, Michael D Byrne, Dana DeBeauvoir, Bryce Eakin, Gail Fisher, Philip Kortum, Neal McBurnett, Julian Montoya, Michelle Parker, and others. STAR-Vote: A secure, transparent, auditable, and reliable voting system. *USENIX Journal of Election Technology and Systems (JETS)*, 1(1):8, 2013.

[75]  Josh Benaloh, Ronald Rivest, Peter Y. A. Ryan, Philip Stark, Vanessa Teague, and Poorvi Vora. End-to-end verifiablity. *Microsoft Research*, November 2016.  URL: `https://www.microsoft.com/en-us/research/publication/` `end-end-verifiablity/`.

[76]  Olivier Pereira and Dan S. Wallach.   Clash Attacks and the STAR-Vote System.   In *Electronic Voting*, Lecture Notes in Computer Science,  pages  228–247.  Springer,  Cham,  October  2017.   URL: `https://link.springer.com/chapter/10.1007/978-3-319-68687-5_14`, `doi:10.1007/978-3-319-68687-5_14`.

[77] Peter Y. A. Ryan. A Variant of the Chaum Voter-verifiable Scheme. In *Proceedings of the 2005 Workshop on Issues in the Theory of Security*, WITS '05, pages 81–88, New York, NY, USA, 2005. ACM. URL: `http://doi.acm.org/10.1145/1045405.1045414`, `doi:10.1145/1045405.1045414`.

[78] David Chaum, Peter Y. A. Ryan, and Steve Schneider. A Practical Voter-Verifiable Election Scheme. In *SpringerLink*, pages 118–139. Springer, Berlin, Heidelberg, September 2005. URL: `https://link-springer-com.proxy.library.adelaide.edu.au/chapter/10.1007/11555827_8`, `doi:10.1007/11555827_8`.

[79] Peter YA Ryan, Steve Schneider, and Vanessa Teague. Prêt à Voter—The Evolution of the Species. In *Real-World Electronic Voting: Design, Analysis and Deployment*, pages 309–342. CRC Press, 2016.

[80] Craig Burton, Chris Culnane, James Heather, Thea Peacock, Peter Ryan, Steve Schneider, Vanessa Teague, Roland Wen, Zhe Xia, and Sriramkrishnan Srinivasan. A Supervised Verifiable Voting Protocol for the Victorian Electoral Commission. *EVOTE 2012*, 2012. URL: `http://orbilu.uni.lu/handle/10993/26567`.

[81] Craig Burton, Chris Culnane, James Heather, Thea Peacock, Peter YA Ryan, Steve Schneider, Vanessa Teague, Roland Wen, Zhe Xia, and Sriramkrishnan Srinivasan. Using Prêt à Voter in Victoria State Elections. *EVT/WOTE*, 2, 2012.

[82] Craig Burton, Chris Culnane, and Steve Schneider. Secure and Verifiable Electronic Voting in Practice: the use of vVote in the Victorian State Election. *arXiv:1504.07098 [cs]*, April 2015. arXiv: 1504.07098. URL: `http://arxiv.org/abs/1504.07098`.

[83] Electoral Matters Committee. Inquiry into electronic voting. Technical report, Parliament of Victoria, May 2017. URL: `https://www.parliament.vic.gov.au/emc/inquiries/article/2825`.

[84] Roland Wenl and Richard Bucklandl. Problems with E-Voting in the 2014 Victorian State Election and Recommendations for Future Elections. Technical report, July 2015.

[85] Ed Gerck, C. Andrew Neff, Ronald L. Rivest, Aviel D. Rubin, and Moti Yung. The Business of Electronic Voting. In *Financial Cryptography*, Lecture Notes in Computer Science, pages 243–268. Springer, Berlin, Heidelberg, February 2001. URL: `https://link.springer.com/chapter/10.1007/3-540-46088-8_21`, `doi:10.1007/3-540-46088-8_21`.

[86] Drew Springall, Travis Finkenauer, Zakir Durumeric, Jason Kitcat, Harri Hursti, Margaret MacAlpine, and J. Alex Halderman. Security analysis of the Estonian Internet voting system. In *Proceedings of the 21st ACM Conference on Computer and Communications Security (CCS '14)*. ACM, 2014. URL: `http://dl.acm.org/citation.cfm?id=2660315`.

[87] iTnews. France withdraws electronic voting over hacking fears. *iTnews*, March 2017. URL: `http://www.itnews.com.au/news/france-withdraws-electronic-voting-over-hacking-fears-455086`.

[88] Kurt Eichenwald. Trump, Putin and the hidden history of how Russia interfered in the U.S. presidential election. *Newsweek*, January 2017. URL: `http://www.newsweek.com/trump-putin-russia-interfered-presidential-election-541302`.

[89] Rhiannon Lucy Cosslett. 12 ways to get even more of Generation Y voting | Rhiannon Lucy Cosslett. *The Guardian*, July 2016. URL: `http://www.theguardian.com/commentisfree/2016/jul/11/12-ways-get-generation-y-voting-referendum`.

[90] Polyas. Increase Voter Turnout: Activate Young Voters, August 2016. URL: `https://www.polyas.com/increase-voter-turnout/activate-young-voters`.

[91] Scytl. Scytl Online Voting. URL: `https://www.scytl.com/en/online-voting/`.

[92] Anne-Marie Oostveen and Peter Van Den Besselaar. Internet Voting Technologies and Civic Participation: The Users' Perspective. *Javnost - The Public*, 11(1):61–78, January 2004. URL: `http://dx.doi.org/10.1080/13183222.2004.11008847`, `doi:10.1080/13183222.2004.11008847`.

[93] D Bochsler. Can Internet Voting Increase Political Participation? Remote Electronic Voting and Turnout in the 2007 Parliamentary Elections. In *Internet and Voting'conference*, 2010.

[94] Priit Vinkel. Internet Voting in Estonia. In Peeter Laud, editor, *Information Security Technology for Applications*, Lecture Notes in Computer Science, pages 4–12. Springer Berlin Heidelberg, October 2011. DOI: 10.1007/978-3-642-29615-4_2. URL: `http://link.springer.com/chapter/10.1007/978-3-642-29615-4_2`.

[95] Ülle Madise and Priit Vinkel. Internet Voting in Estonia: From Constitutional Debate to Evaluation of Experience over Six Elections. In Tanel Kerikmäe,

editor, *Regulating eTechnologies in the European Union*, pages 53–72. Springer International Publishing, 2014. DOI: 10.1007/978-3-319-08117-5_4. URL: `http://link.springer.com/chapter/10.1007/978-3-319-08117-5_4`.

[96] Dan Goodin. Millions of high-security crypto keys crippled by newly discovered flaw, October 2017. URL: `https://arstechnica.com/information-technology/2017/10/crypto-failure-cripples-millions-of-high-security-keys-750k-estonian-ids/`.

[97] Kaspar Korjus. We told you about a potential security vulnerability. Here's our update., October 2017. URL: `https://medium.com/e-residency-blog/we-told-you-about-a-potential-security-vulnerability-heres-our-update-86e0411`

[98] Matus Nemec, Marek Sys, Petr Svenda, Dusan Klinec, and Vashek Matyas. The Return of Coppersmith's Attack: Practical Factorization of Widely Used RSA Moduli. February 2017. URL: `https://acmccs.github.io/papers/p1631-nemecA.pdf`.

[99] John Leyden. Estonia government locks down ID smartcards: Refresh or else, November 2017. URL: `https://www.theregister.co.uk/2017/11/03/estonian_e_id_lockdown/`.

[100] Kaspar Korjus. Digital ID cards now only work with new certificates, November 2017. URL: `https://medium.com/e-residency-blog/digital-id-cards-will-only-work-with-new-certificates-from-midnight-tonight-6`

[101] Ben Adida. Helios: Web-based Open-Audit Voting. In *USENIX security symposium*, volume 17, pages 335–348, 2008.

[102] Helios Voting. Helios Voting: FAQ. URL: `https://heliosvoting.org/faq`.

[103] Nicholas Chang-Fong and Aleksander Essex. The Cloudier Side of Cryptographic End-to-end Verifiable Voting: A Security Analysis of Helios. In *Proceedings of the 32Nd Annual Conference on Computer Security Applications*, ACSAC '16, pages 324–335, New York, NY, USA, 2016. ACM. URL: `http://doi.acm.org/10.1145/2991079.2991106`, `doi:10.1145/2991079.2991106`.

[104] Peter Ryan and Vanessa Teague. Pretty Good Democracy. In Bruce Christianson, James A. Malcolm, Vashek Matyáš, and Michael Roe, editors, *Security Protocols XVII*, number 7028 in Lecture Notes in Computer Science, pages 111–130. Springer Berlin Heidelberg, April 2009. DOI: 10.1007/978-3-642-36213-2_15. URL: `http://link.springer.com/chapter/10.1007/978-3-642-36213-2_15`.

[105] Rui Joaquim, Carlos Ribeiro, and Paulo Ferreira. Veryvote: A voter verifiable code voting system. In *International Conference on E-Voting and Identity*, pages 106–121. Springer, 2009.

[106] James Heather, Peter Ryan, and Vanessa Teague. Pretty Good Democracy for More Expressive Voting Schemes. In Dimitris Gritzalis, Bart Preneel, and Marianthi Theoharidou, editors, *Computer Security – ESORICS 2010*, number 6345 in Lecture Notes in Computer Science, pages 405–423. Springer Berlin Heidelberg, September 2010. DOI: 10.1007/978-3-642-15497-3_25. URL: `http://link.springer.com/chapter/10.1007/978-3-642-15497-3_25`.

[107] Ian Brightwell, Jordi Cucurull, David Galindo, and Sandra Guasch. *An overview of the iVote 2015 voting system*. Tech. rep. New South Wales Electoral Commission, 2015.

[108] J. Alex Halderman and Vanessa Teague. The New South Wales iVote System: Security Failures and Verification Flaws in a Live Online Election. In Rolf Haenni, Reto E. Koenig, and Douglas Wikström, editors, *E-Voting and Identity*, number 9269 in Lecture Notes in Computer Science, pages 35–53. Springer International Publishing, September 2015. DOI: 10.1007/978-3-319-22270-7_3. URL: `http://link.springer.com/chapter/10.1007/978-3-319-22270-7_3`.

[109] David Adrian, Karthikeyan Bhargavan, Zakir Durumeric, Pierrick Gaudry, Matthew Green, J. Alex Halderman, Nadia Heninger, Drew Springall, Emmanuel Thomé, Luke Valenta, Benjamin VanderSloot, Eric Wustrow, Santiago Zanella-Béguelin, and Paul Zimmermann. Imperfect Forward Secrecy: How Diffie-Hellman Fails in Practice. In *Proceedings of the 22Nd ACM SIGSAC Conference on Computer and Communications Security*, CCS '15, pages 5–17, New York, NY, USA, 2015. ACM. URL: `http://doi.acm.org/10.1145/2810103.2813707`, doi:10.1145/2810103.2813707.

[110] Scytl Australia Pty. Ltd. Submission to the Joint Standing Committee on Electoral Matters for the Inquiry into the 2015 NSW state election, August 2015.

[111] New South Wales Electoral Commission. Response from the NSW Electoral Commission to iVote Security Allegations, October 2015. URL: `http://www.elections.nsw.gov.au/about_us/plans_and_reports/ivote_reports/response_from_the_nsw_electoral_commission_to_ivote_security_allegations`.

[112] D. A. Dai Zovi and S. A. Macaulay. Attacking automatic wireless network selection. In *Proceedings from the Sixth Annual IEEE SMC Information Assurance Workshop*, pages 365–372, June 2005. `doi:10.1109/IAW.2005.1495975`.

[113] Hitesh Ballani, Paul Francis, and Xinyang Zhang. A Study of Prefix Hijacking and Interception in the Internet. In *Proceedings of the 2007 Conference on Applications, Technologies, Architectures, and Protocols for Computer Communications*, SIGCOMM '07, pages 265–276, New York, NY, USA, 2007. ACM. URL: `http://doi.acm.org/10.1145/1282380.1282411`, `doi:10.1145/1282380.1282411`.

[114] Rahul Hiran, Niklas Carlsson, and Phillipa Gill. Characterizing large-scale routing anomalies: A case study of the china telecom incident. In *International Conference on Passive and Active Network Measurement*, pages 229–238. Springer, 2013.

[115] Joint Standing Committee on Electoral Matters. Administration of the 2015 NSW Election and Related Matters. Technical report. URL: `https://www.parliament.nsw.gov.au/committees/inquiries/Pages/inquiry-details.aspx?activetab=Reports&pk=1704`.

[116] Western Australian Electoral Commission. 2017 Western Australian State General Election Procedures for Technology Assisted Voting, January 2017. URL: `https://www.elections.wa.gov.au/sites/default/files/media-icons/generic/Approved%20Procedures%20Interent%20Voting%202017%20new.pdf`.

[117] Chris Culnane, Mark Eldridge, Aleksander Essex, and Vanessa Teague. Trust Implications of DDoS Protection in Online Elections. In *Electronic Voting*, Lecture Notes in Computer Science, pages 127–145. Springer, Cham, October 2017. URL: `https://link.springer.com/chapter/10.1007/978-3-319-68687-5_8`, `doi:10.1007/978-3-319-68687-5_8`.

[118] Alastair MacGibbon. Review of the events surrounding the 2016 eCensus. Technical report, Office of the Cyber Security Special Adciser, November 2016. URL: `http://apo.org.au/node/70705`.

[119] Australian Senate: Economics References Committee. 2016 Census: issues of trust. Technical report, November 2016. URL: `http://www.aph.gov.au/Parliamentary_Business/Committees/Senate/Economics/2016Census/Report`.

[120] Igal Zeifman. The Bits and Bytes of Incapsula SSL Support, March 2013. URL: `https://www.incapsula.com/blog/incapsula-ssl-support-features.html`.

[121] Cloudflare Support. SSL FAQ, April 2017. URL: `http://support.cloudflare.com/hc/en-us/articles/204144518-SSL-FAQ`.

[122] Ben Herzberg and Yoav Cohen. How Incapsula Prevents Data Leaks, February 2017. URL: `https://www.incapsula.com/blog/how-incapsula-protects-against-data-leaks.html`.

[123] Yinzhi Cao, Song Li, and Erik Wijmans. (Cross-)Browser Fingerprinting via OS and Hardware Level Features. In *NDSS Symposium 2017*, 2017. URL: `yinzhicao.org/TrackingFree/crossbrowsertracking_NDSS17.pdf`.

[124] Ziv Leyes. How Anycast Works to Bring Content Closer to Your Visitors, February 2017. URL: `https://www.incapsula.com/blog/how-anycast-works.html`.

[125] Western Australian Electoral Commission. How to use iVote, March 2017. URL: `https://www.elections.wa.gov.au/ivote/how-use-ivote`.

[126] R. Kusters, T. Truderung, and A. Vogt. Clash Attacks on the Verifiability of E-Voting Systems. In *2012 IEEE Symposium on Security and Privacy*, pages 395–409, May 2012. `doi:10.1109/SP.2012.32`.

[127] Thomas Vissers, Tom Van Goethem, Wouter Joosen, and Nick Nikiforakis. Maneuvering Around Clouds: Bypassing Cloud-based Security Providers. In *Proceedings of the 22Nd ACM SIGSAC Conference on Computer and Communications Security*, CCS '15, pages 1530–1541, New York, NY, USA, 2015. ACM. URL: `http://doi.acm.org/10.1145/2810103.2813633`, `doi:10.1145/2810103.2813633`.

[128] Ehud Cohen. Making Your Website Invisible to Direct-to-Origin DDoS Attacks, July 2016. URL: `https://www.incapsula.com/blog/make-website-invisible-direct-to-origin-ddos-attacks.html`.

[129] Nick Sullivan. DDoS Prevention: Protecting The Origin, July 2013. URL: `http://blog.cloudflare.com/ddos-prevention-protecting-the-origin/`.

[130] Arlene Fink. *How to Conduct Surveys: A Step-by-Step Guide*. SAGE Publications, November 2012.

[131] Morris Hamburg. *Basic statistics: a modern approach*. Harcourt Brace Jovan-ovich, 1974.

[132] Shanti R Rao and Potluri M Rao. Sample Size Calculator, 2009. URL: `http://www.raosoft.com/samplesize.html`.

[133] Australian Electoral Commission. Size of the electoral roll and enrolment rate 2016, January 2017. URL: `http://www.aec.gov.au/Enrolling_to_vote/Enrolment_stats/national/2016.htm`.

[134] Jon A Krosnick and Stanley Presser. Question and questionnaire design. *Handbook of survey research*, 2(3):263–314, 2010.

[135] Philip Garland. Question Order Matters, March 2011. URL: `https://www.surveymonkey.com/blog/2011/03/02/question-order-matters/`.

[136] Australian Bureau of Statistics. Australian Demographic Statistics - Population by Age and Sex Tables, September 2017. URL: `http://www.abs.gov.au/AUSSTATS/abs@.nsf/Lookup/3101.0Main+Features1Mar%202017?OpenDocument`.

[137] Australian Bureau of Statistics. Education and Work, Australia, May 2016. Technical report, November 2016. URL: `http://www.abs.gov.au/ausstats/abs@.nsf/mf/6227.0`.

[138] Peng Lu. Comparing Distributions, June 2016. URL: `http://homework.uoregon.edu/pub/class/es202/ztest.html`.

[139] Johnny Saldana. *The Coding Manual for Qualitative Researchers*. SAGE, November 2015. Google-Books-ID: ZhxiCgAAQBAJ.

[140] Eliza Laschon. Online voting finally a thing for some electors in WA. *ABC News*, February 2017. URL: `http://www.abc.net.au/news/2017-02-23/wa-election-online-voting-finally-a-thing-for-some-people-in-wa/8294996`.

[141] Paris Cowan. Western Australia gets controversial iVote software. *iTnews*, February 2017. URL: `http://www.itnews.com.au/news/western-australia-gets-controversial-ivote-software-452358`.

[142] Roberto Stefan Foa and Yascha Mounk. The Democratic Disconnect. *Journal of Democracy*, 27(3):5–17, July 2016. URL: `https://muse.jhu.edu/article/623602`, `doi:10.1353/jod.2016.0049`.

[143] Amanda Taub. How Stable Are Democracies? 'Warning Signs Are Flashing Red'. *The New York Times*, November 2016. URL: `https://www.nytimes.com/2016/11/29/world/americas/western-liberal-democracy.html`.

[144] Electoral Commission South Australia. Disability Access and Inclusion Plan, September 2017. URL: `http://www.ecsa.sa.gov.au/about-ecsa/corporate-publications/10-about/494-disability-access-and-inclusion-plan`.

[145] NSW Electoral Commission. iVote Refresh Program - Procurement Strategy. Technical report, June 2017. URL: `http://www.elections.nsw.gov.au/about_us/plans_and_reports/ivote_reports/2017_reports`.

# A   Survey Questionnaire

The full survey questionnaire, including all questions in their default ordering, is located in this appendix. These pages were generated using the "print survey" functionality within the SurveyMonkey software used to conduct the survey.

The embedded videos used for the survey introduction and to describe each voting system are not visible in the printed questionnaire. These videos can be found at the following links.

1. Introduction video:

    `https://www.youtube.com/watch?v=hGKekNHWk2s`

2. The Existing Paper-Based Voting System:

    `https://www.youtube.com/watch?v=oHpL8TYGG4o`

3. The iVote Online Voting System:

    `https://www.youtube.com/watch?v=qJNZzlIxlx0`

4. The vVote Electronic Voting System:

    `https://www.youtube.com/watch?v=T6USNQWy7kI`

# A Survey of Australian Voters: Trust in Voting Systems

## 1. Introduction

Thank you for participating in this survey about electronic voting systems. Your opinions are important and will be used to inform future voting systems which can be trusted by all Australians.

Please watch the brief introduction video below, and then click "next".

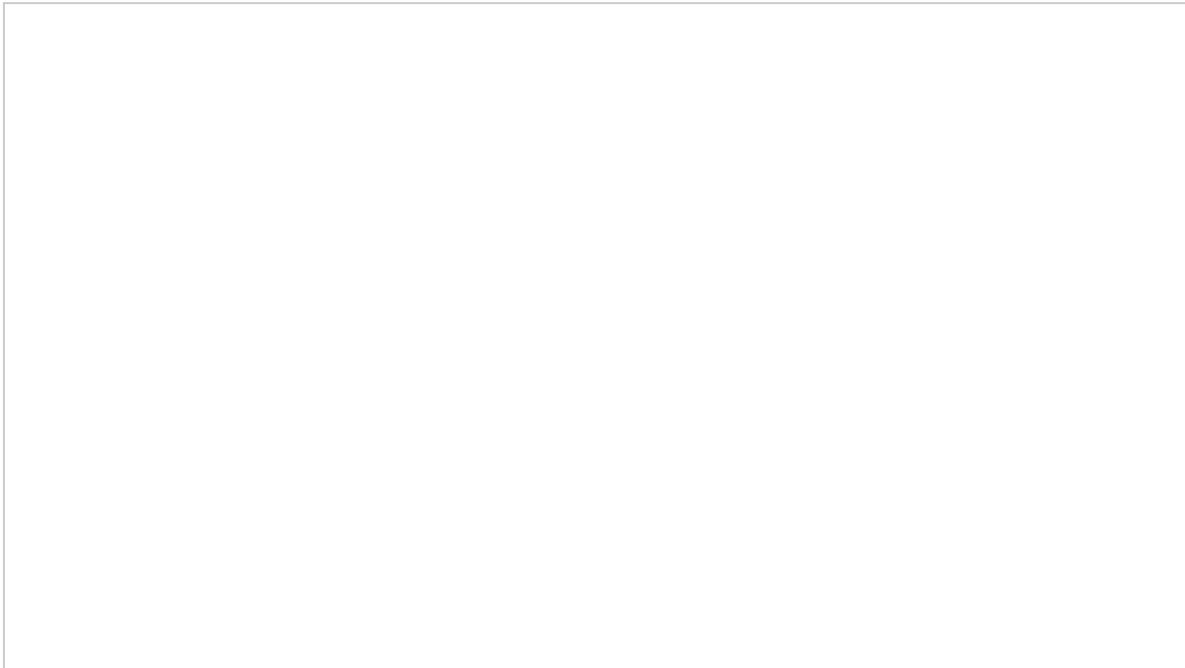

## 2. Demographic Information

**\* 1. Are you an Australian citizen?**

○ Yes

○ No

**\* 2. What is your age?**

○ Under 18 years old

○ 18 to 24 years old

○ 25 to 34 years old

○ 35 to 44 years old

○ 45 to 54 years old

○ 55 to 64 years old

○ over 64 years old

**3. What is the highest level of formal education you have completed?**

○ No formal education

○ Primary

○ Secondary

○ Certificate or Diploma

○ Bachelor's Degree

○ Master's Degree

○ Doctorate/PhD

## 3. The Existing Paper-Based Voting System

These questions relate to the existing paper-based voting system for Australian Federal Elections.

There is a short video below describing how the system works. Please watch the video, and then answer each of the questions.

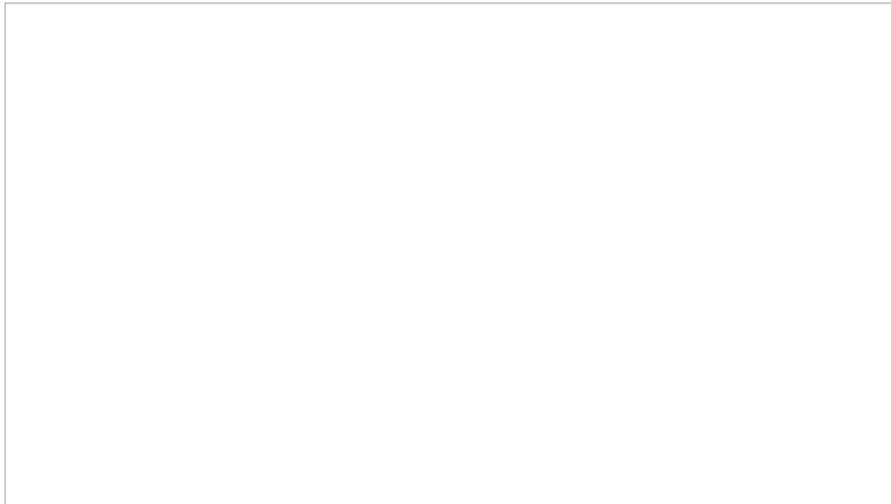

**4. How much would you trust or distrust this system to meet the following requirements for an election?**

|  | Strongly distrust | Distrust | Neither trust nor distrust | Trust | Strongly trust | Don't know |
|---|---|---|---|---|---|---|
| Accurately collect votes? | ○ | ○ | ○ | ○ | ○ | ○ |
| Accurately count votes? | ○ | ○ | ○ | ○ | ○ | ○ |
| Resist attempts to tamper with the election? | ○ | ○ | ○ | ○ | ○ | ○ |
| Protect the privacy of voters? | ○ | ○ | ○ | ○ | ○ | ○ |
| Ensure that all votes are kept anonymous? | ○ | ○ | ○ | ○ | ○ | ○ |
| Be accessible to all voters? | ○ | ○ | ○ | ○ | ○ | ○ |
| Be usable by all voters? | ○ | ○ | ○ | ○ | ○ | ○ |
| Return election results in a timely manner? | ○ | ○ | ○ | ○ | ○ | ○ |

**5. Briefly explain why you gave these ratings.**

**6. How would you rate your level of understanding of how each part of the voting system works?**

|  | Poor understanding | Fair understanding | Good understanding |
|---|:---:|:---:|:---:|
| Voter registration and obtaining ballot papers | ○ | ○ | ○ |
| Completing and casting a ballot | ○ | ○ | ○ |
| Counting the votes and producing a result | ○ | ○ | ○ |
| Supervision of counting by candidates (scrutineering) | ○ | ○ | ○ |
| Verification of election results by voters | ○ | ○ | ○ |

**7. To what extent do you think that your understanding of the system impact your level of trust in the system?**

○ Not at all

○ Very little

○ Somewhat

○ A great deal

**8. If there were errors in counting votes or someone attempted to tamper with the system, do you believe that it would be detected by election officials?**

○ Never

○ Rarely

○ Sometimes

○ Very Often

○ Always

**9. In a few words, please explain your answer to the previous question.**

## 4. The iVote Online Voting System

These questions relate to the iVote Online Voting System.

There is a short video below describing how the system works. Please watch the video, and then answer each of the questions.

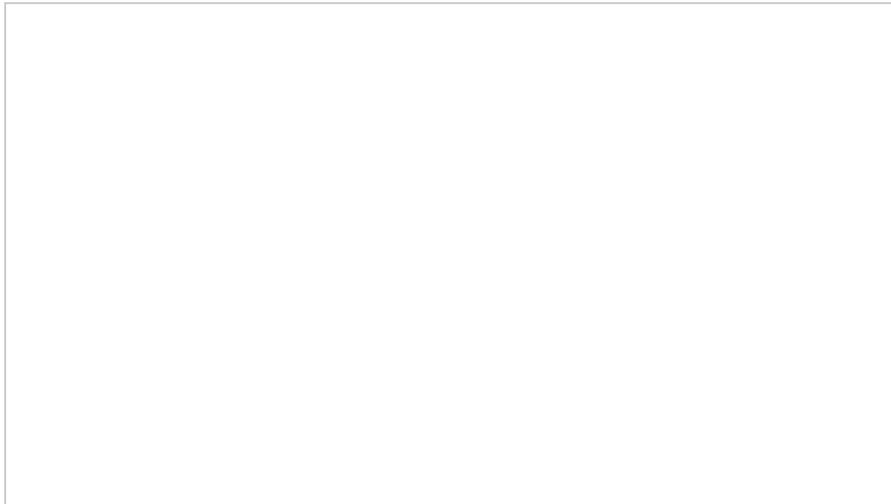

**10. How much would you trust or distrust this system to meet the following requirements for an election?**

|  | Strongly distrust | Distrust | Neither trust nor distrust | Trust | Strongly trust | Don't know |
|---|---|---|---|---|---|---|
| Accurately collect votes? | ◯ | ◯ | ◯ | ◯ | ◯ | ◯ |
| Accurately count votes? | ◯ | ◯ | ◯ | ◯ | ◯ | ◯ |
| Resist attempts to tamper with the election? | ◯ | ◯ | ◯ | ◯ | ◯ | ◯ |
| Protect the privacy of voters? | ◯ | ◯ | ◯ | ◯ | ◯ | ◯ |
| Ensure that all votes are kept anonymous? | ◯ | ◯ | ◯ | ◯ | ◯ | ◯ |
| Be accessible to all voters? | ◯ | ◯ | ◯ | ◯ | ◯ | ◯ |
| Be usable by all voters? | ◯ | ◯ | ◯ | ◯ | ◯ | ◯ |
| Return election results in a timely manner? | ◯ | ◯ | ◯ | ◯ | ◯ | ◯ |

**11. Briefly explain why you gave these ratings.**

**12. How would you rate your level of understanding of how each part of the voting system works?**

|  | Poor understanding | Fair understanding | Good understanding |
|---|---|---|---|
| Voter registration and obtaining ballot papers | ○ | ○ | ○ |
| Completing and casting a ballot | ○ | ○ | ○ |
| Counting the votes and producing a result | ○ | ○ | ○ |
| Supervision of counting by candidates (scrutineering) | ○ | ○ | ○ |
| Verification of election results by voters | ○ | ○ | ○ |

**13. To what extent do you think that your understanding of the system impact your level of trust in the system?**

○ Not at all

○ Very little

○ Somewhat

○ A great deal

**14. If there were errors in counting votes or someone attempted to tamper with the system, do you believe that it would be detected by election officials?**

○ Never

○ Rarely

○ Sometimes

○ Very Often

○ Always

**15. In a few words, please explain your answer to the previous question.**

## 5. The vVote Electronic Voting System

These questions relate to the vVote Electronic Voting System.

There is a short video below describing how the system works. Please watch the video, and then answer each of the questions.

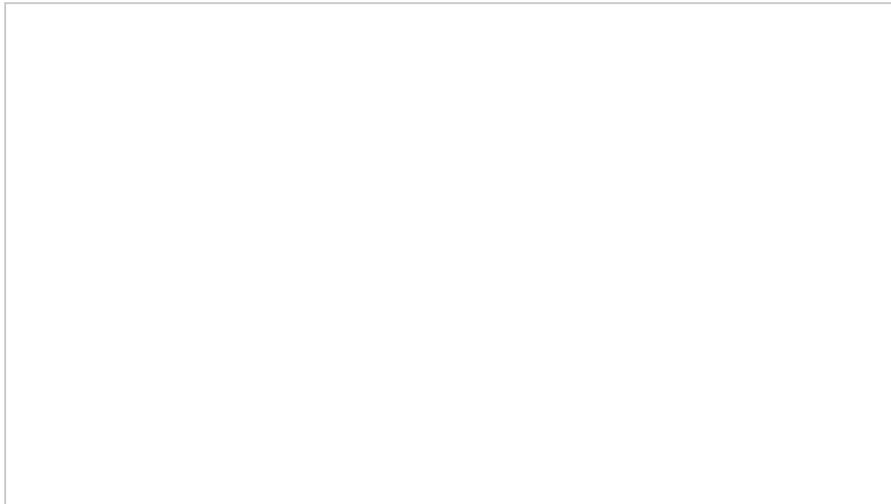

**16. How much would you trust or distrust this system to meet the following requirements for an election?**

|  | Strongly distrust | Distrust | Neither trust nor distrust | Trust | Strongly trust | Don't know |
|---|---|---|---|---|---|---|
| Accurately collect votes? | ○ | ○ | ○ | ○ | ○ | ○ |
| Accurately count votes? | ○ | ○ | ○ | ○ | ○ | ○ |
| Resist attempts to tamper with the election? | ○ | ○ | ○ | ○ | ○ | ○ |
| Protect the privacy of voters? | ○ | ○ | ○ | ○ | ○ | ○ |
| Ensure that all votes are kept anonymous? | ○ | ○ | ○ | ○ | ○ | ○ |
| Be accessible to all voters? | ○ | ○ | ○ | ○ | ○ | ○ |
| Be usable by all voters? | ○ | ○ | ○ | ○ | ○ | ○ |
| Return election results in a timely manner? | ○ | ○ | ○ | ○ | ○ | ○ |

**17. Briefly explain why you gave these ratings.**

**18. How would you rate your level of understanding of how each part of the voting system works?**

|  | Poor understanding | Fair understanding | Good understanding |
|---|:---:|:---:|:---:|
| Voter registration and obtaining ballot papers | ○ | ○ | ○ |
| Completing and casting a ballot | ○ | ○ | ○ |
| Counting the votes and producing a result | ○ | ○ | ○ |
| Supervision of counting by candidates (scrutineering) | ○ | ○ | ○ |
| Verification of election results by voters | ○ | ○ | ○ |

**19. To what extent do you think that your understanding of the system impact your level of trust in the system?**

○ Not at all

○ Very little

○ Somewhat

○ A great deal

**20. If there were errors in counting votes or someone attempted to tamper with the system, do you believe that it would be detected by election officials?**

○ Never

○ Rarely

○ Sometimes

○ Very Often

○ Always

**21. In a few words, please explain your answer to the previous question.**

# B    Qualitative Coding: Pre-set Codes

The pre-set codes using during the initial phase of qualitative coding were:

- Accessibility is good/poor
- Accuracy is good/poor
- Anonyminity is good/poor
- Privacy is good/poor
- Vulnerable/resistant to tampering
- Scrutineering is effective/not effective
- Voter verification is effective/not effective
- Counting speed is fast/slow